\documentclass[twocolumn]{aastex63}

\usepackage{rotating,amsmath,relsize}
\usepackage{blkarray,hhline,multirow}
\usepackage{txfonts,subfigure,colortbl,tcolorbox}
\definecolor{dkbest}  {rgb}{0.10,0.20,0.55}
\definecolor{dkred}    {rgb}{0.80,0.10,0.10}

\definecolor{dkgreen}{rgb}{.95, .40, .76}

\definecolor{maecolor}{rgb}{.50, .06, .90}

\newcommand{\itbf}[1]{\textit{\textbf{#1}}}
\newcommand{\bfsc}[1]{\bf{\scshape{#1}}}

\newcommand{\tablenotemarkeditp}[1]{\tablenotemark{\tiny \it(#1)}}
\newcommand{\tablenotemarkedit}[1]{\tablenotemark{\scriptsize \it(#1)}}

\newcommand{\tablenotetextedit}[2]{\it\tablenotetext{}{\\[-.45cm] \raggedright \scriptsize(#1)~#2}}

\newcommand{\citeedits}[1]{\citeauthor{#1} \citeyear{#1}}

\usepackage{empheq}

\begin{document}


\title{ \large
	 On Neural Architectures for Astronomical Time-series Classification\\
	{with Application to Variable Stars}
}

\author[0000-0002-3929-6668]{Sara Jamal}
	\affiliation{Department of Astronomy, University of California, Berkeley, CA 94720-3411, USA}
\author[0000-0002-7777-216X]{Joshua S. Bloom}
	\affiliation{Department of Astronomy, University of California, Berkeley, CA 94720-3411, USA}
	\affiliation{Lawrence Berkeley National Laboratory, 1 Cyclotron Road, MS 50B-4206, Berkeley, CA 94720, USA}

\received {XX}
\revised {XX}
\accepted{XX} 

\shorttitle {Neural architectures for variable star classification} 
\shortauthors {Jamal and Bloom}


\begin{abstract}{
Despite the utility of neural networks (NNs) for astronomical time-series classification, the proliferation of learning architectures applied to diverse datasets has thus far hampered a direct intercomparison of different approaches. 
Here we perform the first comprehensive study of variants of NN-based learning and inference for astronomical time-series, aiming to provide the community with an overview on relative performance and, hopefully, a set of best-in-class choices for practical implementations.  
In both supervised and self-supervised contexts, we study the effects of different time-series-compatible layer choices, namely the dilated temporal convolutional neural network (dTCNs), Long-Short Term Memory (LSTM) NNs, Gated Recurrent Units (GRUs) and temporal convolutional NNs (tCNNs). 
We also study the efficacy and performance of encoder-decoder (i.e., autoencoder) networks compared to direct classification networks, different pathways to include auxiliary (non-time-series) metadata, and different approaches to incorporate multi-passband data (i.e., multiple time-series per source). 
Performance---applied to a sample of 17,604 variable stars from the MACHO survey across 10 imbalanced classes---is measured in training convergence time, classification accuracy, reconstruction error, and generated latent variables. 
We find that networks with Recurrent NN (RNNs) generally outperform dTCNs and, in many scenarios, yield to similar accuracy as tCNNs. 
In learning time and memory requirements, convolution-based layers are more performant. We conclude by discussing the advantages and limitations of deep architectures for variable star classification, with a particular eye towards next-generation surveys such as LSST, 
{the Roman Space Telescope} 
and ZTF2. 

}\end{abstract}


\section{Introduction} \label{sec:intro}

Time-domain imaging surveys continue to expand access to the photometric phase space of cadence and depth/volume. 
Despite many upcoming projects (e.g., 
	the Rubin Observatory Legacy Survey of Space and Time -- LSST\footnote{\url{https://www.lsst.org/}}, \citeedits{ivezic_lsst_2019}; 
	\textit{Euclid}\footnote{\url{https://www.euclid-ec.org/, https://sci.esa.int/web/euclid}}, \citeedits{laureijs_euclid_2011}; 
	and the {Nancy Grace Roman Space Telescope} -- WFIRST\footnote{\url{https://wfirst.gsfc.nasa.gov/}}, 
	\citeedits{spergel_wide-field_2015}) 
being optimized for transient (supernovae, microlensing) discovery and characterization, the data from these surveys create unprecedented opportunities to broaden our understanding of stellar variability and stellar evolution as well as expand the use of variable stars (VSs) as probes. Detached eclipsing binaries, for example, provide direct measurements of distance (e.g., \citealt{1997vsar.conf..309P}) and fundamental stellar parameters \citep{torres_accurate_2010}, 
while pulsating VSs such as RR Lyrae, Miras and Cepheids due to their accurate period-luminosity relations are considered 
	useful tools to
	trace galactic structures \citep{kraft_galactic_1963, majaess_characteristics_2009,skowron_3D_2019},
	calibrate the cosmic distance ladder \citep{freedman_final_2001, huang_nir_2018, riess_milky_2018, riess_large_2019}  
	as well as standard candles to measure distances to their host-galaxies \citep{carretta_distances_2000, clementini_distance_2003, alves_review_2004}. 

Stellar variability, primarily manifest as changes in brightness and color,  arises from various physical mechanisms. Intrinsic variations arise as flares, rotation, pulsations, and/or violent outbursts due to thermonuclear processes occurring in the surface layers or deeper within. Extrinsic factors that may add to the observed variability include eclipses, relativistic Doppler beaming, mutual interaction in binary systems, and/or gravitational lensing.

The classification of VSs is based usually on brightness variations, typically, at visible wavelengths.
While far from standard, the General Catalog of Variable Stars (GCVS; \citealt{samus_general_2017}) maintains the taxonomy and nomenclature for VSs that distinguishes between subtypes of rotators, pulsators, eruptive variables, cataclysmic variables, eclipsing binaries in addition to other types  such as microlensing sources.  
In general, stellar variability is not expected to fall into a unique type of dynamical behavior, as objects may display a multitude of physical behavior, such as rotational modulation superimposed to pulsation in RCB-type stars, rotational modulation interjected by abrupt episodes of deep minima or a steep increase in brightness in BY\,Dra-type stars, or symbiotic systems with a M-type pulsating Mira star with an accreting white dwarf companion as the R\,Aqr star.
A comprehensive survey of VS classification is provided by \cite{eyer_variable_2008}.   

Large data volumes, the primary strength of massive surveys, also presents an acute challenge: how do we discover and characterize variable stars at scale in streaming, heterogeneous, and noisy time series? Human-free classification, part of a fully automated system to process and analyze survey data, requires the development and deployment of robust and reliable techniques from information-data technology to process large volumes of data and produce tractable and reproducible results.

Traditional machine-learning (ML) approaches for VS classification typically involve ``featurization'' to summarize and encode the raw observables into a set of informative descriptors exploited by a classifier to predict labels. 
Some popular features in the literature include 
	frequency-domain metrics derived from Lomb-Scargle periodograms \citep{scargle_studies_1998,lomb_least-squares_1976}, 
	statistical metrics (e.g., standard deviation, quantiles and skewness), 
	variability indices (e.g., the Stetson indices K and L; \citealt{stetson_automatic_1996}), 
	best model fit parameters
	as well as additional ``metadata'' information from external catalogs (e.g., colors, redshifts and parallaxes measurements).
VS classification using expert-engineered featurization and traditional ML algorithms have been extensively studied and used for decades 
\citep{debosscher_automated_2007, blomme_improved_2011, dubath_random_2011, 
	richards_machine-learned_2011, richards_construction_2012, 
	rimoldini_automated_2012, masci_automated_2014, 
	kim_package_2016, disanto_analysis_2016}.
Specialized libraries for features extracting from astronomical light curves have been made available by the community in open-source software packages such as 
	\texttt{cesium}        \citep{naul_cesium_2016}, 
	\texttt{FATS}          \citep{nun_fats_2015}, 
	\texttt{feets}         \citep{cabral_feets_2018}, 
	\texttt{sncosmo}       \citep{barbary_sncosmo_2016},
	\texttt{gatspy}        \citep{vanderplas_periodograms_2015,vanderplas_gatspy_2016},
	and \texttt{VARTOOLS}  \citep{hartman_vartools_2016}.
	
Ideally, featurization produces a uniform dimensionality reduced representation of the observations that adequately captures the intrinsic properties of the data needed for classification. 
However, a known issue in hand-coded  feature-based classification lies in the fact that the generated low-dimensional representation may overlook subtleties in higher-order systems and restrict such complexity into a set of low-level descriptors tailored for specific use-case applications. Furthermore, developing domain-specific features can be time-consuming, computationally expensive, highly dependent of expert-knowledge, and may show a strong dependency on survey characteristics.

Representation learning (RL) techniques offer an alternatively possibility to process raw observables without  traditional feature engineering. 
The benefit of fully-automating the classification task using RL lies in the ability to reach a higher level of abstraction and capture complex structures embedded in the data. 
Distinct approaches in RL to automate features extraction from astronomical time-series have already been introduced in a broad range of studies. Used techniques include 
	unsupervised learning algorithms \citep{armstrong_k2_2016}, 
	dimensionality reduction techniques,  
	data transformations \citep{johnston_variable_2020},
	autoencoders \citep{naul_recurrent_2018}
	or dictionary learning \citep{pieringer_algorithm_2019}.

In the recent years, VS classification using deep learning (DL)/neural architectures has been explored in several works that achieved satisfactory classification performance and thus demonstrated the ability of DL systems to learn stellar variability types from light-curves measurements and auxiliary metadata. 
Among mostly-used DL architectures, recurrent neural networks (RNNs) proved to be highly-performant for 
	periodic VS classification \citep{naul_recurrent_2018,tsang_deep_2019},
	supernovae (SNe) classification \citep{charnock_deep_2017} 
	and online transient events detection \citep{muthukrishna_rapid_2019, moller_supernnova_2019}.
Convolutional neural networks (CNNs) have also proven comparably performant, with a better training convergence time and lower memory allocation requirements in comparison to RNNs in various applications such as exoplanet transit detection \citep{shallue_identifying_2018, schanche_ML_2019, ansdell_scientific_2018}, SNe binary classification \citep{pasquet_pelican_2019} and Cepheid classification \citep{dekany_into_2019}.
A review on the recent contributions of DL techniques in SNe classification is given by \cite{ishida_machine_2019}.

To classify astronomical time-series, two main approaches have emerged, 
	either (1) design an automated system to encode the photometric observables into a set of features (semi- or self-supervised learning) that constitute the entry point to traditional algorithms (e.g., support-vector machines, NNs or tree-based classifiers),
	or (2) develop a DL architecture to find an optimal mapping between the photometric observables and the labels through a supervised learning scheme.

The objective of this paper is twofold: first is to provide an overview of the current ML/DL techniques for variable stars classification, and second to discuss through a test example the applicability of a selected set of NNs architectures and the importance of data representation for classification of stellar variables. This paper also investigates approaches for variable stars classification using multiband photometric data.
The paper is organized as follows. 
Section \ref{sec:archi_dl} first introduces the \textit{state-of-the-art} of ML/DL techniques for VS classification then proceeds to present our proposed architectures for classification. 
In Section \ref{sec:application}, we present variants of NNs architectures and discuss their performances through a test example using public data from the MACHO VS database.
The networks performances are evaluated in term of training convergence time, classification accuracy, properties of the generated latent representations and light-curves reconstruction.
Finally, we conclude in Section \ref{sec:conclusions}.



\begin{figure*}[hbt!]
	\centering
	\includegraphics[width=0.75\textwidth]{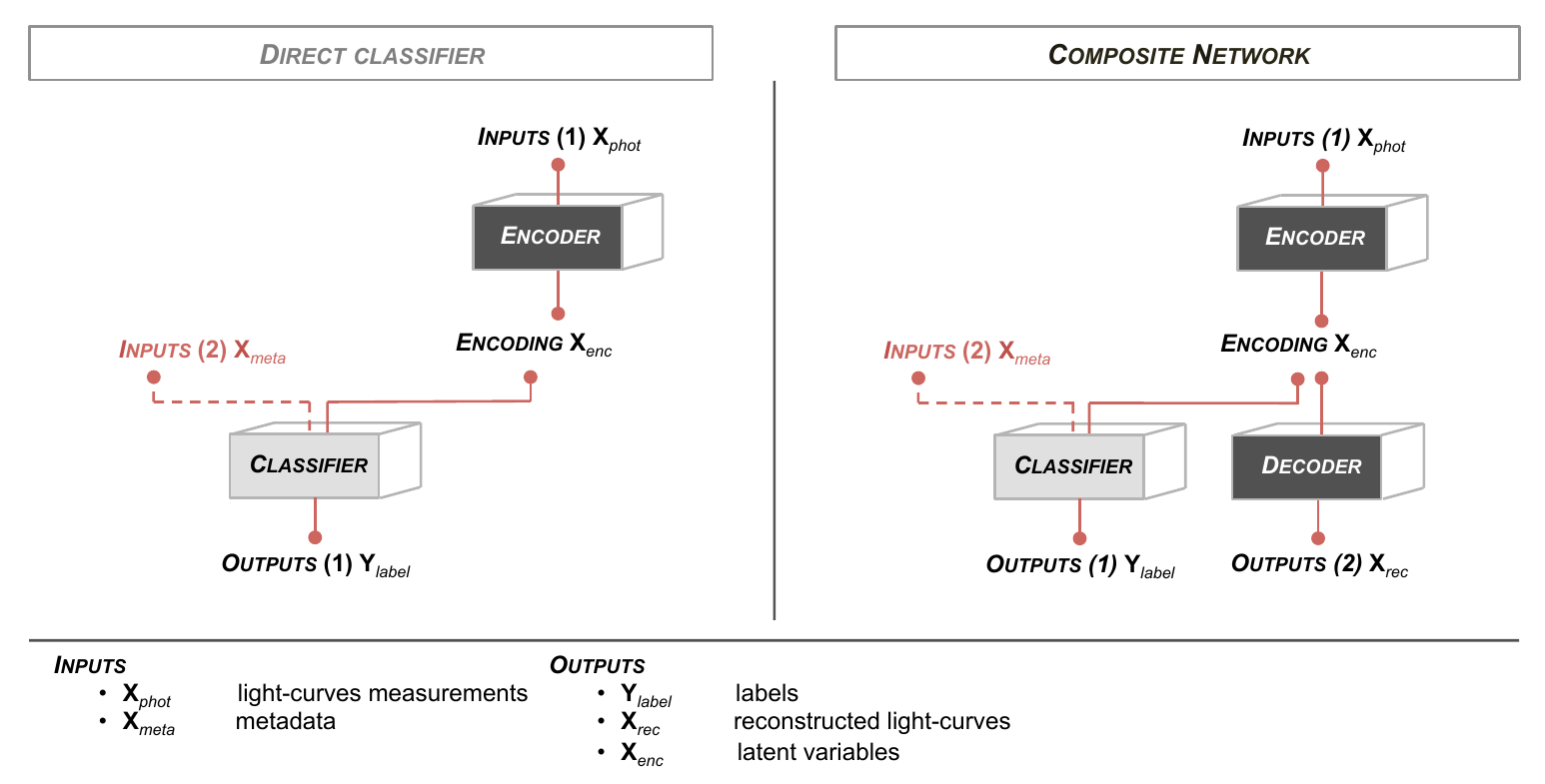}
	\caption{High-level architecture of the direct classifier (left) and composite (right) networks. In direct classifiers, the time-series data 
	($\mathbf{X}_{phot}$) and metadata ($\mathbf{X}_{meta}$) are combined through a series of neural layers and 
	concatenations, leading to classification predictions ($\mathbf{Y}_{label}$) on which the loss function is optimized. Composite networks use $\mathbf{X}_{phot}$ and a bottleneck/decoder to predict a reconstructed light curve ($\mathbf{X}_{rec}$). The bottleneck layer along with $\mathbf{X}_{enc}$ is also used to predict $\mathbf{Y}_{label}$. Both $\mathbf{X}_{rec}$ and $\mathbf{Y}_{label}$ are used in the loss function of composite networks.}
	\label{fig:structure_singleband}
\end{figure*}

\section{Deep learning architectures for VS classification} \label{sec:archi_dl}

\subsection{State-of-the-art} \label{subsec:stateart}

Classification of variable stars can be performed in feature space or in data space. 
The first approach consists of finding an optimal mapping between the labels, a vector $\itbf{Y}$, and an (encoded) feature set, a matrix $\itbf{X}_{enc}$ derived from the direct observables $\itbf{X}$; the second approach focuses on finding a direct mapping between 
{the observables $\itbf{X}$ and the labels $\itbf{Y}$.}

In feature space classification, traditional approaches require hand-coded feature extraction to compute a set of informative descriptors using domain knowledge.
For variable stars, the most discriminant features of stellar variability depends strongly on the specific class or subclass. 
For instance, the dominant modes of the pulsation mechanism can be well characterized in the frequency domain, found through Fourier decomposition or periodogram analysis.
In eclipsing binaries, the shape, duration, and relative phase of the eclipses in their light-curves inform the type of the interaction (contact binaries, detached binaries or semi-detached binaries) and their properties such as the mass and radii ratios.
Cataclysmic variables are described through the morphology of their light-curves at maximum light, the decline shape, the duration of the burst event, the event occurrence in time distinguishing between recurrent outbursts and final explosions, the quiescent state of the star post-event and observed spectral features during outburst. 
For eruptive variables, their light-curves and spectra show distinctive variations as sudden brightening or dimming episodes over extended periods of time due to flares and violent processes taking place in the corona or the chromosphere of these stars.

In traditional approaches, feature engineering relies on domain knowledge while feature learning automates the extraction procedure from the data using   
	dimensionality reduction techniques,
	self-supervised networks (e.g., autoencoders), 
	dictionary learning, 
	or unsupervised algorithms.
The extracted features constitute a discretized set of encoded information exploited by feature-based classifiers to predict labels.
Among notable references, the work by \cite{armstrong_k2_2016} exploits the unsupervised learning algorithm SOMs (\textit{Self-Organizing Maps}) to encode photometric light-curves into a set of features processed, along with additional descriptors, by the tree-based classifier, the random forest (RF, \citealt{breiman_random_2001}), to predict labels for periodic VSs.
The work of \cite{naul_recurrent_2018} presents a bidirectional RNN autoencoder to discretize the photometric observables into a set of latent variables exploited, along with ancillary metadata, by a RF classifier to predict labels for periodic variables.

For classification in data space, common techniques exploit DL to identify embedded characteristics in the data and find a direct mapping between the input observables and the output labels.
Applications using DL techniques for astronomical time-series classification include 
	(1) SNe classification using RNN architectures to process multiband photometric data and auxiliary metadata (e.g., redshift measurements) \citep{charnock_deep_2017, muthukrishna_rapid_2019, moller_supernnova_2019},
	(2) online transient events detection using RNN architectures to compute timely class-predictions for early-observed light-curves in order to forecast potential pre-SN outbursts and prompt follow-up procedures before the event reaches its maximum light \citep{muthukrishna_rapid_2019, moller_supernnova_2019},
	and (3) exoplanetary transit detection using a composite convolutional networks that analyzes the full light curve and the eclipses to discern between planetary transits and stellar eclipsing binaries \citep{shallue_identifying_2018, ansdell_scientific_2018, schanche_ML_2019}.

More recently, composite architectures has been introduced for astronomical time-series classification in the form of NNs composed of different submodules designed for specific tasks.
Notable references include the work by \cite{pasquet_pelican_2019} for SNe classification where the authors propose a DL architecture (\texttt{PELICAN}) with three modules: an autoencoder branch to generate the embeddings at the bottleneck level by optimal reconstruction, a classifier for label predictions and a contrastive module designed to reduce the discrepancy between the Test and Train sets.
Binary classification of type Ia SNe is performed using multiband photometric data and ancillary metadata (redshift measurements of the host galaxies).
The work by \cite{tsang_deep_2019} proposes a similar approach for periodic VS classification, initially derived from the DAGMM (\textit{Deep Autoencoding Gaussian Mixture Model}) network in \cite{zong_deep_2018}. The authors propose a network composed of two modules: an autoencoder branched out to a classifier module at the bottleneck level. 
\\\\

\subsection{Architectures}\label{subsec:architectures}

This section presents selected architectures for VS classification.
We distinguish between  two types of architectures: direct classifiers and composite networks, as shown in Figure \ref{fig:structure_singleband}.
Both architectures are composed of an encoder and classifier module. Composite networks are supplemented with a decoder connected at the bottleneck level.
The encoder-decoder combination (i.e., autoencoder) aims to learn a latent representation $\mathbf{X}_{enc}$ of the input photometric data $\mathbf{X}_{phot}$ by optimal reconstruction, while the classifier maps the encodings to the labels $\mathbf{Y}_{label}$.
By connecting the autoencoder to the classifier module, the system ideally learns a latent representation (ie., compressed summary) that is closely correlated to different stellar variability types in the training data.
The high-level design in Figure \ref{fig:structure_singleband} is adaptable: modules can be adjusted to the application and data as needed.
In the current work, the decoder module corresponds to a mirror-image of the encoder, which is a common choice for autoencoder architectures. 
We evaluate the encoder for different types of NNs such as recurrent neural networks, convolutional neural networks and their variants, while the classifier module is set to a 2-layer MLP (\textit{Multi-Layer Perceptron}) for all networks to predict labels.
A short description of the neural networks used in this work is provided in Appendix \ref{sec:nets_intro}.

The majority of applications for VS classification using DL exploit auxiliary features that complement the photometric observables such as redshift measurements, detectable frequencies for periodic variables, amplitudes and colors.
The photometric information in the light-curves constitutes a fundamental description of the evolutionary state of the star over time but does not contain the entirely of the available information; unless the light curve of a certain class is demonstrably different than other classes, additional metadata can be expected to improve classification accuracy.
Typically, classification tasks require an upstream phase of data preparation. However, preprocessing transformations applied to the photometric observables may inadvertently remove discriminating features linked to the stellar variability types, thus altering the quality of the information necessary for classification.
For instance, light-curves of periodic variables are preprocessed through phase-folding and data normalization. 
The phase-folding procedure transforms the periodic data into a compact representation in phase of one to two cycles by stacking multiple observations, thus removing the periodicity information over time.
On the other hand, data normalization via minmax normalization yields to rescaled magnitude measurements, amplitudes and errors. As a direct result, similarly shaped light-curves with different peak-to-peak amplitudes cannot be distinguished from one another.

In the current work, we investigate the importance of the metadata in two scenarios in which the network classifies the data solely based on (preprocessed) light-curves without auxiliary metadata as opposed to supplementing metadata to the system as a secondary input for label predictions.

In VS classification, multiband photometry can be processed either
	by transforming the photometric passband measurements into a single entity fed to the network
	or by jointly processing individual encodings from each passband measurements for label predictions.
A simplified representation of the aforementioned {approaches}, identified in this work as the \textit{merged} and \textit{hybrid} {approaches}, is provided in the Appendix Figure \ref{fig:structure_multiband}.
In the \textit{merged} approach, multi-passband measurements are combined into a unique observable $\mathbf{X}_{phot, merged}$ processed by the encoder module to compute the latent representation $\mathbf{X}_{enc}$.
%
{
	The preprocessed light-curves per band can be combined through distinct representations as the variants 
	presented in the Appendix \ref{sec:structures_multiband}.
}
%
Whereas, the \textit{hybrid} approach independently encodes the multi-passband measurements into individual features combined at a latter time into a compact encoded representation $\mathbf{X}_{enc}$. 
In both scenarios, the classifier module exploits the generated encodings $\mathbf{X}_{enc}$, along with metadata $\mathbf{X}_{meta}$, for label predictions.
Composite networks differ from the direct classifiers by the addition of a decoder module connected at the bottleneck level to the encoder to generate the embeddings by optimal reconstruction.


\section{Application} \label{sec:application}

This section presents a set of NN-based architectures for VS classification.
We discuss through a test example the performance of these networks in terms of training convergence time, label predictions, reconstruction and generated latent representations.

\subsection{Data}\label{subsec:data}

As an exemplar, for all architectures, we use public photometric data and labels from the MACHO survey \citep{alcock_macho_1996_a}. 
The MACHO project carried-out a long-term photometric monitoring of stars in the Magellanic Clouds and the galactic bulge from 1992 to 1999 in search for rare microlensing events, observable in theory if the dark matter is composed of massive compact halo objects (MACHOs).
The large collection of data from the survey has allowed over the years a rare insight into a variety of stellar populations such RR Lyrae, Cepheids, LPVs (\textit{Long-Period Variables}) and eclipsing binaries \citep{cook_variable_1995}. 
In MACHO, the LPVs are categorized into four subtypes -- namely the four Wood sequences A, B, C and D -- referring to the parallel sequences identified from the period-luminosity relation of these red variables \citep{wood_macho_1999}.
Photometric data in MACHO consists of magnitude measurements in two photometric filters (the MACHO red and blue filters), the associated 1-$\sigma$ error measurements and the observation epochs expressed in MJD. 
We exploit the multiband photometric data of the public MACHO VS database\footnote{\url{http://macho.nci.org.au/}} to test our selected NNs architectures, and we perform further checks on the data. 
In particular, we retain the confirmed set of eclipsing binaries with corrected periods from \cite{derekas_eclipsing_2007}.
The full process yields 17604 periodic variables from the initial count of 24k periodic VS in the MACHO database to test our architectures: 1806 Cepheid variables, 9163 RR Lyrae, 2965 long-period variables (LPVs) and 3670 eclipsing binaries (cf.\ Table \ref{table:macho_tab}).

The majority of ML/DL open-source software exploits input data in the form of a fixed-size input tensors and few DL architectures are able to process observables with different lengths, e.g., using generator functions on an iterable list of input data.
In our approach, the architecture of the direct classifier can process fixed-size data in a batch mode as well as a list of observables with different lengths using generator functions.
Composite networks, however, require fixed-size inputs to comply with the implementation specifications of the decoder module. 
To meet such requirements, we reduce the data into a fixed-size format.
Typically, data reduction can be achieved using padding, rebinning, interpolation (e.g., splines or polynomials) or model fit and prediction.  
More recently, model prediction using Gaussian Processes (GPs) have been used used on astronomical time-series 
\citep{ambikasaran_fast_2016, foreman-mackey_fast_2017, pruzhinskaya_anomaly_2019, boone_avocado_2019}.
To describe the stellar variability in the MACHO periodic light-curves, a GP model with a quasi-periodic covariance function is fitted to each object using the open source code \cite{exoplanet:exoplanet_zenodo}. 
The selected GP model is a mixture of stochastically-driven damped oscillators (SHOs), briefly discussed in the Appendix \ref{sec:gp_celerite}.
{
	For each object, a GP model is fitted. Using a MAP (Maximum {\textit{a posteriori}}) estimate, 
}
fixed-size light-curves are generated by model prediction on a reduced time-frame sampled within the range of the observed epochs.
{ 
	The GP predictions mean and error correspond to the reduced photometric data exploited in the rest of this work.}
%
%
At a latter stage, the reduced photometric light-curves are normalized and phase-folded to span over 2 cycles. 
{
	Phase-folding of periodic light-curves allows a better visualization of the cyclic behavior of the variables.
	For instance, short-period variables as RR\,Lyrae pulsate over timescales ranging from $\sim$0.3 to 1 day. 
	To fully observe the pulsation profile over a complete cycle, the observation cadence has to exceed the variability 
	frequency to cover a full cycle. 
	In practice, given realistic survey cadences, any one individual cycle will be sparsely observed if at all.
	By combining the observations through period-folding, however, even high-frequency variables such as RR\,Lyrae stars can have dense phase coverage.
	Still, properties of long-period variables that evolves over longer timescales, such as the Mira-type stars with pulsation 
	periods ranging from $\sim$80 to 1000 days, can be distinguishable in the initial time-frame without phase-folding.
	In the current application, we exploit a dataset of periodic VSs with light-curves from short-period pulsators, 
	long-period pulsators and eclipsing binaries and apply the phase-folding procedure to all variables.
	Observed shortcoming from the preprocessing are discussed in the result section (cf. Section \ref{subsec:perfostudy}).
	In the majority of ML applications, normalization is applied to the data to improve the numerical stability and reduce the training 
	time. 
	%
	In this work, the inputs $\textbf{X}_{phot}$ to the networks correspond to preprocessed light-curves, obtained after data reduction, 
	phase-folding and normalization.
	The metadata  $\textbf{X}_{meta}$ consists of the amplitudes, averaged magnitudes and colors extracted from the raw data in 
	addition to the primary periods.
	}
%

\renewcommand{\arraystretch}{1.2}
\begin{table}[htp]
	\footnotesize        
	\caption{Selected dataset from the MACHO VS database.}
	\label{table:macho_tab}
	\begin{center}{   
	\begin{tabular}{l l} 
	        \hline\hline
			\sc{Class Labels}	 & \sc{\# in Class}	\\  
	       \hline
	       Cepheids FU	    	& 1143 	\\ 	
	       Cepheids FO		& 663	\\ 	
	       RR Lyrae (type $ab$)	& 7147 	\\ 	
	       RR Lyrae (type $c$)	& 1716 	\\ 	
	       RR Lyrae (type $e$)	& 300 	\\ 	
	       LPV (Wood seq. A)	& 310  	\\	
	       LPV (Wood seq. B)	& 799  	\\	
	       LPV (Wood seq. C)	& 1100 	\\	
	       LPV (Wood seq. D)	& 756  	\\	
	       Eclipsing binaries	& 3670\tablenotemarkedit{a} \\	
	       \hline
	\end{tabular}
	\tablenotetextedit{a}{\centering confirmed binaries in \cite{derekas_eclipsing_2007}}
	}\end{center}
\end{table}



\subsection{Design and implementation}\label{subsec:desing}
We experiment with a variety of NN architectures and discuss the absolute and relative performances.
Classification using multiband photometry is evaluated using the approaches introduced in Section \ref{subsec:architectures}, i.e. the \textit{merged} and \textit{hybrid} approaches.
%
{
	The \textit{merged} approach exploits the second representation introduced in Appendix \ref{sec:structures_multiband} combining the individual measurements of preprocessed light-curves into a 2-d tabular format.
	In the current application, normalization and phase-folding are applied to each band measurements independently, 
	prior to generating the merged representation.
	Error measurements are used solely as weights in the autoencoder loss function. Incorporating error mesurements within the network can be part of further development of such networks.
}
 We investigate {as well} the use of auxiliary metadata as a secondary input on the classification accuracy for the direct classifier and composite networks via two scenarios, in which the network classifies the data using the information from preprocessed light-curves without metadata as opposed to supplementing auxiliary metadata as a secondary input for label predictions.
 Data entries for the different use case scenarios are summarized on the Appendix Table \ref{table:networks_config0}.
The autoencoder is evaluated for different NN: RNN with LSTM cells, RNN with GRU cells, temporal CNN (tCNN) and dilated TCN (dTCN). 
	
For VS classification, the objective of NNs is to empirically determine a mapping between 
the inputs, 
	the photometric observables $\mathbf{X}_{phot} = [ \itbf{x}_{1}, \cdots, \itbf{x}_{N_t} ]^{T}$ 
	and auxiliary metadata $\mathbf{X}_{meta}= [ \itbf{d}_{1}, \cdots, \itbf{d}_{N_t}]^{T}$, 
and the output labels $\mathbf{Y}_{label}=[ y_{1}, \cdots, y_{N_t} ]^{T}$, where $N_t$ designates the total number of objects. 
In the current application, the datavectors $(\itbf{x}_{i})_{i: 1\rightarrow N_t}$ refer to the reduced photometric measurements -- magnitudes, observation epochs and error measurements-- spanning over $N_p$ datapoints for the $i^{th}$ object 
and the metadata $(\itbf{d}_{i})_{i: 1\rightarrow N_t}$ is composed of $N_f$ features such as the periods, the amplitudes, the averaged magnitudes and the colors.
The elements $(y_{i})_{i: 1\rightarrow N_t}$ refer to scalar values encoding the labels. For categorical classification, the labels are encoded into codewords transcribing the membership of the object to a variability class $\{j\}_{j:1\rightarrow N_C}$ with $N_C$ the total number of classes. A standard approach in categorical classification consists in transcribing class memberships into binary codewords $\{0,1\}$.
In online transient event detection applications, a similar methodology is used to map the observables into the output space.  
Class predictions are computed on the pixel level, which corresponds to a prediction vector $(\itbf{y}_{i})_{i: 1\rightarrow N_t}$ allowing to monitor the evolution over time of the label predictions and forecast potential pre-SN outbursts and trigger follow-up observations within a rapid decision time before the event reaches its maximum light.
In contrast, the \textit{static}-type prediction approach adopted for VS classification consists in associating the static label predictions to fully observed light-curves. The approach is used in this study aiming to map the observables $\{\mathbf{X}_{phot}, \mathbf{X}_{meta}\}$ into scalar values $\mathbf{Y}_{labels}$. 
Nonetheless, our architectures can be adapted to perform online predictions on the pixel level by adjusting the inputs-outputs format.

Detailed representations of the proposed architectures are shown in the Appendix Figures \ref{fig:archidet_RNN} to \ref{fig:archidet_dTCN}.
After preprocessing (i.e., data reduction, phase-folding and data normalization of periodic light-curves), the networks proceed as follows.
In the RNN architectures, the first part of the structure following the input layer is a stack of RNN layers that generates a sequence data of fixed length. 
The direct classifiers and composite networks differ in the last layer of the encoder module in which composite nets are supplemented with a fully-connected layer -- a dense layer with a linear activation function -- to transform the sequence data from the last RNN layer into an encoded representation $\mathbf{X}_{enc}$ of a fixed size $N_{enc}$. 
If auxiliary metadata $\mathbf{X}_{meta}$ is supplemented as a secondary input, the classifier exploits the augmented features to predict the labels $\mathbf{Y}_{label}$. 
The classifier module corresponds to a MLP of two dense layers with a rectified linear unit (ReLu) and softmax activation functions.
In composite networks, the decoder transforms the encoded variables into reconstructed representations similar to \cite{naul_recurrent_2018}. The decoder proceeds by duplicating the embeddings in a number of times equivalent to the length of the input data, joins the epochs and passbands information and feeds the augmented embeddings into a stack of RNNs. 
The last component of the decoder module consists of a fully-connected layer that generates the final sequence data of the reconstructed light-curves.
To prevent overfitting, regularization is added to the model through the dropout method \citep{srivastava_dropout_2014} that randomly removes units during training. 

Following the input layer, the tCNN architecture is composed of a series of convolutional layers with a number of filters and fixed kernel sizes for convolutions. 
The activation function for each convolutional layer is set to the hyperbolic tangent function and dropout is used for regularization.
In the encoder module, the output of the last convolutional layer is flattened to generate a vector output. The latter corresponds to the final encodings for direct classifiers, while composite networks exploits an additional fully-connected layer to generate fixed-size embeddings.
Similar to the RNN architectures, the encodings are fed to the classifier module, along with metadata when applicable.
In the decoder, the data is processed through a reshaping substructure, stacks of transposed convolutional layers and a final fully-connected layer to generate the outputs.
The reshaping substructure emulates the RNN decoder in order to reframe the encodings into a fixed-length format matching the input datapoints.

The dTCN architecture closely follows the TCN architecture \citep{lea_temporal_2017} initially derived from the \texttt{Wavenet} architecture \citep{oord_wavenet_2016}.
Our design adds dropout functions after the causal convolutions to prevent overfitting.
Following the input layer, the first component of the dTCN consists of a causal convolutional layer connected to interconnected stacks of residual blocks with dilated convolutions and gated activation units.
Following the \texttt{Wavenet} design, a ReLu activation function and two convolutional layers are used after the stacks.
The encoder outputs are transformed into a vector format and a fully-connected layer is supplemented in composite networks.
The generated encodings are then fed into the classifier module and augmented with metadata when applicable.
The decoder module is composed of a reshaping substructure, a TCN unit mirrored to the encoder module and a final fully-connected layer.

The NNs are tested with a different set of hyperparameters (cf.\ Appendix Table \ref{table:networks_config1}). The number of parameters per model are provided in the Appendix Table \ref{table:network_sizes}. 
All models are trained using the \texttt{Adam} optimization algorithm \citep{kingma_adam_2017} with a learning rate of 5$\times10^{-4}$, a fixed batch size per gradient-update in optimization, a dropout fraction and an early-stopping procedure to prevent the networks from overfitting.
%
{ 
	For all networks, a validation-based early-stopping procedure interrupts the training when an optimal solution is 
	found (i.e., convergence) before reaching the maximum number of training epochs.
	The standard approach monitors the evolution of a scoring metric, typically the validation loss, over a period lag and 
	terminates the training if the error estimated at the current time exceeds the last-verified value 
	implying a large variance in the overfitting regime.
	}
%
An improved search for the hyperparameters space can be performed as an upstream stage of the final classification procedure. However, we choose in this study to empirically evaluate different sets of hyperparameter configurations, discuss the networks performances, and identify the best-performing models.
In training, the autoencoder (i.e., encoder-decoder) aims to minimize the reconstruction loss, 
{the weighted MAE (\textit{Mean-Absolute Error}) } function, and the classifier branch is set to minimize the categorical cross-entropy loss. 
%
{
	Networks are trained to minimize the total loss $\textnormal{L}_{tot}$.
	\begin{equation}
		\begin{split}
		\textnormal{direct classifier}
			\quad & \textnormal{L}_{tot} = \textnormal{L}_{ec},\\
		\textnormal{composite network}
			\quad &\textnormal{L}_{tot} = w_{ec}\,\textnormal{L}_{ec}  + w_{ed}\,\textnormal{L}_{ed},
	\end{split}	
	\label{eq:total_loss}
	\end{equation}
	where, $\{w_{ec}, w_{ed}\}$ respectively refer to the loss weights of the encoder-classifier and the encoder-decoder branches. 
	In the current work, the individual losses $\textnormal{L}_{ec}$ and $\textnormal{L}_{ed}$ correspond to the following.
	\begin{align}
		&\textnormal{L}_{ec} =
			 \frac{1}{N_t} \sum_{i=1}^{N_t} \bigg(
			 		  -\sum_{j=1}^{N_C} y_{true, i}^{(j)} \; \log{\Big(y_{pred, i}^{(j)} \Big)} \bigg),\\
		&\textnormal{L}_{ed}=
			 \frac{1}{N_t} \sum_{i=1}^{N_t} \bigg( 
			 		\frac{1}{N_p} \sum_{k=1}^{N_p} w_{i}^{(k)} \Big|x_{phot, i}^{(k)} - x_{rec, i}^{(k)}\Big| \bigg),   
	\label{eq:total_loss_branch}
	\end{align}
	where, of the $i^{th}$ object, $\itbf{w}_{i}$ corresponds to the sample weights of the autoencoder branch 
	computed using the inverse of the error measurements $\sigma_{i}$.  
	The losses, averaged across the sample of $N_t$ objects, are computed by,  
		on one hand, averaging the differences between the input photometric light-curves $\mathbf{X}_{phot}$ 
		and the decoder reconstructions $\mathbf{X}_{rec}$ across the $N_p$ datapoints, 
		and, on the other hand, computing the crossentropy between the true labels $\mathbf{Y}_{true}$ 
			and the classifier predictions $\mathbf{Y}_{pred}$ over the $N_C$ classes. 
	%
	Using the expression of the total loss in Eq \ref{eq:total_loss}, tested NNs are trained to minimize at each epoch 
	the weighted sum of the individual losses with weights $\{w_{ec}, w_{ed}\}$ chosen equal to unity, as to depict a 
	similar contribution from the individual branches into the total loss.
	The weighting scheme and cost functions can be revised for different applications and data types.
}
%
Best-performing models are identified from minimum loss obtained on a Test set, i.e. the subset of data neither used in training nor validation. 
Our tests used up to 2$-$4 cores on a CPU model \texttt{Intel Xeon E5-2643v3} on the UC Berkeley Savio Linux cluster.
The NN architectures are implemented in the \texttt{keras} \citep{chollet2015keras} and \texttt{Tensorflow} \citep{tensorflow2015_whitepaper, abadi_tensorflow_2016} programming frameworks, and the RNN autoencoder branch is partly adapted from the architecture in \cite{naul_code_data_2017_zenodo}.
To facilitate reuse and reproducibility, our benchmarking codebase is provided in an open source 
{repository}\footnote{\url{https://github.com/sarajamal57/deepnets_vs}}. 


\subsection{Performance study}\label{subsec:perfostudy}

This section presents the results obtained on the public MACHO VS dataset.
Computations are performed through a classical training-validation-test scheme with 80\% of data for training-validation and 20\% of unseen objects in the test prediction phase.
Performance metrics are computed for all models, with a particular emphasis on the performances reached on the Test set.
Metrics include the system total loss, the classification accuracy as well as averaged precision, recall and F1-score (cf.\ Appendix \ref{sec:metrics_clf}). 
The total loss corresponds to the classification loss (categorical cross-entropy) evaluated on the encoder-classifier branch, supplemented in composite networks with the weighted MAE evaluated on the encoder-decoder branch.
The classification accuracy is obtained through a direct comparison between the true labels $\mathbf{Y}_{true}$ and the network predictions $\mathbf{Y}_{pred}$. 
We also discuss the classification performances within three main types of stellar variability: short-period pulsators including the RR Lyrae and Cepheids, long-period pulsators (LPVs) and eclipsing binaries.
The autoencoder performances are assessed through the quality of the reconstructed light-curves $\mathbf{X}_{rec}$, and the embeddings $\mathbf{X}_{enc}$ are projected onto a 3-d representation using a data reduction algorithm. 
The degree of separation (or lack thereof) in the reduced latent space is discussed.

\subsubsection{Training convergence time} \label{subsubsec:perfo_time}

The training convergence time is reported for all networks in the Appendix Table \ref{table:cputime_training}.
The network type, size and hyperparameters influence on the total time required to reach convergence, with an average training time for RNNs (LSTM and GRU cells) scaling higher in comparison with the convolutional networks (tCNNs and dTCNs) due to higher memory requirements for RNNs that entail a longer time in training.
As expected, increasing the network size (i.e., number of parameters) correlates with a longer time in training. 
For instance, direct classifiers without metadata ($c_{F}$) corresponding to the hyperparameter set configuration $\it(6)$ (the largest models for LSTM, GRU and dTCN) converges in training after approximately
	4$-$5 hours for RNNs, 
	{22} min for dTCNs  			
	and {$\sim$6} minutes for tCNNs 
using one photometric band on CPU.
Networks processing multiband data converge at a slower rate due to larger data entries.
After training, the prediction step is extremely fast: 0.5$-$3 ms per object for tCNNs, 1$-$10 ms per object for dTCNs and up to 3$-$20 ms per object for the RNNs on a CPU.

To track the evolution of the total loss and the accuracy during the training and validation stage, the Appendix Figures \ref{fig:loss_acc_display_lstm_df}$-$\ref{fig:loss_acc_display_lstm_dfmeta} report the performances of the LSTM composite networks $d_{F}$ and $d_{F,meta}$ {using wMAE loss on the autoencoder branch and the categorical cross-entropy on the classifier branch.}
Over the training epochs, the loss function decreases and converges asymptotically to a constant value whereas the accuracy increases and stabilizes when the system reaches convergence.
The system converges to reach at best {$\sim$73\%} for the best-performing LSTM $d_{F}$ without metadata. 
By supplementing the metadata as secondary input, the accuracy increases up to $\sim91\%$. 
%
{
	During the validation step, the loss values decrease and moderately exceed the training losses which reinforces 
	the ability of the networks to generalize the learned mapping from the training to the unseen validation data, 
	as a lack of generalization would correspond to a larger gap between the training and validation losses.
	Generally, overfitting is detectable from a high variance in the model and a divergence in the validation loss function 
	across training epochs despite the continuing decrease of the training loss.
	To prevent from such limitation, our models are trained using regularization through dropout functions in addition to a 
	validation-based early-stopping procedure.
	Early-stopping monitors the validation loss values and interrupts the training before reaching the total number of training epochs
	if an optimal solution is found (i.e., convergence) or if the system oversees an high increase in the validation losses indicating 
	a large variance.
} 

{
	We experimented with NNs classification using raw (i.e., without normalization) phase-folded light-curves. 
	Models converge at a slower rate in an unstable pattern across the training epochs.
In what follows, we focus our analysis on the results obtained using preprocessed (i.e., phase-folded and normalized) 
	light-curves and associated metadata for MACHO periodic variables stars.
}

\renewcommand{\arraystretch}{1.2}
\begin{table}[htp]
	\footnotesize
	\caption{
		Classification accuracy obtained on the Test set for the best-performing networks (see text for a description) across 
		three different datasets (B-band only [top], and two variants of the combination of R- and B-bands [middle and 
		bottom]). 
		Best performances among all nets are highlighted.
		} 
	\label{table:perfoBest}
	{\begin{center}{ 
 	\begin{tabular}{ c | c | cccc}
		\hline\hline 
			\multirow{5}{*}{\centering{\rotatebox[origin=c]{90}{\itbf{\color{blue!35!black} MACHO - B$_{band}$ }}}}
			& {\sc{Id net}}
				&\multicolumn{1}{c}{LSTM} 
				&\multicolumn{1}{c}{GRU} 
				&\multicolumn{1}{c}{tCNN}
				&\multicolumn{1}{c}{dTCN} 
				\\
		\hhline{~-----} 
		&$c_{F}$            
			& 0.749		    
			& 0.781		    
			& 0.732		    
			& 0.675		    
			\\
		\hhline{~-----} 
		&$c_{F,meta}$       
			& \bf 0.916	    
			& 0.907	        
			& 0.887		    
			& 0.786	    	
			\\
		\hhline{~-----} 
		&$d_{F}$            
			& 0.730		    
			& 0.739		    
			& 0.701		    
			& 0.689		    
			\\
		\hhline{~-----} 
		&$d_{F,meta}$       
			&  \bf 0.905	
			& 0.886         
			& 0.900	        
			& 0.802		    
			\\
		\hline  
	\end{tabular}
	}\end{center}}
	{\begin{center}{	
	\begin{tabular}{c | c | cccc}
		\hline\hline 	
			\multirow{5}{*}{\centering{\rotatebox[origin=c]{90}{\itbf{\color{cyan!50!black}MACHO - RB$_{merged}$}}}} 
			& {\sc{Id net}}
				&\multicolumn{1}{c}{LSTM} 
				&\multicolumn{1}{c}{GRU} 
				&\multicolumn{1}{c}{tCNN}
				&\multicolumn{1}{c}{dTCN} 
				\\
		\hhline{~-----} 
		&$c_{F}$            
			& 0.737		    
			& 0.780		    
			& 0.722		    
			& 0.667	    	
			\\
		\hhline{~-----} 
		&$c_{F,meta}$       
			& 0.890		    
			& \bf 0.910	    
			& 0.815		    
			& 0.747		    
			\\
		\hhline{~-----} 
		&$d_{F}$            
			& 0.726		    
			& 0.738		    
			& 0.691		    
			& 0.686		    
			\\
		\hhline{~-----} 
		&$d_{F,meta}$       
			& 0.906	        
			& 0.883		    
			& \bf 0.912	    
			& 0.814		    
			\\
		\hline 	
	\end{tabular}
	}\end{center}}
	{\begin{center}{
	\begin{tabular}{c | c | cccc}
		 \hline\hline 					
			\multirow{5}{*}{\centering{\rotatebox[origin=c]{90}{\itbf{\color{violet!60!black}MACHO - RB$_{hybrid}$}}}}
			&{ \sc{Id net}}	
				&\multicolumn{1}{c}{LSTM} 
				&\multicolumn{1}{c}{GRU} 
				&\multicolumn{1}{c}{tCNN}
				&\multicolumn{1}{c}{dTCN} 
			\\
		\hhline{~-----} 
		&$c_{F}$            
			& 0.776	        
			& 0.789	        
			& 0.744	        
			& 0.696	        
			\\
		\hhline{~-----} 
		&$c_{F,meta}$       
			&  \bf 0.917	
			& 0.905	        
			& 0.845		    
			& 0.768	    	
			\\
		\hhline{~-----} 
		&$d_{F}$            
			& 0.748		    
			& 0.749		    
			& 0.706		    
			& 0.726		    
			\\
		\hhline{~-----} 
		&$d_{F,meta}$       
			& \bf 0.905	    
			& 0.880	    	
			& 0.904	        
			& 0.818		    
			\\
		\hline 	
	\end{tabular}
	}\end{center}}	
\end{table}

\subsubsection{Labels predictions} \label{subsubsec:perfo_clf}

Total loss and classification accuracy for all trained models are reported on the Appendix Tables \ref{table:totloss_perfoAll}$-$\ref{table:classif_perfoAll}.
We identify the best-performing models for each architecture type (LSTM, GRU, tCNN and dTCN) from minimum loss on the Test set, {neither used in training nor validation}.
As previously mentioned, the total loss corresponds to the loss evaluated on the encoder-classifier branch (categorical cross-entropy) supplemented in composite networks to the loss  the encoder-decoder branch (weighted MAE).
%
{
	For RNNs, the losses computed on the Validation set are close to the values obtained on the Train set across all datasets.
	However, a larger difference is seen in a few configurations of the tCNN direct classifier processing the multiband data in 
	addition to the configurations of the dTCN direct classifiers.
	The gap between the validation and the training losses is significant for the dTCNs, which emphasizes the lack of 
	generalization of these type of networks due to a higher complexity (large number of parameters) of the network inconsistent 
	with the type of data in hand (1-d phase-folded light curves).
	Complex data would certainly benefit from higher-level network design as in dTCNs.
	Conversely, composite networks indicate a better stability, due to the addition of the autoencoder contribution to the total loss.
	We, also, notice an increase in the classification accuracy for the dTCN composite networks compared to their direct classifier 
	counterparts. 
	In the current application, best classification performances are achieved by RNNs and tCNNs.
}
%
Appendix Table \ref{table:classif_perfoBestID} reports the identifiers of the hyperparameters set configurations associated to the best-performing networks.
%
{
	The identifiers corresponds to the configurations set identifiers described in the Appendix Table 
	\ref{table:networks_config1} with varying number of layers or stacks \{1,2,3\} and sizes \{16,32\}.
	Using this naming scheme, the identifiers of the best-performing models appear to not fall into a unique 
	hyperparameters set configuration, as the architectures using LSTM layers performed well using 1 to 3 
	layers of different sizes. 
	From Appendix Table \ref{table:classif_perfoAll}, the classification results indicate about only a $\sim$1$-$6\% 
	dispersion within the individual results per layer type in each of the architectures, which implies that the  
	hyperparameters of the tested networks have a low-to-medium influence on the overall performances.
	Nonetheless, we notice for most LSTM architectures with 1 layer ($n_L=1$) underperform when processing 
	multi-passband data. Deeper networks ($n_L>1$) appear necessary in processing larger datasets.
}
%
In the current application, the qualification of {best-performing} models is solely based on an empirical search in the network hyperparameters space. 
Nonetheless, an improved search through the hyperparameters space can constitute a future work to investigate the effect of hyperparameters selection on the network performances and optimization landscape during training. %

\begin{figure*}[htp]
	\centering
	\includegraphics[width=1.0\textwidth]{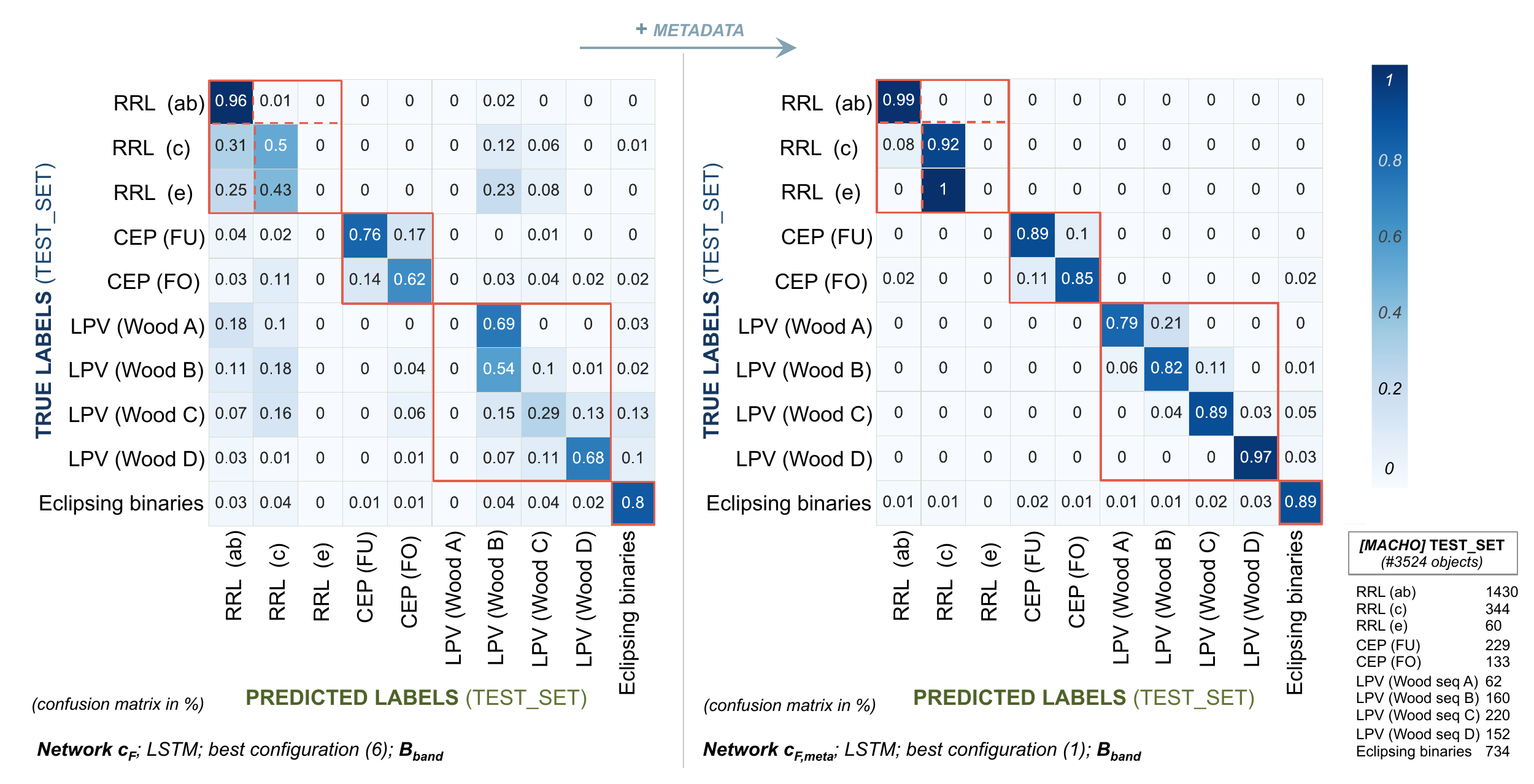}  
	\caption{
		Confusion matrices obtained on the Test set predictions for the best-performing LSTM direct classifiers 
		$c_{F}$ and $c_{F,meta}$ on the B-band. 
		The values in each box correspond to the number of predictions versus the initial count of true labels 
		(indicated on the right).
		Main stellar variability groups are highlighted in red, respectively the RR Lyrae, the Cepheids, the LPVs and 
		the eclipsing binaries.
		}
	\label{fig:confmat_lstm_classifier_recap}
\end{figure*}

Table \ref{table:perfoBest} summarizes the classification accuracy of the best-performing models reached on Test set.
For the best-performing models, the overall accuracy ranges between 67\% to 79\% for models processing the light-curves information without metadata.
In particular, without metadata , the dTCNs achieve a {classification} accuracy of 67$-${72\%} whereas the accuracy of RNNs and tCNNs jointly scales higher with {70}$-$79\% fraction of correct predictions.
By incorporating the metadata $-$colors, amplitudes, averaged magnitudes and periods$-$ in the models $c_{F,meta}$ and $d_{F,meta}$, the accuracy improves by 10$-$20\% to reach at best $\sim$88$-$92\% for RNNs, $\sim$82$-$92\% for tCNNs and only $\sim$75$-${82}\% for dTCNs.

The NN architectures learned on single band and multiband light-curves (the blue band $B$ versus the multiband $RB$) show comparable classification accuracy.
%
{
	By comparing the individual results in Table \ref{table:perfoBest} for each layer type, the addition of the MACHO red band data 
	injects a moderate effect on the classification accuracy.
	In particular, the accuracy of the best-performing direct classifiers decreases by 0.1$-$3\% for RNNs 
	and 1$-$7\% for tCNNs and dTCNs when using the multiband RB data in the \textit{merged} approach 
	in comparison with the B-band.
	Whereas, networks processing the multiband data via the \textit{hybrid} approach mainly achieve a 1$-$4\% increase 
	in the classification accuracy compared to the B-band data.
}
%
Moreover, the results {on best-performing models} do not indicate a significant difference in the {individual performances} reached by networks processing {the} multiband photometry $RB$ via the \textit{merged} and \textit{hybrid} approaches, {as individual results only indicate a 0.3$-$4\% differences}, due to the photometric bands (red and blue) in MACHO equally containing informative characteristics on the stellar variability types.
%
{
	The moderate disparity may indicate the need for a different strategy order when combining the data in the \textit{merged} 
	approach, as we combine in the current application the preprocessed (i.e., phase-folded and normalized) individual 
	measurements per band.
	A different strategy can consist of combining the light-curves prior to normalization.
}
%
{In general}, we would expect that classification using sparsely-observed multi-passband photometry would benefit from the availability of several sources of information on stellar variability. 
In such a case, the advantage of the \textit{hybrid} approach would be more prevalent for systems sequentially processing the multi-passband photometry in an optimized scheme, whereas the \textit{merged} approach may call for higher memory requirements in terms of CPU/GPU usage. 

Classification performances are better summarized on a confusion matrix that reports the fraction of predicted labels compared to the true class labels. Optimal results correspond to a diagonal matrix with a fraction of true positives per class (i.e., diagonal elements) close to unity.
Figure \ref{fig:confmat_lstm_classifier_recap} shows the performance obtained on the Test set for the best-performing LSTM direct classifiers $c_{F}$ and $c_{F,meta}$ that respectively reached an overall accuracy of 75\% and 92\%. 
Overall, the confusion matrices tell a similar story: the main stellar variability groups, highlighted in red, are recovered to a fair accuracy despite the overlap between subtypes and the misclassifications.
The network $c_F$ provides class predictions based on the information from preprocessed (i.e., normalized and phase-folded) light-curves without metadata. The observed degeneracy is to be expected between classes of objects sharing a similar shape of light-curves.
By supplementing the metadata to the network, the accuracy-per-class for $c_{F,meta}$ increases and the number of false positives (i.e., off-diagonal elements in confusion matrices) is significantly diminished.
Moreover, the observed porosity between adjacent classes in the confusion matrix appears to remain within the main stellar variability types.

From the confusion matrices, label predictions for the eclipsing binaries in our sample appear to overlap with other variability groups despite the use of the metadata.  
These misclassifications are {possibly} due to some degree of (true) label noise 
{ 
	 	that impairs/affects the reliability of the mapping generated from training.
}
In the current tests, class predictions of few subtypes remain erroneous despite the use of metadata. The second overtone RRL pulsators (type $e$) are inaccurately predicted as first-overtone RRL (type $c$).
To avoid confusion within subtypes, a proposed solution would be to introduce a weighting scheme on the feature contributions $\{\mathbf{X}_{enc}, \mathbf{X}_{meta}\}$ injecting some prior knowledge regarding the features importance in VS classification.

To further investigate the classification performances, we compute additional metrics (cf.\ Appendix Table \ref{table:perfoBest_othermetrics}). 
The recall (i.e., the true positives rate or sensitivity) characterizes the ability of the network to properly retrieve the true labels and the precision depicts the level of agreement between the predictions per class and the true labels. 
The F1-score corresponds to a combination of both metrics.
We report in Appendix Table \ref{table:exp_metrics} the averaged metrics for the best-performing LSTM direct classifiers $c_{F}$ and $c_{F,meta}$ that respectively showcased a classification accuracy of 75\% and 92\%. 
The averaged precision, recall and F1-score are boosted respectively to 77\%, 80\% and 78\% from an initial $\sim$51\% rate. 
Individual metrics per class highlight a better improvement.
However, the inability of the network to predict a few subtypes such as the second-overtone RRL pulsators affect the recall and precision {averaged across all observations in the Test set.}
Appendix Table \ref{table:perfoBest_othermetrics} reports the averaged metrics for the best-performing models {for the GRU, tCNN and dTCN}. Similar conclusions can be reached regarding the performances of the RNNs and tCNNs outperforming the dTCNs in the current tests.

\begin{table*}[htp!]
	\scriptsize
	\setlength{\tabcolsep}{.15cm}  
	\caption{Selected subset of objects from the MACHO VS database for display.}\vspace{-.5cm}
	\label{table:macho_displays_rec}
	\begin{center}{   
	\begin{tabular}{c l l CCRCCCCCCCc}  
	         \hline \hline
         		& \multirow{2}{*}{\sc{Object id} \tablenotemarkeditp{a}}
			& \multirow{2}{*}{\sc{Variability Type}} 
			& \alpha 
			& \delta 
			& \textsc{Period} 
			& \langle R \rangle 
			& \langle B \rangle  
			& \langle K_V\rangle 
			& \langle K_R\rangle 
			& \langle K_V\rangle - \langle K_R\rangle  
			& \textsc{Amplitude}  
			& \textsc{Amplitude} 
			& \multirow{2}{*}{SSR\tablenotemarkeditp{b}} 
			\\
		&&& \it(rad) & \it(rad) & \it(days) & \it(mag) & \it(mag) & \it(mag) & \it(mag) & \it(mag) & \textit{(in R)} &\it \textit{(in B)}& \\
		\hline  
			   n1 	& 82.9138.798  &  RRL (type ab) & 1.46723 & -1.20004 & 0.539  
			    	& -4.342  &  -4.198 & 20.096  & 19.692  & 0.403  & 0.729  & 1.053  & 0.416 \\
			   n2 	& 1.3449.1187  &  RRL (type c) & 1.31759  & -1.20170 & 0.337  
			    	& -4.829  &  -4.906  & 19.427  & 19.244 & 0.183  & 0.258  & 0.399  & 0.652 \\
			   n3 	& 1.4052.2961  &  RRL (type e) & 1.33093  & -1.20398  & 0.264  
			    	& -4.849 &  -4.848 & 19.471  & 19.21 & 0.261  & 0.226  & 0.396  & 1.005 \\
			   n4 	& 1.3441.45  &  CEP (FU)  & 1.31567  & -1.21106 & 3.362  
			    	& -8.420  &  -8.226  & 16.059  & 15.605 & 0.454  & 0.505  & 0.715  & 0.063 \\
			   n5 	& 81.9241.38 &  CEP ( FO)  & 1.47038 & -1.22092 & 1.973  
			    	& -7.797  &  -7.420  & 16.832  & 16.195  & 0.637  & 0.217  & 0.320  & 0.208 \\
			   n6 	& 1.4046.1610  &  LPV (Wood seq. A) & 1.33270 & -1.21091 & 45.150 
			    	& -9.351  & -8.220  & 15.896 & 14.505  & 1.391  & 0.085  & 0.086  & 0.928 \\
			   n7 	& 5.4407.15  & LPV (Wood seq. B) & 1.34311 & -1.21385  & 131.004 
			    	& -9.397  & -7.895  & 16.154  & 14.392   & 1.762  & 0.169  & 0.251  & 0.959 \\
			   n8 	& 82.9134.20 &  LPV (Wood seq. C) & 1.46758 & -1.20433 & 267.152  
			    	& -7.578  &  -5.348  & 18.570 & 16.080   & 2.490  & 1.462  & 2.424  & 0.271 \\
			   n9 	& 1.3689.30  &  LPV (Wood seq. D) & 1.32271 & -1.20431& 826.788
			    	& -8.949   &  -7.259  & 16.756 & 14.806  & 1.949  & 1.086  & 1.346  & 0.467 \\
			   n10 & 1.3442.233 &  Eclipsing binary & 1.31643 & -1.21027 & 1.640  
			    	& -6.332  & -6.725 & 17.665  & 17.798  & -0.132  & 0.930  & 1.090  & 0.208 \\ 
		\hline
	\end{tabular}
	\tablenotetextedit{a}{the standard three-integer identifier in the MACHO photometry database (\textit{field.tile.sequence}).}
	\tablenotetextedit{b}{model fit residuals from \texttt{Supersmoother} applied to the reduced photometric B-band data.}
	}\end{center}
\end{table*}

We, also, investigate the accuracy within three main stellar variability groups -- short-period pulsators (group 1), eclipsing binaries (group 2) and long-period variables (group 3), and report the classification accuracy {achieved on} the Test set for the best-performing models per group in the Appendix Table \ref{table:perfoBest_acc_pergroup}.
{Examining the results of the best-perfoming RNNs and tCNN, the classification accuracy increases significantly after incorporating the metadata}.
For short-period pulsators, the accuracy of the best-performing networks reaches 78$-$85\% without metadata solely based on the distinctive shape of these stars light-curves as the characteristic asymmetry and steep luminosity increase observed in fundamental mode pulsators.
With metadata, the accuracy increases up to 92$-$94\%.
The classification of the eclipsing binaries in our sample shows a comparable improvement with a 91$-$92\% correct predictions at best when using metadata. 
Similarly, the LPVs sample benefits from the use of metadata as a secondary input as best-performing networks {achieve} a 65$-$88\% fraction of true positives at best from an initial 30$-$54\% {without metadata}.

{
	To characterize the need for metadata and photometric observables for classification, we performed a supplementary classification 
	test using only the metadata as inputs to the classifier module.
	Results are reported on the Appendix Table \ref{table:exp_metrics_metaonly}.
	Using the metadata (i.e., the amplitudes, averaged magnitudes, periods and colors), the network achieves a 83\% accuracy 
	(i.e., fraction of true positives). 
	However, the classification metrics per class indicate a high number of false predictions in addition to the inability to predict some 
	subtypes, which corresponds to a lower precision, recall and F1-score compared for instance to the performances of the 
	best-performing RNN direct classifier and composite network in Appendix Table \ref{table:perfoBest_othermetrics}.
	Overall, combining the information encoded in the light-curves and the metadata allows a better mapping characterizing the 
	stellar variability types.
}


\begin{figure*}[htp!]
	\centering
	\includegraphics[width=0.7\textwidth]{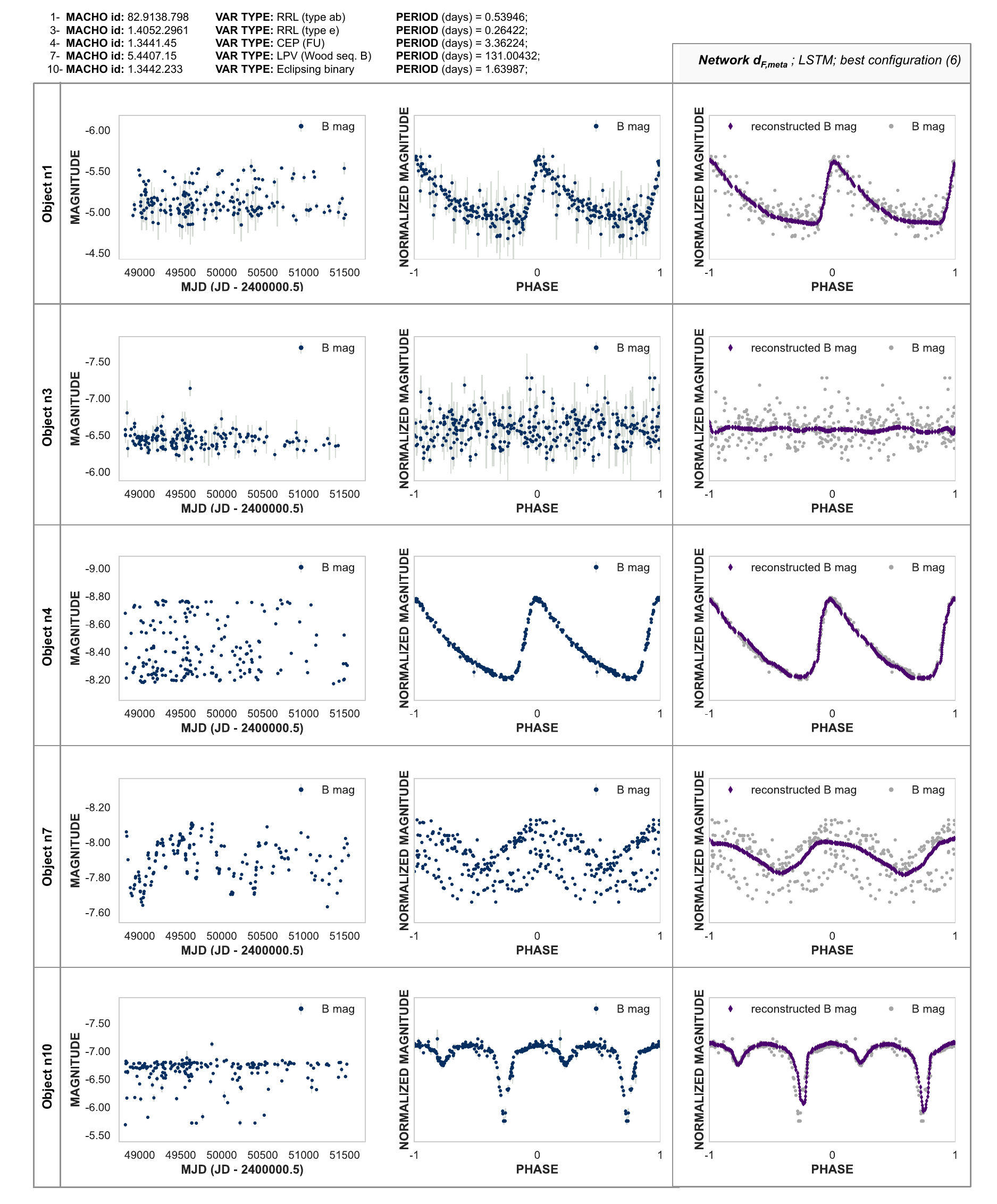}
	\caption{
		Reconstructed light-curves from the Test set for the best-performing LSTM composite network $d_{F,meta}$.
	}
	\label{fig:reconstruction_displays_subset}
\end{figure*}

\begin{figure*}[htp]
	\centering 
	\includegraphics[width=0.55\textwidth]{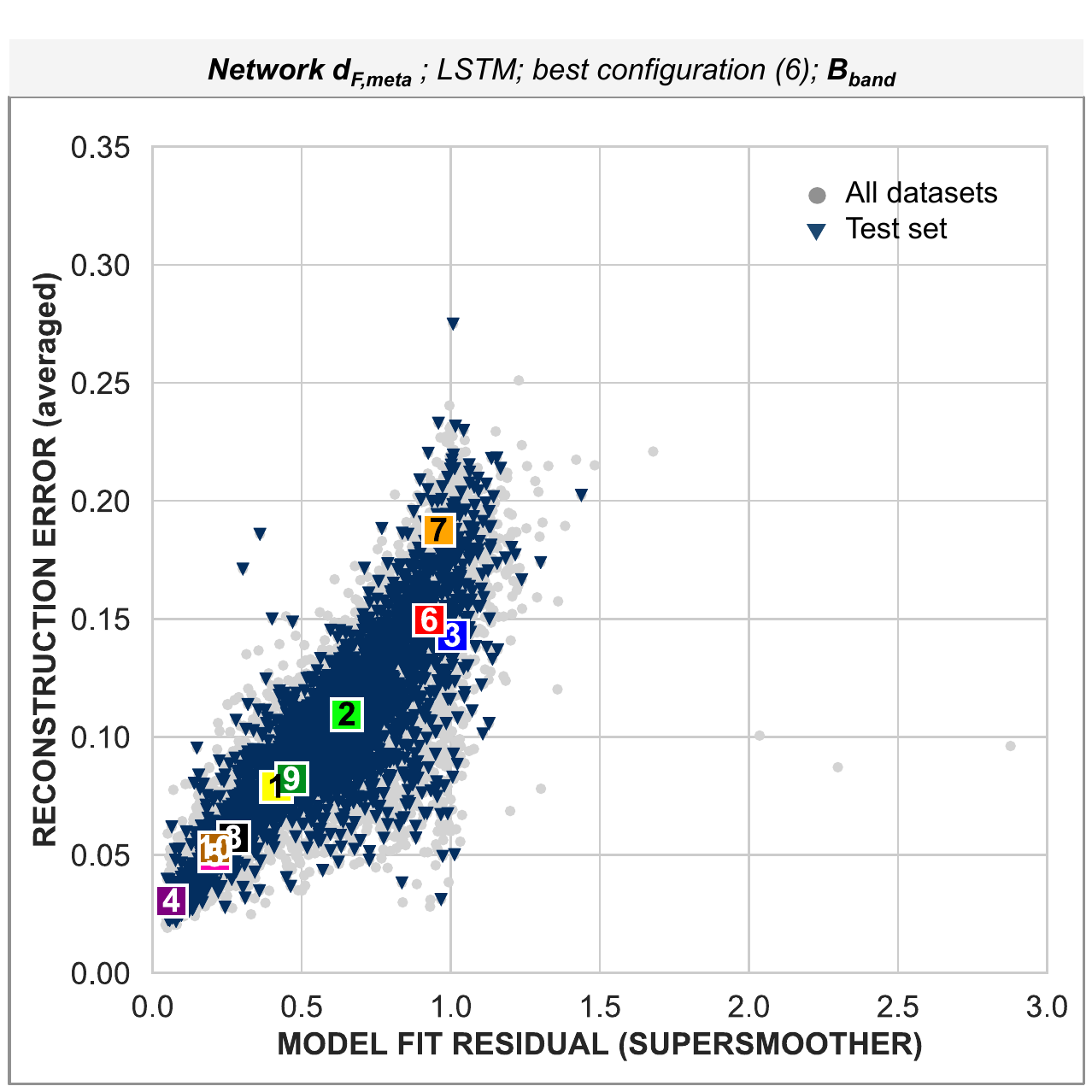}  
	\caption{
			Reconstruction error (MAE) as a function of the model fit residuals from the \texttt{SuperSmoother} algorithm 
			\citep{friedman_variable_1984} for the best-performing LSTM $d_{F,meta}$ on the B-band.
			The highlighted numbers (1 to 10) refer to the subset of selected objects from the Test set used to showcase the 
			reconstruction quality of the autoencoder branch.
			}
	\label{fig:rec_error_distribution_lstm}
\end{figure*}

\begin{figure*}[htp]
	\centering
	\includegraphics[width=1.0\textwidth]{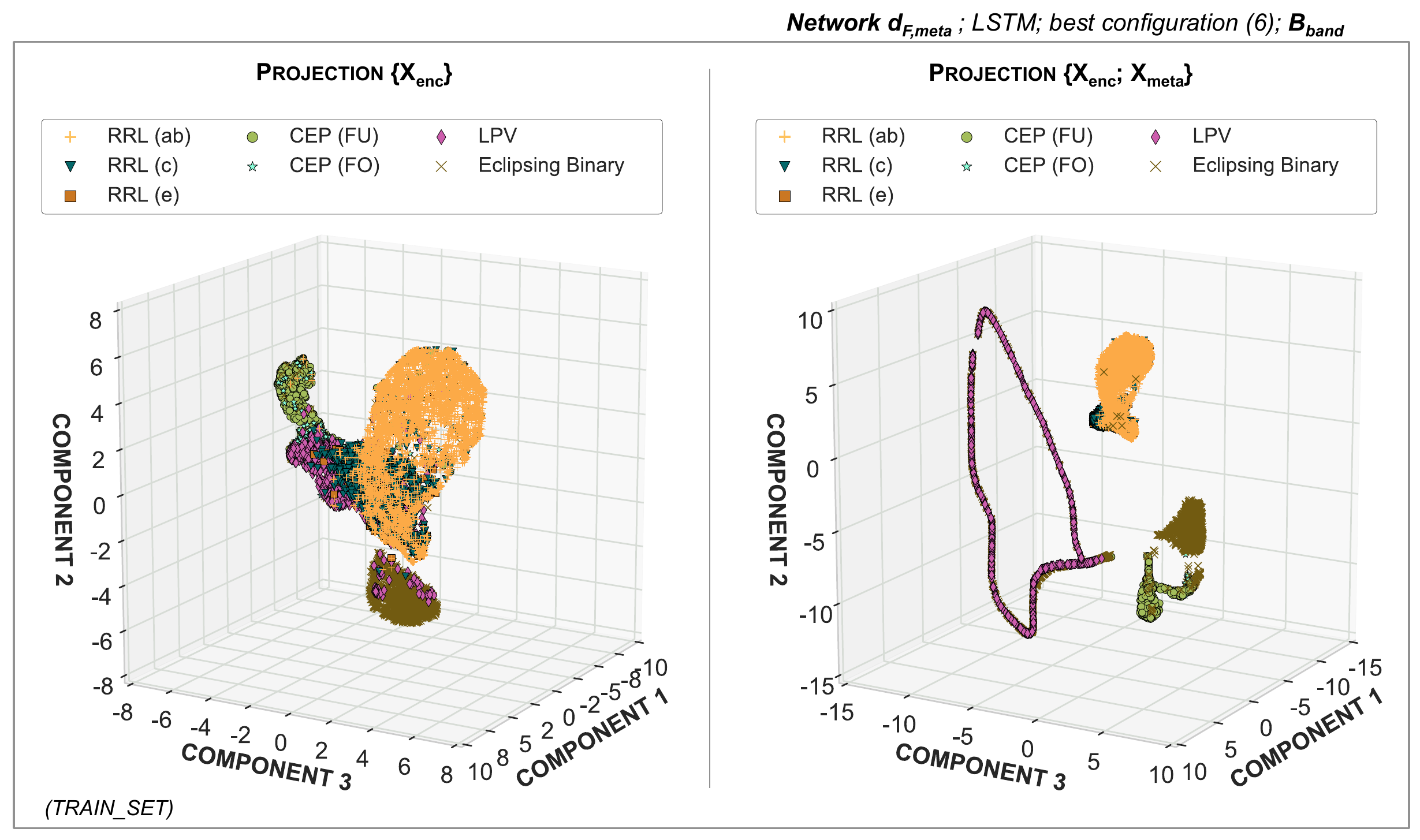}
	\caption{
		3-d representation of encoded features from the best-performing LSTM {composite network} 
		$d_{F,meta}$ on the B-band. Generated encodings are projected into a reduced 3-d representation using the UMAP 
		algorithm.
		}
	\label{fig:umap_projection_lstm}
\end{figure*}

\subsubsection{Reconstructed light-curves} \label{subsubsec:perfo_ae}

In this section, the autoencoder performance is assessed through the quality of reconstructed light-curves $\mathbf{X}_{rec}$.
An ideal reconstruction would preserve the embedded (denoised) structure of the input data $\mathbf{X}_{phot}$.
To visually assess the reconstruction quality, we select a sample of objects for display (cf.\ Table \ref{table:macho_displays_rec}) and show the reconstructed light-curves for the best-performing composite network $d_{F,meta}$ in the Appendix Figures \ref{fig:reconstruction_displays_set1}--\ref{fig:reconstruction_displays_set2}.
A subset for reconstructed light-curves for the best-performing LSTM composite network is also shown in Figure \ref{fig:reconstruction_displays_subset}.

Overall, the reconstruction results indicates a smoothing effect on the magnitude measurements as well as a correlation between the decoder performance and the input data quality. 
In particular, low signal-to-noise levels in the data limits the ability of the network to recover real structures and the resulting $\mathbf{X}_{rec}$ appears exaggeratedly smoothed out with no distinct features (such as the characteristic pulsation profile or peaks at maximum light).
Furthermore, some pulsating variables can exhibit irregularity or low-frequency modulations that may evolve on timescales longer than the observed time-range, noticeably seen for Long-Secondary Periodic stars (LSP) in the Wood sequence D \citep{wood_macho_lsp_2004}. 
In such cases, the folded light curves using the primary period appear as a mismatched superimposition of multiple cycles, as seen in the irregular LPV displayed in Figure \ref{fig:reconstruction_displays_subset}.
To prevent such limitations, detection of multiperiodicity, irregularity and low-frequency modulations should be considered to potentially isolate those objects that may require a different preprocessing strategy and classification approach.

To assess the global performances reached on the Test set, we evaluate the reconstruction error as a function of a data-quality indicator (SSR) corresponding to the model fit residuals computed from the \texttt{SuperSmoother} algorithm in \cite{friedman_variable_1984}.
{ 
	The \texttt{SuperSmoother} performs a component-wise linear smoothing of the time-serie data using adaptive 
	bandwidths. 
	The residuals obtained from averaged differences between the time-series and the regression fits. 
	In the current application, we use the SSR as a direct indicator of the data-quality in order to discuss the decoder reconstructions 
	in the tested neural networks.
}
%
{We expect the} light-curve measurements with low signal-to-noise, irregular variations or 
		extended low-frequency modulations {in time to be} associated with a high SSR.
The distribution of the reconstruction error in Figure \ref{fig:rec_error_distribution_lstm} suggests a distinct trend: the system is able to {moderately} recover the morphology of objects associated with low SSR (e.g., the Cepheids n4 and n5 and the Mira star n8) as opposed to noisier and irregular light-curve profiles (e.g., objects n3, n6 and n7 which are RRL of type $e$ and LPVs from the Wood sequences A and B).

A comparable analysis on the performances of the best-performing GRU, tCNN and dTCN composite models $d_{F,meta}$ reaches a similar conclusion {on} the distribution of the reconstruction error (cf.\ Appendix Figure \ref{fig:rec_error_distribution_}).
{From} the reconstructed profiles of the selected objects (cf.\ Appendix Figures \ref{fig:reconstruction_displays_set1}$-$\ref{fig:reconstruction_displays_set2}), {networks achieve overall }comparable performances despite few noticeable differences, such as the {reconstruction of the} shockwave propagating before the maximum light in the fundamental model RR Lyrae (object n1) that is recovered by the convolutional networks but heavily smoothed out in the LSTM and GRUs. The RNNs also appear to overly smooth out the modulations of the LPV star (object n9, LPV Wood sequence D), while convolutional nets partially restore discontinuous plateaus. 
{Furthermore}, the composite LSTM model appears to preserve the shape of the primary and secondary eclipses in the detached eclipsing binary (object n10), while convolutional nets (dTCNs and tCNNs) partially recover the depth of the eclipses and the GRU disproportionately smoothing out these features.
%
{ 
	Based on the reconstructed profiles of selected objects, the moderate smoothing of small-scale features by the RNNs can 
	be interpreted by these models focusing on an overall data structure propagated through the RNN cells. 
	The signal-to-noise of these features plays a role as well in the reconstruction quality.
	Convolutional NN variants, given the choice of kernel sizes for convolutions, exhibit a similar behavior through a localized 
	smoothing in the reconstructed profiles.}

	{From the reconstruction error distributions, the reconstruction performances of best-performing LSTM, GRU, tCNN and dTCN 
	show a comparable reconstruction ability.
	In our current designs, the decoder outputs reconstructed magnitudes. 
	To further characterize the ability of the decoder modules to unfold the generated embeddings into reconstructed profiles 
	close to the input data, the development of novel designs to output reconstructed profiles along with the associated prediction 
	errors is left for future study.
}


\subsubsection{Latent space exploration} 	\label{subsubsec:perfo_latent}

Using dimensionality reduction algorithms, encoded features of Train set can be projected into a reduced (2 or 3-d) representation.
The results for the best-performing LSTM composite network $d_{F,meta}$ are shown in Figure \ref{fig:umap_projection_lstm}. 
For better visualization, projections are separated for the three main variability groups in the Appendix Figures \ref{fig:umap_projection_short} to \ref{fig:umap_projection_eclipsing}.
We limit the analysis of the latent representation to the training dataset as it corresponds to the learned partitioning.

The encoded features generated from the best-performing composite network are projected onto a three-dimensional representation using the UMAP algorithm \citep{mcinnes_umap_2018} 
{described in Appendix \ref{sec:umap_intro}.
}
The degree of separation of the clusters in the reduced representation space characterizes the type of information fed to the classifier network. 
Without metadata, solely based on the encoded morphology of the light-curves, the projection {outlines} (cf.\ Figure \ref{fig:umap_projection_lstm} on the left) a large fraction of RR Lyrae, Cepheids, eclipsing binaries and LPVs isolated {to some extend} in the projected space.
The level of separability of these clusters is limited by the overlap between classes of objects sharing a similar shape of (preprocessed) light-curves, as {noticed for} the majority of LPVs, overtone pulsating RR Lyrae, Cepheids and few eclipsing binaries in our sample.
In the detailed representations (cf.\ Figures \ref{fig:umap_projection_short} to \ref{fig:umap_projection_eclipsing}), the encoded features of eclipsing binaries are clustered into composition of a compact aggregate, a dispersed set and outliers, 
while short-period pulsators cluster into a compact aggregate of fundamental mode pulsators and an overlapping blend of overtone pulsators given their similar light-curves profiles. 
In the latent space, the LPVs sample clusters into a compact aggregate without a clear delineation between the different subtypes; this suggests a lack of discriminating features in $\mathbf{X}_{enc}$ that would be necessary to distinguish the different subtypes.

When metadata is supplemented, the projection of the augmented features set \{$\mathbf{X}_{enc}, \mathbf{X}_{meta}$\} shows a better separability in the reduced latent space (cf.\ Figure \ref{fig:umap_projection_lstm} on the right); this separability also coincides with the improvement in the classification accuracy for the networks utilizing metadata to complement the encoded photometry.
In particular, the detailed representations for the main variability types highlight well-separated clusters {for} the short-period pulsators, while the sample of eclipsing binaries clusters into a composition of a compact aggregate, a filamentary structure and dispersed outliers.
The diversity in substructures in the eclipsing binaries sample can be explained by the different categories of binaries merged in the MACHO database (e.g., contact, detached and semi-detached stellar binaries) in addition to possible label contamination.
Similarly, the LPVs sample is well-separated from others variability groups and is projected onto a filamentary structure with an enhanced separability between the different subtypes. 
The best-performing GRU, tCNN and dTCN composite networks give comparable projection results.

{
	 To investigate the properties of the embeddings obtained from composite networks versus direct classifiers, 
	 a projection of the generated encodings obtained by the best-performing LSTM direct classifier is reported 
	 in Appendix Figure \ref{fig:umap_projection_classifier}.
	 Compared to the composite network, the direct classifier network generates an embedding layer $\textit{X}_{enc}$ by 
	 propagating the information between an encoder and a classifier module.  
	 In the latent representation, distinct clusters are noticeable along with the expected overlap of objects 
	 with a similar light-curves shape.
	From the 3-d representation, the main difference between the embeddings generated from 
		the LSTM direct classifier (cf.\ Appendix Figure \ref{fig:umap_projection_classifier}) 
		and the LSTM composite network (cf. Figure \ref{fig:umap_projection_lstm}) 
	lies in the clusters (intraclass) dispersion.
	The encoder-decoder combination in the composite network appears to narrow the clusters in the latent representation; this effect would be useful for anomaly detection to locate potential outliers at the outskirts of identified compact aggregates or the intersection of adjacent clusters. 
}

To summarize, composite NN architectures are able to encode the photometric observables into substructures associated with the stellar variability classes.
Without metadata, the encodings generated from the photometric data cluster into distinct aggregates despite the overlap between classes of objects sharing a similar morphology of preprocessed light-curves.
By supplementing the metadata as a secondary input to complement the encoded photometry, the level of separation in the latent space is enhanced, which aligns with the overall increase in classification accuracy.
An examination of the nature of the overlap regions of the latent space, as well as the properties of the different aggregates, are left for future study. We would expect similar studies using larger VS datasets will help test the potential universality of the latent space, and also reveal  potential outliers that could constitute new subclasses. 


\section{Conclusions} \label{sec:conclusions}

In this work, we explored the use of NN architectures for VS classification through various use-case scenarios.
These architectures allowed us to generate higher-abstraction encodings of the photometric data without the need for hand-coded feature engineering.

Two types of architectures, identified as direct classifiers and composite networks, are tested. Both networks are composed of an encoder module to transform the data into a reduced representation and a classifier to predict labels. Composite networks include a decoder module to define an encoded representation of the input data by optimal reconstruction.
In our analysis, VS classification using multi-passband photometric data can be performed in two approaches, either by encoding a merged representation of all passband measurements (\textit{merged} approach) or jointly processing individual encodings (\textit{hybrid} approach). Sparsely observed multi-passband photometry would benefit from adopting the latter approach.

In this work, we also experimented with a variety of NN architectures and investigated the effect of ancillary metadata on classification performance. 
{
	Through an empirical search on different hyperparameters set configurations, best-performing models were identified.
	Models exploiting hyperparameters tuning through optimized ML approaches or bayesian optimization are left for future work. 
}
In our work, we found that systems solely exploiting the time-series data are able to reach a $\sim$70\% accuracy for the best-performing models. By supplementing the metadata as a secondary input, a net increase in the classification accuracy is observed across all network types, reaching at best a $\sim$91\% accuracy for the best-performing LSTMs and temporal CNNs models.
Misclassifications for the best-performing networks are primarily restricted to the main stellar variability types, which provides a strong incentive for a multiple-stage architecture for label predictions, to first predict the main stellar variability type followed by subtypes prediction.
On a computational level, the training convergence time for RNNs models was found to be longer, due in part to larger memory allocation costs.

For composite networks, the reconstruction quality of the decoder module appears highly contingent on the input data properties (ie., the signal-to-noise level and smoothness of the light-curves profiles).
Variable stars exhibiting multiperiodicity, irregularity or modulations over time appear in their phase-folded representation as a mismatch of superimposed cycles that a network is unable to learn and overly smooths out in reconstruction. 

The exploration of the learned encodings indicated a clear clustering linked to the stellar variability types.
Without metadata, clusters of variables appear isolated despite the overlap noticed for objects sharing similar light-curves profiles.
Supplementing the metadata to the encoded information predictably lessens such degeneracy and enhances the separation between the classes; this in turn, accounts for the increase in the fraction of correct predictions.  
We would expect the latent representations to highlight interesting properties in the data and pinpoint to potential outliers, unknown stellar variability types or new subtypes within known variability classes.

To conclude, various NN architectures are able to capture low-dimensional data representations and reach achieve excellent classification accuracy without the need for hand-coded featurization. The best-performing networks in our tests are primarily LSTM- and tCNN-based models, with the latter benefiting from smaller training convergence time and smaller memory footprints.

As a future direction, developing a baseline for an automated system able to learn a wider range of stellar variability traits should be explored.
The need for general architectures is strongly motivated from the fact that massive surveys are set to produce large datasets with a blend of different types of stellar variables such as aperiodic VSs (e.g., cataclysmic stars and microlensing events) as well as periodic and quasi-periodic variables (e.g., pulsators, rotators and eclipsing binaries).
Automated classification  for periodic VSs presented here exploits phase-folded representations as well as the information from the frequency domain, while classification of quasi-periodic variables require the use of a combination of multiple data representations -- phase-folded light-curves, periodograms, O$-$C diagrams and the time-series -- to produce reliable class predictions.
To meet the need for a general framework, one proposed design would consist of a multi-stage architecture with different components specialized to distinguish distinct stellar variability traits, to first discriminate between the three categories of periodic, quasi-periodic and aperiodic VSs then follow with the classification into the stellar variability types and subtypes.

{
	Despite our analysis being focused on applications for periodic variable stars classification, all arguments presented in the 
	scope of this work extend to other types of astronomical time-series. 
	NNs can be built with a comparable architecture for supernovae classification and transient detection, with adaptations in 
	the  of the input data representation, the preprocessing strategy and the necessary metadata (e.g., redshift measurements
	 and spectral features).
	Similarly, our approaches in processing multi-passband photometric data for classification can be generalized to other 
	variable objects.
	On the network design, the complexity of the networks should conform with the type of the data. 
	In particular, higher-level multi-dimensional data would require deeper and complex architectures compared to a 1-d 
	information such as phase-folded light-curves.
	Using the presented methodology on different datasets, we would expect, on one hand, an increase of the classification accuracy 
	when supplementing the metadata, and on the other hand, a significant improvement in classification when combing sparse 
	observations across multiple photometric bands. 
	For sparsely observed light-curves, the networks processing multiband data would very likely exceed the classification results 
	obtained with a single photometric band. 
	We expect the landscape of the latent space representation to differ from current results. Further analysis of latent representations 
	obtained from a larger scope of variability types is left for future study.
	%
}


\acknowledgments {
\small 
This research made use of the Savio computational cluster resource provided by the Berkeley Research Computing program at the University of California, Berkeley (supported by the UC Berkeley Chancellor, Vice Chancellor for Research, and Chief Information Officer).
SJ, JSB were partially supported by a Gordon and Betty Moore Foundation Data-Driven Discovery grant.
{The authors would like to thank the referee for helpful comments and multiple suggestions that significantly improved this manuscript.}
We, also, would like to acknowledge and thank J. Mart\'{i}nez-Palomera for his careful readings and helpful comments.
%
This paper utilizes public domain data obtained by the MACHO Project, jointly funded by the US Department of Energy through the University of California, Lawrence Livermore National Laboratory under contract No. W-7405-Eng-48, by the National Science Foundation through the Center for Particle Astrophysics of the University of California under cooperative agreement AST-8809616, and by the Mount Stromlo and Siding Spring Observatory, part of the Australian National University.
}

\newpage
\software{
\small
	\texttt{NumPy} \citep{oliphant_guide_2006, walt_numpy_2011},
	\texttt{scikit-learn}  \citep{scikitlearn_11},
	\texttt{TensorFlow} \citep{tensorflow2015_whitepaper,abadi_tensorflow_2016},
	\texttt{keras} \citep{chollet2015keras},
	and \texttt{exoplanet} package \citep{exoplanet:exoplanet_zenodo} and its dependencies for GP model prediction:  
	       \texttt{celerite}	\citep{foreman-mackey_fast_2017, foreman-mackey_fast_2018}, 
	       \texttt{pymc3}	\citep{exoplanet:pymc3} and
	       \texttt{theano}	\citep{exoplanet:theano}.
}

\newpage~\newpage
\bibliographystyle{aasjournal}
\bibliography{current_lib_red,jsb}{}

\appendix
\restartappendixnumbering

\section{Classification using multiband photometric data} \label{sec:structures_multiband}{

\begin{figure*}[htp]
	\centering
	\includegraphics[width=1.0\textwidth]{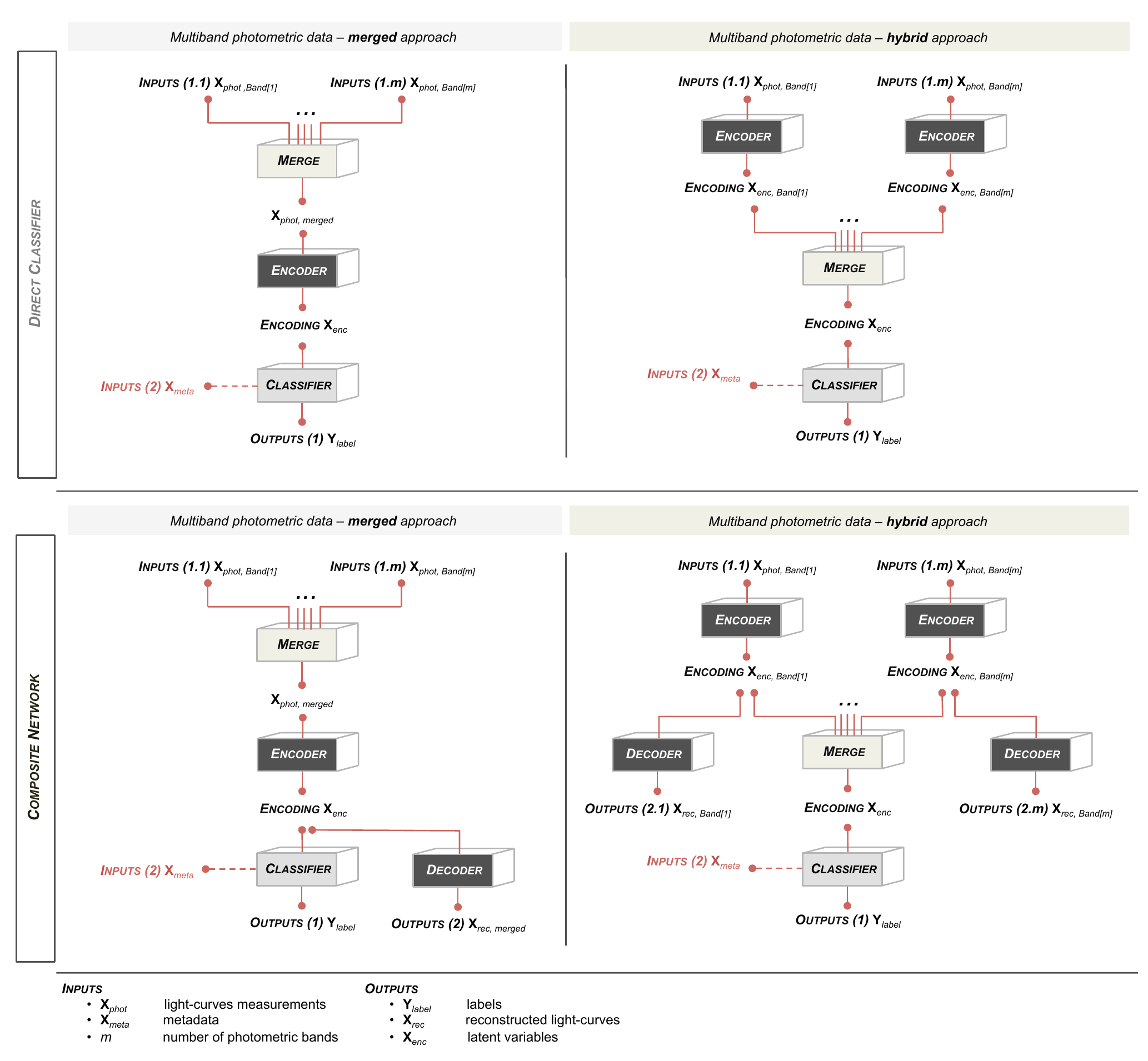}
	\caption{High-level architecture of the direct classifier (top) and composite (bottom) networks across two variants of the combination of multi-passbands photometric data, identified as \textit{merged} and  \textit{composite} approaches [left and right].
	In the \textit{merged} approach, multi-passband measurements are merged into a unique observable $\mathbf{X}_{phot, merged}$ processed by the encoder module to compute the latent representation $\mathbf{X}_{enc}$.
Whereas, the \textit{hybrid} approach independently encodes the multi-passband measurements into individual features merged at a latter time into a compact encoded representation $\mathbf{X}_{enc}$. 
In both scenarios, the classifier module exploits the generated encodings $\mathbf{X}_{enc}$, along with metadata $\mathbf{X}_{meta}$, for label predictions.
Composite networks differ from the direct classifiers by the addition of a decoder module connected at the bottleneck level to the encoder to generate the embeddings by optimal reconstruction.
	}
	\label{fig:structure_multiband}
\end{figure*}

{ 
In the \textit{merged} approach, the multi-passband photometric data can be combined into a 2-d tabular data of dimension $\Big(N_{p, merged}^*\times\big(2m+1\big)\Big)$, in which the $m$ photometric band magnitude and associated error measurements are provided along with the observation epochs.\\
The $i^{th}$ observation in $\mathbf{X}_{phot, merged}$ corresponds to the matrix representation:
\begin{align}
        \itbf{x}_i 
        = 
       \begin{pmatrix}
			t_{i}^{(1)}, 	
				& \Big[	\textnormal{mag}_i^{(1)}, \sigma_i^{(1)} \Big]_{\textnormal{band}\,1},
				& \cdots
				& \Big[	\textnormal{mag}_i^{(1)}, \sigma_i^{(1)} \Big]_{\textnormal{band}\,m},
				\\
			\vdots&\vdots&&\vdots \\
			t_{i}^{(P^*)}, 	
				& \Big[	\textnormal{mag}_i^{(P^*)},  \sigma_i^{(P^*)} \Big]_{\textnormal{band}\,1}
				& \cdots
				& \Big[	\textnormal{mag}_i^{(P^*)},  \sigma_i^{(P^*)} \Big]_{\textnormal{band}\,m}	\\
	\end{pmatrix}
,\end{align}
where, $m$ refers to the number of photometric bands and ${P^*}$ designates the total count of datapoints $N_{p, merged}^*$ for the $i^{th}$ observation.
In this representation, the sparsity level of the matrix depends on the observation times across the different photometric bands. 
\\\\
Alternatively, multi-passband data can be combined into a 2-d tabular data of dimension $\Big(N_{p, merged}\times4\Big)$, in which an auxiliary vector encoding the photometry band is provided along with the observation epochs and the $m$ band magnitude and associated error measurements.
The encoding vector associates each photometric band type to a dictionary item (numerical or qualitative variables).
\setlength\fboxrule{0.1pt}
\begin{align}
        	 \itbf{x}_i 
	= 
       \begin{pmatrix} 
			\textnormal{\bf dict}_{\; \textnormal{band}\,1} ,
				&\boxed{
					\big[\textbf{t}_i, 
						\textnormal{\bf mag}_i, 
						\boldsymbol{\sigma}_i 
					\big]}_{\; \textnormal{band}\,1} 	\\			
			\vdots							 \\
			\textnormal{\bf dict}_{\; \textnormal{band}\,m} ,
				&\boxed{
					\big[	\textbf{t}_i, 
						\textnormal{\bf mag}_i, 
						\boldsymbol{\sigma}_i 
					\big]}_{\; \textnormal{band}\,m} 	\\
	\end{pmatrix}
,\end{align} 
where, the enclosed measurements per band $\{b\}_{b:1\rightarrow m}$ correspond to a matrix of $\Big(P_b\times4\Big)$ dimension as follows.
\begin{align}	
	\boxed{
		\big[   \textbf{t}_i, 
			\textnormal{\bf mag}_i, 
			\boldsymbol{\sigma}_i 
		\big]}_{\; \textnormal{band}\,b}
		=		
		\begin{pmatrix}
			\big[	t_i^{(1)}, 
				\textnormal{mag}_i^{(1)}, 
				\sigma_i^{(1)} 
			\big]_{\; \textnormal{band}\,b}	\\
			\vdots 					\\
			\big[	t_i^{(P_b)}, 
				\textnormal{mag}_i^{(P_b)}, 
				\sigma_i^{(P_b)} 
			\big]_{\; \textnormal{band}\,b}	\\
		\end{pmatrix} 
,\end{align} 
\\
where, $\{P_b\}_{b: 1\rightarrow m}$ refers to the number of datapoints per band.\\
}

}

\newpage
\setcounter{table}{0} \setcounter{figure}{0}
\section{Description of NNs, UMAP and gaussian process modeling} \label{sec:algos_shortintro}{

\subsection{RNN, CNN and TCN} \label{sec:nets_intro}{
Recurrent neural network (RNN) refers to a neural network architecture composed with interconnected nodes through a directed graph (cyclic or acyclic) along a temporal sequence. 
In the standard architecture, the fully-connected RNN layer is constructed such that each node is interlinked to the adjacent units. Each node in the standard architecture corresponds to a neural unit with an activation function and a weight.
The LSTM (\textit{Long-Short Term Memory}, \citealt{hochreiter_long_1997}) and the GRU (\textit{Gated Recurrent Network}, \citealt{cho_learning_2014}) are variants of the RNN with a higher level node structure composed with multiple subunits acting as internal regulators to the propagated information within the network. 
Both the LSTM and GRU cells exploit a forget gate and an input gate, and the LSTM utilizes an additional output gate.
RNNs applications on astronomical time-series include 
	SNe classification \citep{charnock_deep_2017}
	VS classification using autoencoders \citep{naul_recurrent_2018},
	{
		gravitational-waves signal denoising \citep{shen_denoising_2019},  
		}
	periodic VS classification \citep{tsang_deep_2019},
	and online transient events detection \citep{muthukrishna_rapid_2019, moller_supernnova_2019}.

Convolutional neural network (CNN) refers to a NN architecture with convolutional layers that applies a series of convolutions through overlapping windows, allowing to capture spatial correlations in the data.
The standard CNN architecture utilizes pooling layers after convolutions to downsize the data in addition to fully-connected layers.
The performances of convolutional-based neural networks has been demonstrated in a broad range of astronomical data applications such as 
	galaxy classification 
		\citep{dieleman_cnn_gal_2015, aniyan_cnn_radiogal_2017,  kim_stargalaxy_2017, dominguez_sanchez_cnn_gal_2018},
	VS classification using asteroseismology 	
		\citep{hon_cnn_asteroseis_2017},  
	supernovae classification 
		\citep{cabrera-vives_cnn_sne_2016,  pasquet_pelican_2019, brunel_cnn_2019},
	photometric redshifts estimation 				
		\citep{hoyle_dnn_convolutional_zphot_2016, disanto_photometric_2018, pasquet_cnn_zphot_2019}, 
	cosmological parameters inference				
		 \citep{ntampaka_hybrid_2020},
	parameters estimation from 21-cm tomography 	
		\citep{gillet_deep_2019},
	strong lensing detection						
		 \citep{lanusse_cmu_2018, jacobs_finding_2019},
	gravitational-waves signal detection 
		\citep{gebhard_convwave_gw_2017, gabbard_gw_cnn_2018, george_cnn_gw_2018, 
				fan_applying_2019, gebhard_cnn_gw_2019}
	{
		\citep{george_deep_2018} 
		},
	generator models for weak lensing convergence maps 
		\citep{mustafa_cosmogan_2019},
	cosmic rays modeling \citep{erdmann_deep_2018},
	%
	{
		detection of damped Ly$\alpha$ systems in quasar spectra
			\citep{parks_deep_2018},
		fast radio bursts classification
			\citep{connor_applying_2018},
		and classification of supernovae spectra
			\citep{muthukrishna_dash_2019}.
	}
	
Temporal convolutional neural network (TCN) refer to a NN architecture in \cite{lea_temporal_2017}, initially derived from the \texttt{Wavenet} architecture \citep{oord_wavenet_2016}.
The network is a composition of a serie of dilated convolutions and residual blocks used to expand the filters receptive fields and reduce the training convergence time.  A detailed description of deep learning techniques is available in specialized computer science publications, such as the overview on DL techniques by \cite{schmidhuber_deep_2015}. 

}

\subsection{UMAP}\label{sec:umap_intro}{
The UMAP (\textit{Uniform Manifold Approximation and Projection}) algorithm is a nonlinear dimensionality reduction algorithm introduced by \cite{mcinnes_umap_2018}. Assuming a uniform distribution of the data on a locally connected Riemannian manifold, the algorithm computes a low-dimensional representation by optimizing a fuzzy set cross-entropy between the simplicial set representations of the data and the target embeddings.
The UMAP has gained interest and use recently for astronomical data applications such as the SDSS DR15 spectroscopic data classification in \cite{clarke2019_sdss_umap} and anomaly detection in SDSS galaxy samples in \cite{reis2019_sdss_umap}.

}

\subsection{Model prediction using Gaussian Processes} \label{sec:gp_celerite}{

{Part of} preprocessing, data reduction is performed using Gaussian Processes (GPs) model to generate fixed-length representations of each source.
\citet{foreman-mackey_fast_2017} provides GP kernels suitable for astronomical time-series, with various applications including radial velocity fitting and transit modeling.
For periodic VSs, we select a GP kernel based on a composition of stochastically driven damped harmonic oscillators (SHOs) with a quasi-periodic covariance term. 
For each SHO model, the associated power spectral density $S(\omega)$ is defined as following:
\begin{align} 
	S(\omega)		=\mathlarger{\sum}_{j=1}^{N_{SHO}}{S_j(\omega)} ,		\\
	S_j(\omega)	=\sqrt{\frac{2}{\pi}} \cfrac{ S_{0,j} \; \omega_{0,j}^{4}}  
					{(\omega^2 - \omega^2_{0,j} )^2 + \omega^2 \omega^2_{0,j} / Q^2_{j} } ,
	\label{eq:sho_psd}
\end{align} 
where, $\omega_{0,j}$ and $Q_{j}$ respectively refer to the frequency of the undamped oscillator and the associated quality factor of the $j^{th}$ oscillator.
The parameter $S_{0,j}$ is proportional to the resonance (i.e., $ \omega=\omega_{0,j}$) power.

Our data reduction approach is a two-fold process: first, we fit a GP model on the observed data  
	{$\itbf{x}_{obs}$}
and second, we use the model to predict 
{ 
	a representation of the time-serie over a reduced time-frame $\mathbf{T}_{red}$ along with the uncertainties from the GP fit.
}
\\
For periodic VSs, we use a GP model corresponding to a mixture of $N_{SHO} = 2$ SHOs. 
{
	We based our selection on the applications of GP modeling to data showcasing periodicity patterns as transit light-curves or stellar variables in \cite{foreman-mackey_fast_2017}. The GP model with 2 SHOs is chosen in order to take into consideration of multiperiodicity often encountered in variables stars.
	The current parametrization and kernel type is proper to the current work realized on a dataset of pulsating variables and eclipsing binaries.
}
 
In the model parameterization, the periodicity (due to pulsation or binarity) is captured by the resonance frequency, such that:
\begin{align}
	\omega_{0,j} 	= \frac{4 \pi Q_{j} }{ P_j+ \sqrt{4 Q_{j}^2 - 1}},		\\
	S_{0,j} 		= \frac{A_j}{\omega_{0,j} \; Q_{j} }.
\end{align} 
Here, $P_j$ and $A_j$ refer respectively to the period and amplitude of the variability per SHO model $j$.  
For each observed light-curve, an independent GP model is fitted.
A full description of the GP modeling can be found in \cite{foreman-mackey_fast_2017} and the available open source  \cite{exoplanet:exoplanet_zenodo}.

Using the MAP (\textit{Maximum-a-posteriori}) solution, model prediction is performed on a time frame $\mathbf{T}_{red}$ sampled within the range of observed epochs $\mathbf{T}_{obs}$. 
For unevenly sampled data, the GP predictions located in large time gaps are associated to a high uncertainty of the model. 
To prevent such limitation, we developed an approach to generate a random time range within the observed $\mathbf{T}_{obs}$ outside significant time gaps. 
Using the unsupervised K-means algorithm applied to the time differences $\Delta \mathbf{T}_{obs}$, observations are clustered based on their proximity in time. Significant time gaps $\mathbf{T}_{gaps}$ are identified within the group of large time differences, and an optional rejection criterion is supplemented to refine the detected time gaps to span, at least, higher than $n$ cycles of the primary period of the light-curve.
The reduced time frame $\mathbf{T}_{red}$ is generated via random sampling of time values within the observed time range outside the identified time gaps $\mathbf{T}_{set}=\{\mathbf{T}_{obs}\mathsmaller{\setminus}\mathbf{T}_{gaps}\}$.
For each detected subset $j$ of clustered observations, time values are obtained either by randomly generating values within the range of $\mathbf{T}_{set, j}$ or by randomly selecting of values in $\mathbf{T}_{set, j}$ shifted by a random $\delta_t>0$. The second approach generates a time frame $\mathbf{T}_{red}$ close to the observed $\mathbf{T}_{set}$. 

The GP model fitted to the data is a mixture of 2 SHOs with the following parameters.
\begin{equation}
	\begin{blockarray}{llll}
		& P_1 = P,\; 				P_2 = P/2,\;				
			A_1 = \exp^{logA},\;  		A_2 = m_A\times \exp^{logA},\; 
			Q_1=Q_0+\Delta Q,\;	Q_2=Q_0,\;
		\\
		& m_A \sim \mathcal{U}(0,1),\;								
			logA \sim \mathcal{N}(\mu_{A},\,\sigma_{A}^{2}),\;		
			Q_0>\frac{1}{2},\;
			 \Delta Q\sim\mathcal{N}(\mu_{Q},\,\sigma_{Q}^{2}),
	\end{blockarray}
\end{equation}
with, $P$ is the primary period of the time-serie,
$(\mu_{A},\,\sigma_{A}^{2})$ refer to the amplitude of the time series and associated error (or a fixed variance),
and $(\mu_{Q},\,\sigma_{Q}^{2})$ are strictly positive values set to separate the two oscillation modes.

To illustrate the results of data reduction using the aforementioned GP model and prediction scheme, a display of MACHO light-curves is provided in Figure \ref{fig:gp_preds_exp}.

\begin{figure}[htp!]
	\centering
	\includegraphics[width=0.95\textwidth]{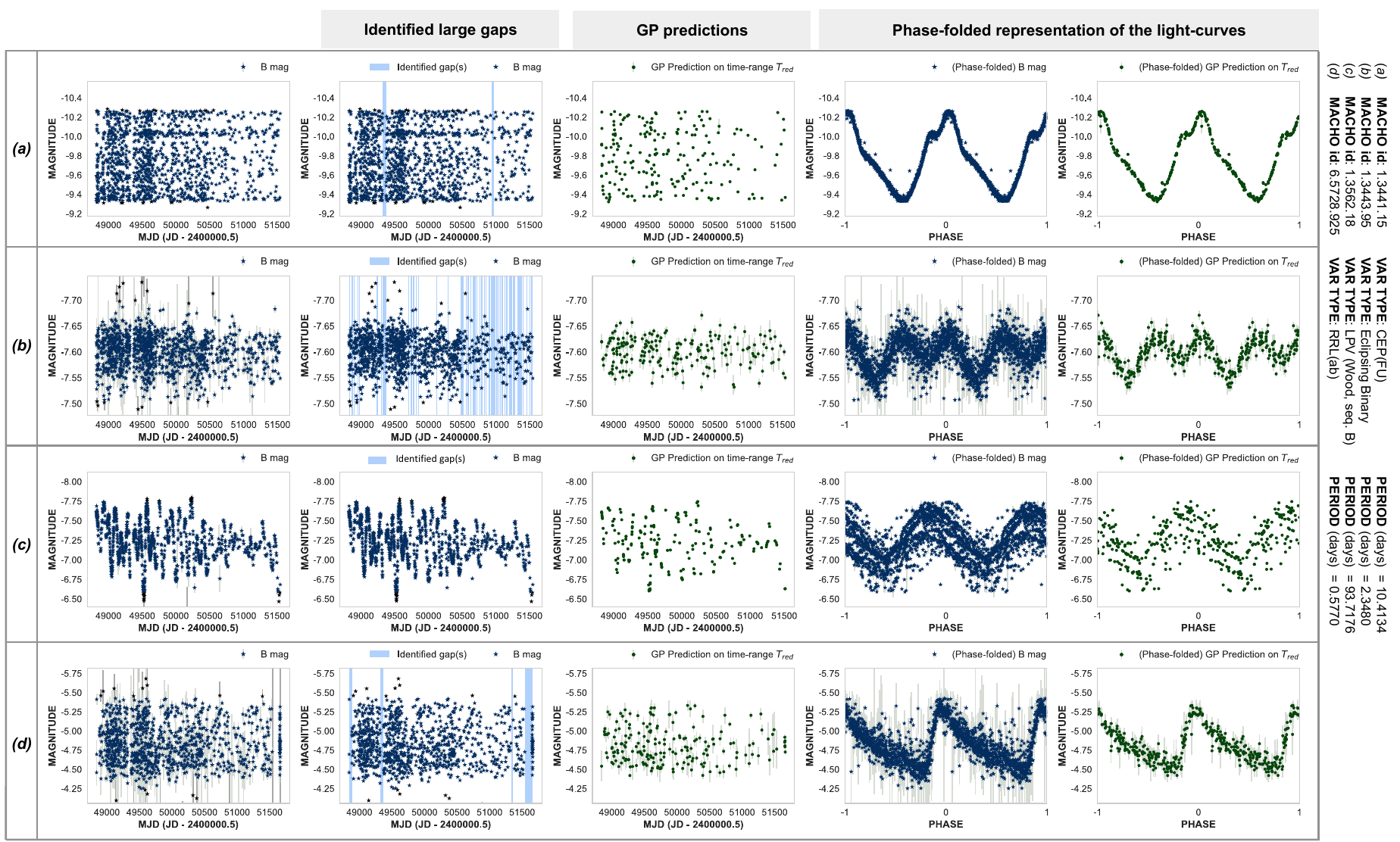}
	\caption{Displays of reduced MACHO light-curves using gaussian processes (GPs) model fit and prediction. 
	A GP model is fitted to each observed light-curve and the best model is associated to the MAP (\textit{Maximum-a-posteriori}) solution.
	The predictions --magnitudes and associated errors-- are computed on a reduced time-range $\mathbf{T}_{red}$ of 200 datapoints obtained by a random selection of time values within the observed time-range $\mathbf{T}_{set}$ (outside the identified large gaps highlighted in light blue) and shifted with a random lag $\delta_t\in[0,\Delta T_{set}]$. 
	Phase-folded representations of light-curves are given only as a reference. GP predictions are computed on the initial time-series.} 
	\label{fig:gp_preds_exp}
\end{figure}

}
}

\newpage
\setcounter{table}{0} \setcounter{figure}{0}
\section{Metrics for multiclass classification} \label{sec:metrics_clf}{
     
To quantify the performances in multiclass classification, the following metrics are computed:
\renewcommand{\arraystretch}{2}
\begin{table}[htp]{
	\footnotesize
	\caption{Classification metrics}
	\label{tab:metrics_eq}
	\setlength{\tabcolsep}{.32cm}
	{\begin{center}{ 
	 	\begin{tabular}{ l | ll | ll}
			\bfsc{Metrics}
			&\multicolumn{2}{c|}{ \sc Per class $\{j\}_{j:1\rightarrow N_C}$}
			&\multicolumn{2}{c}{ \sc Macro-averaging} 
			\\
			\hline
			\sc Precision
				& $\text{Precision}_{(j)}$ 	&$= \cfrac{\text{TP}_{(j)}} {\text{TP}_{(j)} \,+\, \text{FP}_{(j)}}$
				& $\text{Precision}_{M}$ 	&$=  \cfrac{1}{N_C} \mathlarger{\sum}_{j=1}^{N_C} \,\text{Precision}_{(j)} $ 
			\\[.5em]
			\sc Recall (Sensitivity)    
				& $\text{Recall}_{(j)}$ 		&$=\cfrac{\text{TP}_{(j)} } {\text{TP}_{(j)} \,+\, \text{FN}_{(j)} }$ 
				& $\text{Recall}_{M}$ 		&$=\cfrac{1}{N_C} \mathlarger{\sum}_{j=1}^{N_C} \,\text{Recall}_{(j)} $ 
			\\[.5em]
			\sc F1-score 
				& $\text{F1-score}_{(j)}$	&$= 2\times \cfrac{\text{Recall}_{(j)} \, \times \, \text{Precision}_{(j)} } 
											{\text{Recall}_{(j)} \,+ \, \text{Precision}_{(j)} }$ 
				& $\text{F1-score}_{M}$	&$= \cfrac{1}{N_C} \mathlarger{\sum}_{j=1}^{N_C} \,\text{F1-score}_{(j)} $ 
			\\[.5em]
		\hline
			\sc Accuracy  
				& $\text{Accuracy}$ 	& \multicolumn{3}{l}{$= \cfrac{1}{N_s} \mathlarger{\sum}_{j=1}^{N_C}\,\text{TP}_{(j)} $} 
			\\[.5em]
		\hline
		\end{tabular}
		}\end{center}}
}\end{table}

with $N_C$ the total number of classes, $N_s$ the number of true samples and TP$_{(j)}$, TN$_{(j)}$, FP$_{(j)}$ and FN$_{(j)}$ respectively referring to the true positives, the true negatives, the false positives and the false negatives computed from the predictions on objects from the true class $j$.

}

\newpage
\setcounter{table}{0} \setcounter{figure}{0}
\section{Networks} \label{sec:nets_architectures}{
     
\renewcommand{\arraystretch}{1.2}
\begin{table}[htp]{
	\footnotesize  
	\caption{Networks configuration -- Entries.} 
	\label{table:networks_config0}	
	\setlength{\tabcolsep}{.25cm}
 	{\begin{center}
	\begin{tabular}{lll}
		\hline\hline   
				\sc{Network type}	
				&\sc{Inputs}
				&\sc{Outputs}\\
		\hline						
		classifier $c_F$                   & $\mathbf{X}_{phot}$ & $\mathbf{Y}_{label}$                                         \\    \cline{1-3}
		classifier $c_{F, meta}$       & $\{\mathbf{X}_{phot}, \mathbf{X}_{meta}\}$ & $\mathbf{Y}_{label}$      \\    \cline{1-3}
		composite $d_F$                & $\mathbf{X}_{phot}$	& $\{\mathbf{Y}_{label}, \mathbf{X}_{rec}\}$      \\     \cline{1-3}
		composite $d_{F,meta}$     & $\{\mathbf{X}_{phot}, \mathbf{X}_{meta}\}$ & $\{\mathbf{Y}_{label}, \mathbf{X}_{rec}\}$	 \\
		\hline		
	\end{tabular}
	\end{center}}
}\end{table}

\renewcommand{\arraystretch}{1.2}
\begin{table}[htp]{
	\footnotesize 
	\caption{Networks configuration -- Hyperparameters.} 
	\label{table:networks_config1}
	\setlength{\tabcolsep}{.32cm}
 	{\begin{center}
	\begin{tabular}{ccccl}
		\hline\hline
				\sc{Layer type} 	
				&\sc{Configuration id}
				&\sc{Nb layers} ($n_{L}$)
				&\sc{Size} 	($n_{S}$)
				&\sc{Common hyperparameters} \\	
		\hline	
			\multirow{6}{*}{RNN \{LSTM; GRU\}}
				& \it(1)	&1	& \multirow{3}{*}{16}	& bidirectional network;	
				    \\	\cline{2-3}
				& \it(2)	&2	&& categorical classification; 		
				    \\	\cline{2-3}
				& \it(3)	&3	&& drop fraction=0.25;	 	
				    \\	\cline{2-4}
				& \it(4)	&1	& \multirow{3}{*}{32}	& batch size=128; maximum training epochs=200;		
				    \\	\cline{2-3}
				& \it(5)	&2	&& \texttt{Adam}\tablenotemarkedit{a} optimizer; optimizer learning rate=5$\times10^{-4}$; 	
				    \\	\cline{2-3}
				& \it(6)	&3	&& [composite nets: embedding dimension $n_{E}$=8; loss weights\tablenotemarkedit{b}=\{1:1\}]  \\
		\hline		
	\end{tabular}
	\end{center}}
	\setlength{\tabcolsep}{.385cm} 
 	{\begin{center} 
	\begin{tabular}{ccccl}
		\hline\hline
				\sc{Layer type} 	
				&\sc{Configuration id}
				&\sc{Nb layers} ($n_{L}$)
				&\sc{Size} 	($n_{S}$)
				&\sc{Common hyperparameters} \\			
		\hline
			\multirow{6}{*}{temporal CNN}	
				&\it(1) 	&1	&\multirow{3}{*}{16}	& convolution kernel size $n_{K}$=5;		
				    \\	\cline{3-3}
				&\it(2)	&2 	&& categorical classification;  						
				    \\	\cline{3-3}
				&\it(3)	&3	&& drop fraction=0.25;	   							
				    \\	\cline{3-4}
				&\it(4)	&1	&\multirow{3}{*}{32}	& batch size=128; maximum training epochs=200;	
				\\	\cline{3-3}
				&\it(5)	&2	&& \texttt{Adam}\tablenotemarkedit{a} optimizer; optimizer learning rate=5$\times10^{-4}$;	
				    \\	\cline{3-3}
				&\it(6)	&3	&& [composite nets: embedding dimension $n_{E}$=8; loss weights\tablenotemarkedit{b}=\{1:1\}] \\
		\hline	 
	\end{tabular} 
	\end{center}}
	\setlength{\tabcolsep}{.42cm}
 	{\begin{center}
	\begin{tabular}{ccccl}
		\hline\hline
				\sc{Layer type} 	
				&\sc{Configuration id}
				&\sc{Nb stacks} ($n_{L}$)
				&\sc{Size} 	 ($n_{S}$)
				&\sc{Common hyperparameters} \\	
		\hline	 						
			\multirow{6}{*}{dilated TCN}	
				&\it(1)	&1	&\multirow{3}{*}{16} & dilation rate\tablenotemarkedit{c} $r$=2; convolution kernel size $n_{K}$=3; 		
				    \\	\cline{3-3}
				&\it(2)	&2	&& categorical classification;  						
				    \\	\cline{3-3}
				&\it(3)	&3	&& drop fraction=0.25;	 							
				    \\ 	\cline{3-4}
				&\it(4) 	&1	&\multirow{3}{*}{32}	& batch size=128; maximum training epochs=200;					
				    \\	\cline{3-3}
				&\it(5)	&2	&& \texttt{Adam}\tablenotemarkedit{a} optimizer; optimizer learning rate=5$\times10^{-4}$;	
				    \\	\cline{3-3}
				&\it(6)	&3	&& [composite nets: embedding dimension $n_{E}$=8; loss weights\tablenotemarkedit{b}=\{1:1\}] \\
		\hline	
	\end{tabular}
	\tablenotetextedit{a}{\texttt{Adam} optimization algorithm \citep{kingma_adam_2017}.}
	\tablenotetextedit{b}{Loss weights report the contributions of individual branches $\{w_{ec}, w_{ed}\}$ in the composite networks, respectively the encoder-classifier and encoder-decoder branches.}
	\tablenotetextedit{c}{Dilation factors in dTCN correspond to $\{2^{j-1}\}_{j:1\rightarrow r}$, with $r>1$ the dilation rate.}
	\end{center}}
}\end{table}

\begin{figure}[htp!]
	\centering
	\includegraphics[width=0.67\textwidth]{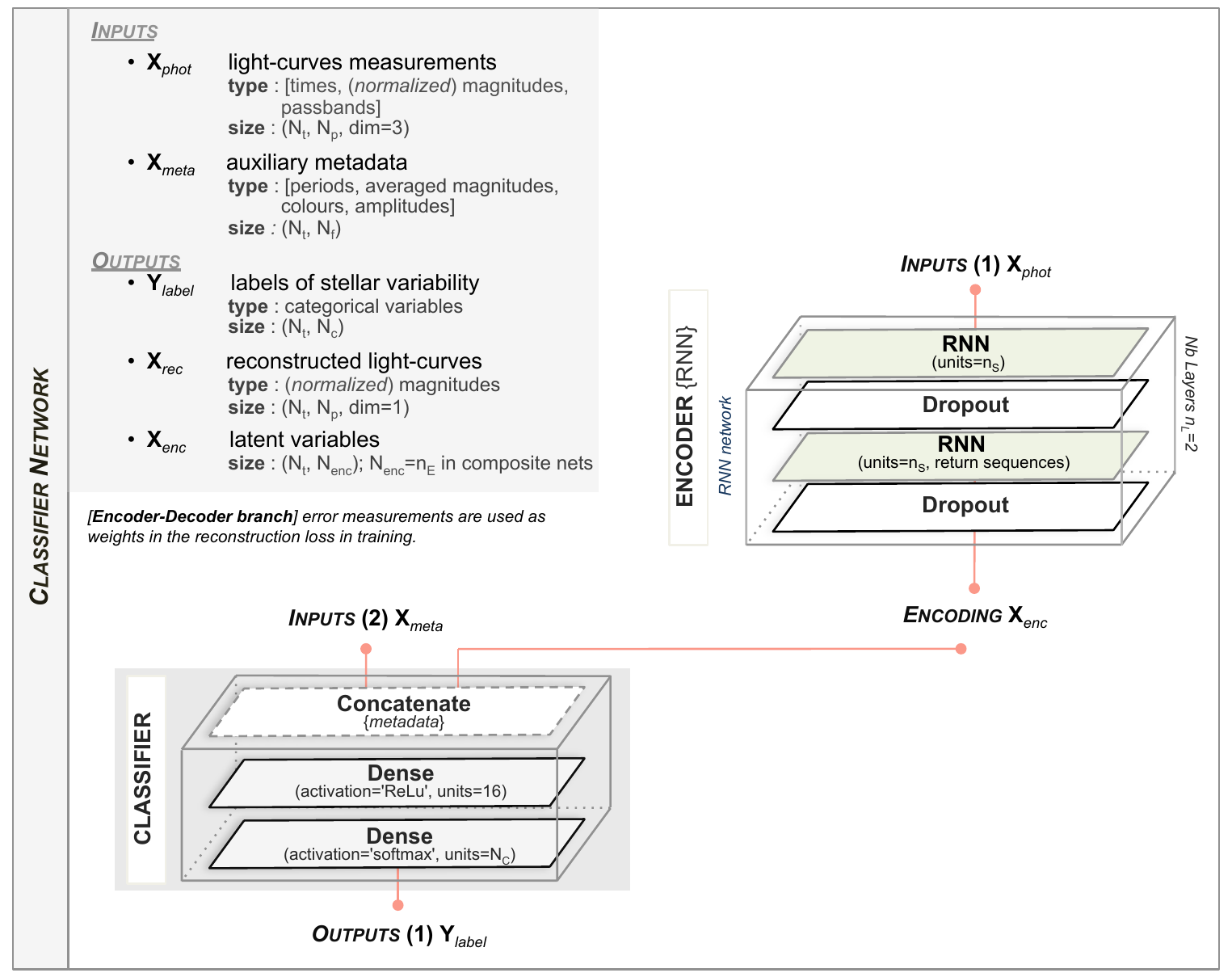}
	\includegraphics[width=0.67\textwidth]{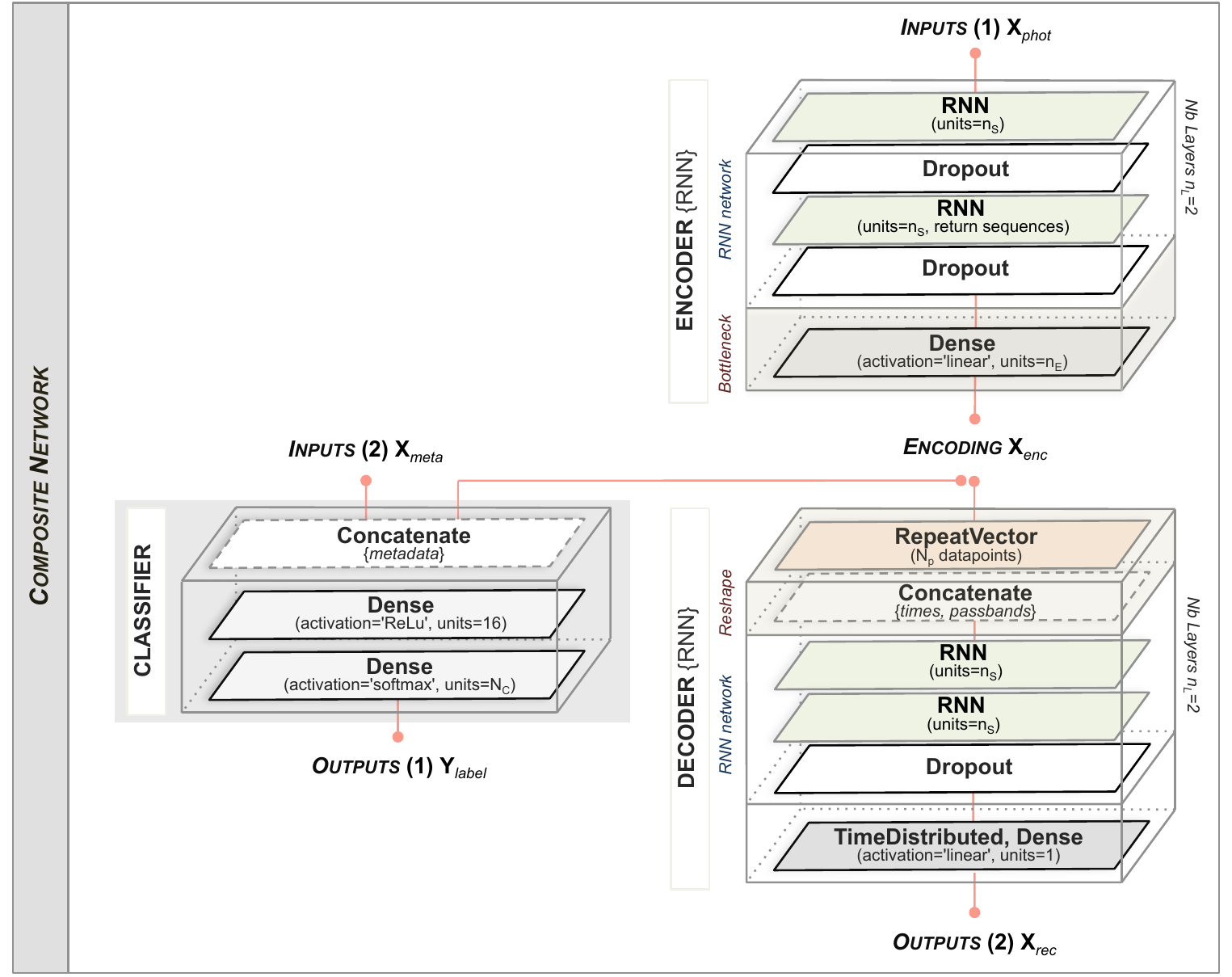}
	\caption{Architectures of the direct classifier and composite RNNs.
			Naming convention follows the implementation in \texttt{keras}.} 
	\label{fig:archidet_RNN}
\end{figure}

\begin{figure}[htp!]
	\centering
	\includegraphics[width=0.67\textwidth]{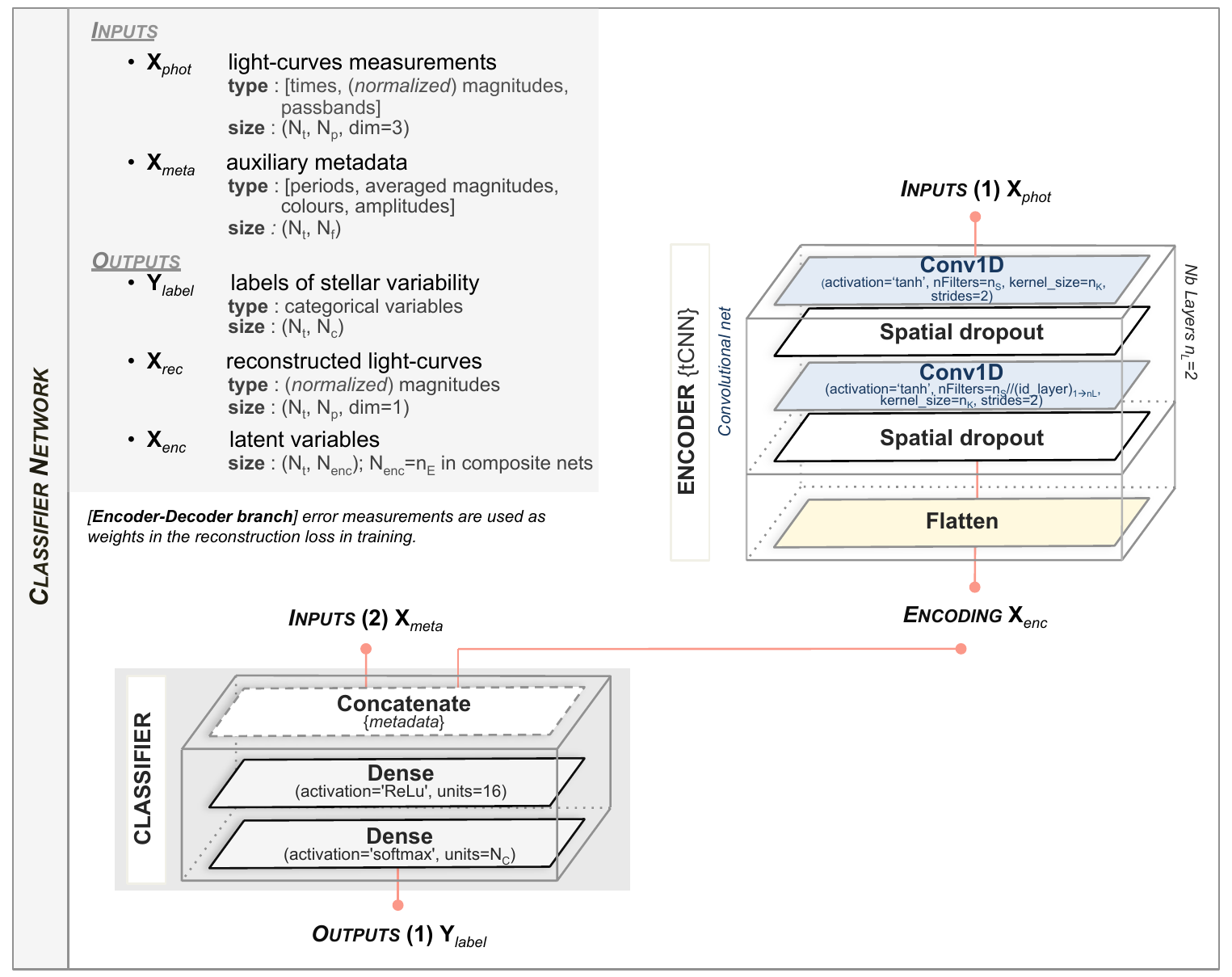}
	\includegraphics[width=0.67\textwidth]{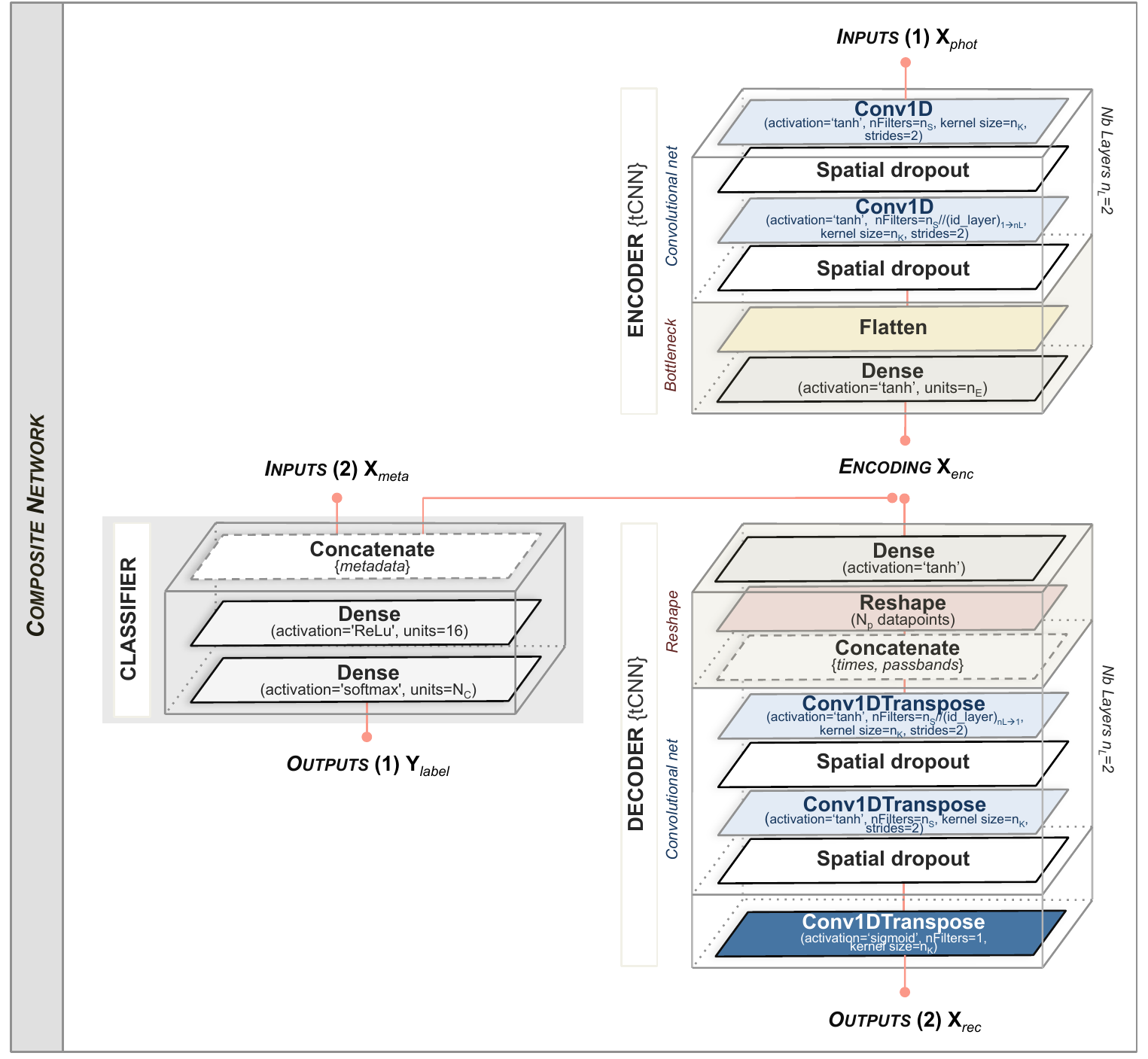}
	\caption{Architectures of the direct classifier and composite tCNNs.
			Naming convention follows the implementation in \texttt{keras}.} 
	\label{fig:archidet_tCNN}
\end{figure}

\begin{figure}[htp!]
	\vspace{-1.2cm}
	\centering	
	\includegraphics[width=0.65\textwidth]{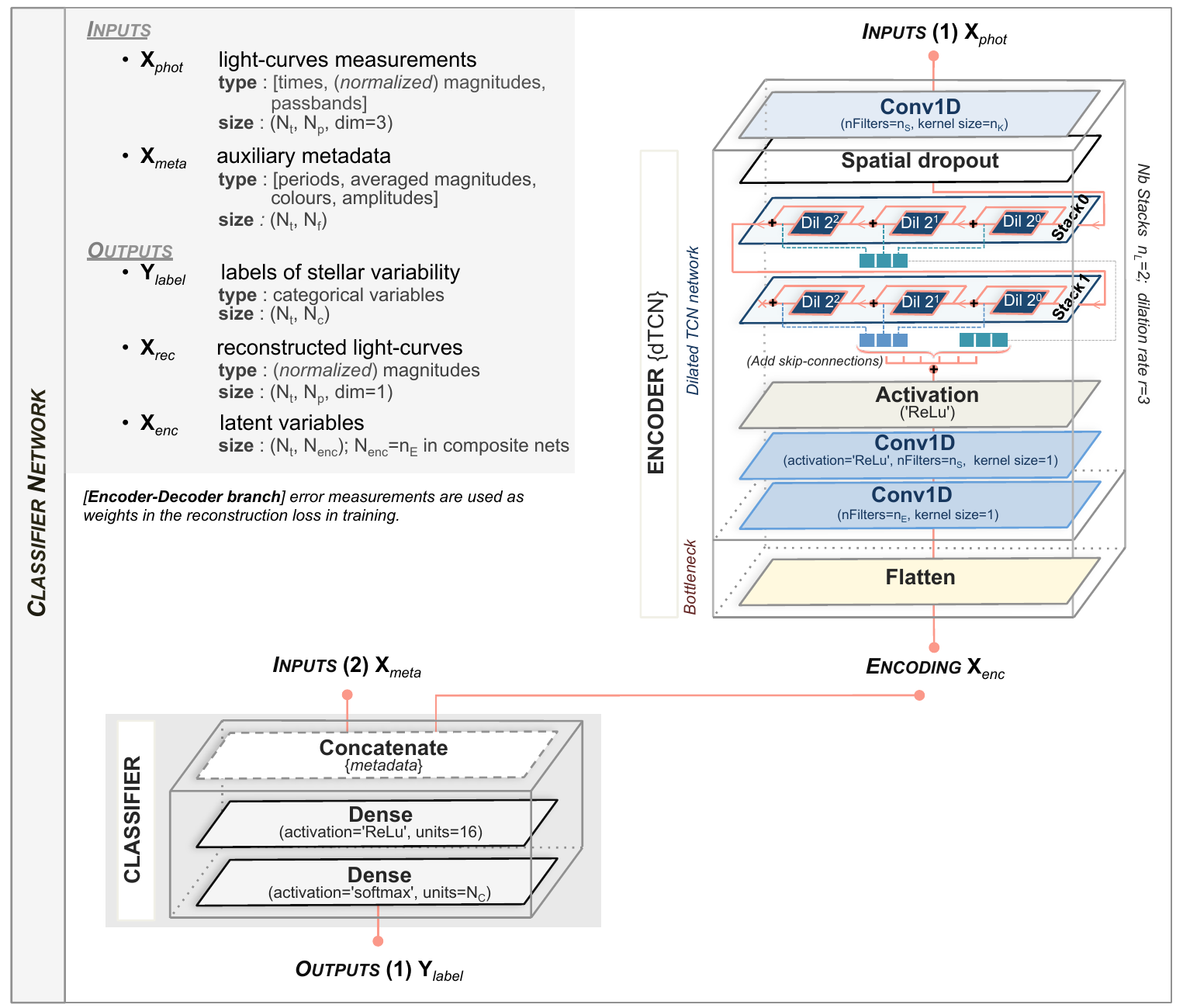}
	\includegraphics[width=0.65\textwidth]{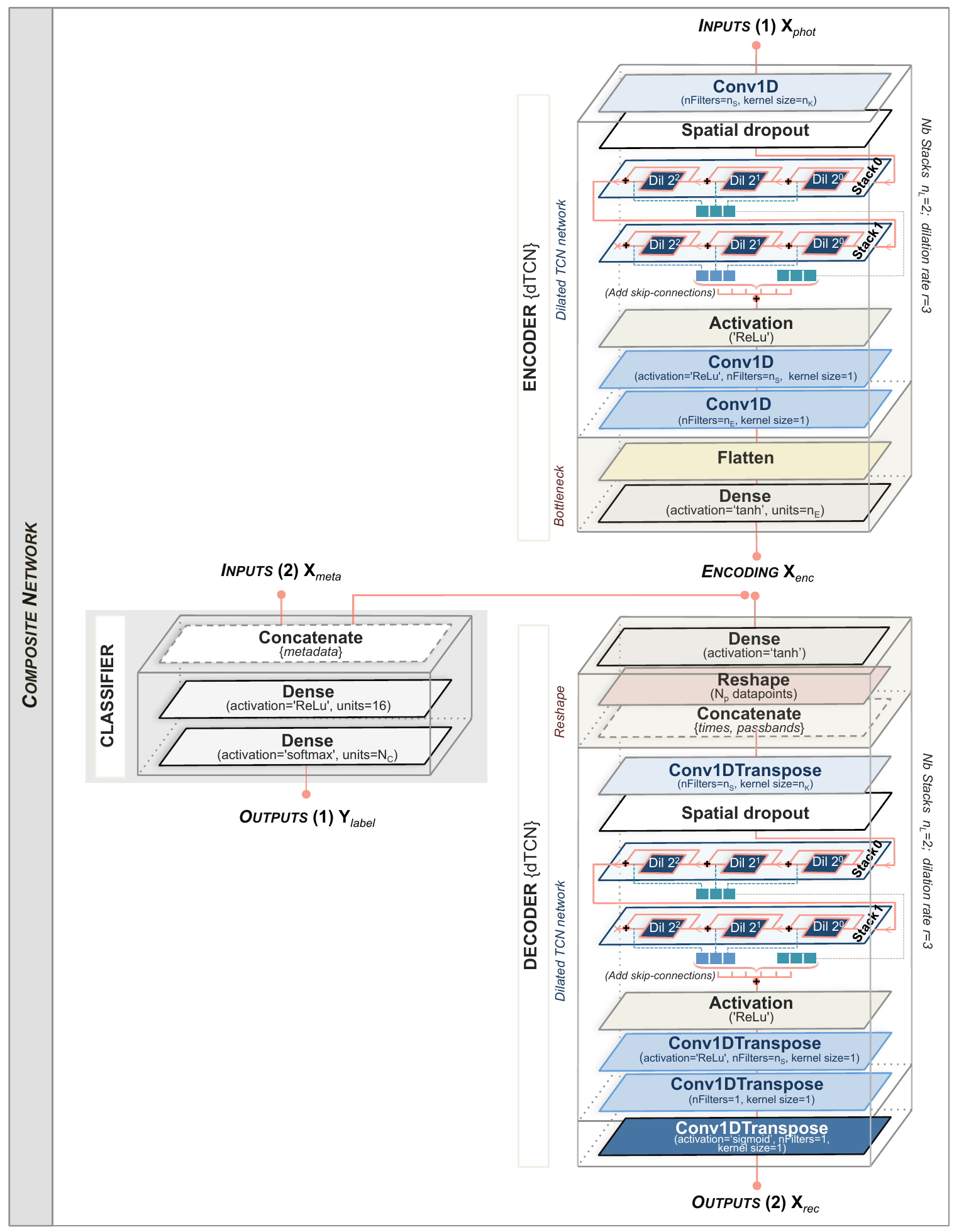}
	\vspace{-.25cm}
	\caption{{** Corrected figures ** }
			Architectures of the direct classifier and composite dTCNs. 
			The series of dilated convolutions and residual stacks follow the \texttt{Wavenet} architecture \citep{oord_wavenet_2016} 
				augmented with additional dropout functions to prevent overfitting.
			Naming convention follows the implementation in \texttt{keras}.} 
	\label{fig:archidet_dTCN}
\end{figure}

\newpage
\begin{table}[h]{
	\scriptsize  
	\caption{Network sizes corresponding to the total number of parameters per model across three different datasets (B-band only [top], and two variants of the combination of R- and B-bands [middle and bottom]). 
	The identifiers of the hyperparameters set configurations \textit{(1)} to \textit{(6)} are ordered by increasing size. 
	Largest networks for the RNNs, tCNNs and dTCNs correspond respectively to the identifiers \textit{(6)}, \textit{(4)} and \textit{(6)} and the smallest configurations to \textit{(1)}, \textit{(3)} and \textit{(1)}.
	The metadata in the current application ($N_f=6$) adds 96 parameters.}
	\label{table:network_sizes}	 
	\parbox{.5\linewidth}
	{\begin{center} \setlength{\tabcolsep}{.13cm}\hspace{-1.5cm}
	\begin{tabular}{c|c|c|c|c|c}	
		\hline\hline 
			\multirow{24}{*}{\centering{\rotatebox[origin=c]{90}{\itbf{\color{blue!35!black} MACHO - B$_{band}$}}}}	
				&{\sc Network}
					&  \multirow{2}{*}{$c_{F}$} 
					&  \multirow{2}{*}{$c_{F,meta}$} 
					&  \multirow{2}{*}{$d_{F}$}
					&  \multirow{2}{*}{$d_{F,meta}$} \\
				&\sc type&&&\\
		\hhline{~*{5}{-}} \hhline{~*{5}{-}} \hhline{~*{5}{-}}
		&\multirow{6}{*}{LSTM}
			& 3292;  \it(1)& 3388;  \it(1)& 6661;  \it(1)& 6757;  \it(1)\\
			&& 9564;  \it(2)& 9660;  \it(2)& 19205;  \it(2)& 19301;  \it(2)\\
			&& 10460;  \it(4)& 10556;  \it(4)& 21157;  \it(4)& 21253;  \it(4)\\
			&& 15836;  \it(3)& 15932;  \it(3)& 31749;  \it(3)& 31845;  \it(3)\\
			&& 35292;  \it(5)& 35388;  \it(5)& 70821;  \it(5)& 70917;  \it(5)\\
			&& 60124;  \it(6)& 60220;  \it(6)& 120485;  \it(6)& 120581;  \it(6)\\
		\hhline{~*{5}{-}} \hhline{~*{5}{-}} \hhline{~*{5}{-}}
		&\multirow{6}{*}{GRU}
			& 2652;  \it(1)& 2748;  \it(1)& 5157;  \it(1)& 5253;  \it(1)\\
			&& 7356;  \it(2)& 7452;  \it(2)& 14565;  \it(2)& 14661;  \it(2)\\
			&& 8156;  \it(4)& 8252;  \it(4)& 16101;  \it(4)& 16197;  \it(4)\\
			&& 12060;  \it(3)& 12156;  \it(3)& 23973;  \it(3)& 24069;  \it(3)\\
			&& 26780;  \it(5)& 26876;  \it(5)& 53349;  \it(5)& 53445;  \it(5)\\
			&& 45404;  \it(6)& 45500;  \it(6)& 90597;  \it(6)& 90693;  \it(6)\\
		\hhline{~*{5}{-}} \hhline{~*{5}{-}} \hhline{~*{5}{-}}
		&\multirow{6}{*}{tCNN}
			& 5329;  \it(3)& 5425;  \it(3)& 8371;  \it(3)& 8467;  \it(3)\\
			&& 12118;  \it(6)& 12214;  \it(6)& 15925;  \it(2)& 16021;  \it(2)\\
			&& 13924;  \it(2)& 14020;  \it(2)& 16465;  \it(6)& 16561;  \it(6)\\
			&& 28908;  \it(5)& 29004;  \it(5)& 26853;  \it(5)& 26949;  \it(5)\\
			&& 51676;  \it(1)& 51772;  \it(1)& 41349;  \it(1)& 41445;  \it(1)\\
			&& 103132;  \it(4)& 103228;  \it(4)& 67941;  \it(4)& 68037;  \it(4)\\
		\hhline{~*{5}{-}} \hhline{~*{5}{-}} \hhline{~*{5}{-}}
		&\multirow{6}{*}{dTCN}
			& 54100;  \it(1)& 54196;  \it(1)& 60335;  \it(1)& 60431;  \it(1)\\
			&& 56212;  \it(2)& 56308;  \it(2)& 64559;  \it(2)& 64655;  \it(2)\\
			&& 58324;  \it(3)& 58420;  \it(3)& 68783;  \it(3)& 68879;  \it(3)\\
			&& 61380;  \it(4)& 61476;  \it(4)& 75119;  \it(4)& 75215;  \it(4)\\
			&& 69700;  \it(5)& 69796;  \it(5)& 91759;  \it(5)& 91855;  \it(5)\\
			&& 78020;  \it(6)& 78116;  \it(6)& 108399;  \it(6)& 108495;  \it(6)\\

		\hline 
	\end{tabular}
	\end{center}}	
	\parbox{.5\linewidth} 
	{\begin{center} 	\setlength{\tabcolsep}{.13cm}\hspace{-1.5cm}
	\begin{tabular}{c|c|c|c|c|c}	
		\hline\hline 
			\multirow{24}{*}{\centering{\rotatebox[origin=c]{90}{\itbf{\color{cyan!50!black}MACHO - RB$_{merged}$}}}} 
					&{\sc Network}
					&  \multirow{2}{*}{$c_{F}$} 
					&  \multirow{2}{*}{$c_{F,meta}$} 
					&  \multirow{2}{*}{$d_{F}$}
					&  \multirow{2}{*}{$d_{F,meta}$} \\
				&\sc type&&&\\
		\hhline{~*{5}{-}}  
		&\multirow{6}{*}{LSTM}
			& 3292 ; \it(1) & 3388 ; \it(1) & 6661 ; \it(1) & 6757 ; \it(1) \\
			&& 9564 ; \it(2) & 9660 ; \it(2) & 19205 ; \it(2) & 19301 ; \it(2) \\
			&& 10460 ; \it(4) & 10556 ; \it(4) & 21157 ; \it(4) & 21253 ; \it(4) \\
			&& 15836 ; \it(3) & 15932 ; \it(3) & 31749 ; \it(3) & 31845 ; \it(3) \\
			&& 35292 ; \it(5) & 35388 ; \it(5) & 70821 ; \it(5) & 70917 ; \it(5) \\
			&& 60124 ; \it(6) & 60220 ; \it(6) & 120485 ; \it(6) & 120581 ; \it(6) \\
		\hhline{~*{5}{-}} \hhline{~*{5}{-}} \hhline{~*{5}{-}}
		&\multirow{6}{*}{GRU}
			& 2652 ; \it(1) & 2748 ; \it(1) & 5157 ; \it(1) & 5253 ; \it(1) \\
			&& 7356 ; \it(2) & 7452 ; \it(2) & 14565 ; \it(2) & 14661 ; \it(2) \\
			&& 8156 ; \it(4) & 8252 ; \it(4) & 23973 ; \it(3) & 16197 ; \it(4) \\
			&& 12060 ; \it(3) & 12156 ; \it(3) & 16101 ; \it(4) & 24069 ; \it(3) \\
			&& 26780 ; \it(5) & 26876 ; \it(5) & 53349 ; \it(5) & 53445 ; \it(5) \\
			&& 45404 ; \it(6) & 45500 ; \it(6) & 90597 ; \it(6) & 90693 ; \it(6) \\
		\hhline{~*{5}{-}} \hhline{~*{5}{-}} \hhline{~*{5}{-}}
		&\multirow{6}{*}{tCNN}
			& 9329 ; \it(3) & 9425 ; \it(3) & 13971 ; \it(3) & 14067 ; \it(3) \\
			&& 20118 ; \it(6) & 20214 ; \it(6) & 24065 ; \it(6) & 24161 ; \it(6) \\
			&& 26724 ; \it(2) & 26820 ; \it(2) & 29525 ; \it(2) & 29621 ; \it(2) \\
			&& 54508 ; \it(5) & 54604 ; \it(5) & 46853 ; \it(5) & 46949 ; \it(5) \\
			&& 102876 ; \it(1) & 102972 ; \it(1) & 81349 ; \it(1) & 81445 ; \it(1) \\
			&& 205532 ; \it(4) & 205628 ; \it(4) & 133541 ; \it(4) & 133637 ; \it(4) \\
		\hhline{~*{5}{-}} \hhline{~*{5}{-}} \hhline{~*{5}{-}}
		&\multirow{6}{*}{dTCN}
			& 105300 ; \it(1) & 105396 ; \it(1) & 114735 ; \it(1) & 114831 ; \it(1) \\
			&& 107412 ; \it(2) & 107508 ; \it(2) & 118959 ; \it(2) & 119055 ; \it(2) \\
			&& 109524 ; \it(3) & 109620 ; \it(3) & 123183 ; \it(3) & 123279 ; \it(3) \\
			&& 112580 ; \it(4) & 112676 ; \it(4) & 129519 ; \it(4) & 129615 ; \it(4) \\
			&& 120900 ; \it(5) & 120996 ; \it(5) & 146159 ; \it(5) & 146255 ; \it(5) \\
			&& 129220 ; \it(6) & 129316 ; \it(6) & 162799 ; \it(6) & 162895 ; \it(6) \\		
		\hline 
	\end{tabular}
	\end{center}}	
	
	\vspace{-.3cm}	
	{\begin{center} \setlength{\tabcolsep}{.13cm}\hspace{-1.5cm}
	\centering
	\begin{tabular}{c|c|c|c|c|c}	
		\hline\hline 
			\multirow{24}{*}{\centering{\rotatebox[origin=c]{90}{\itbf{\color{violet!60!black}MACHO - RB$_{hybrid}$}}}} 	
					&{\sc Network}
					&  \multirow{2}{*}{$c_{F}$} 
					&  \multirow{2}{*}{$c_{F,meta}$} 
					&  \multirow{2}{*}{$d_{F}$}
					&  \multirow{2}{*}{$d_{F,meta}$} \\
				&\sc type&&&\\
		\hhline{~*{5}{-}} 
		&\multirow{6}{*}{LSTM}
			& 6364 ; \it(1) & 6460 ; \it(1) & 13110 ; \it(1) & 13206 ; \it(1) \\
			&& 18908 ; \it(2) & 19004 ; \it(2) & 38198 ; \it(2) & 38294 ; \it(2) \\
			&& 20700 ; \it(4) & 20796 ; \it(4) & 42102 ; \it(4) & 42198 ; \it(4) \\
			&& 31452 ; \it(3) & 31548 ; \it(3) & 63286 ; \it(3) & 63382 ; \it(3) \\
			&& 70364 ; \it(5) & 70460 ; \it(5) & 141430 ; \it(5) & 141526 ; \it(5) \\
			&& 120028 ; \it(6) & 120124 ; \it(6) & 240758 ; \it(6) & 240854 ; \it(6) \\
		\hhline{~*{5}{-}} \hhline{~*{5}{-}} \hhline{~*{5}{-}}
		&\multirow{6}{*}{GRU}
			& 5084 ; \it(1) & 5180 ; \it(1) & 10102 ; \it(1) & 10198 ; \it(1) \\
			&& 14492 ; \it(2) & 14588 ; \it(2) & 28918 ; \it(2) & 29014 ; \it(2) \\
			&& 16092 ; \it(4) & 16188 ; \it(4) & 31990 ; \it(4) & 32086 ; \it(4) \\
			&& 23900 ; \it(3) & 23996 ; \it(3) & 47734 ; \it(3) & 47830 ; \it(3) \\
			&& 53340 ; \it(5) & 53436 ; \it(5) & 106486 ; \it(5) & 106582 ; \it(5) \\
			&& 90588 ; \it(6) & 90684 ; \it(6) & 180982 ; \it(6) & 181078 ; \it(6) \\
		\hhline{~*{5}{-}} \hhline{~*{5}{-}} \hhline{~*{5}{-}}
		&\multirow{6}{*}{tCNN}
			& 10438 ; \it(3) & 10534 ; \it(3) & 16530 ; \it(3) & 16626 ; \it(3) \\
			&& 24016 ; \it(6) & 24112 ; \it(6) & 31638 ; \it(2) & 31734 ; \it(2) \\
			&& 27628 ; \it(2) & 27724 ; \it(2) & 32718 ; \it(6) & 32814 ; \it(6) \\
			&& 57596 ; \it(5) & 57692 ; \it(5) & 53494 ; \it(5) & 53590 ; \it(5) \\
			&& 103132 ; \it(1) & 103228 ; \it(1) & 82486 ; \it(1) & 82582 ; \it(1) \\
			&& 206044 ; \it(4) & 206140 ; \it(4) & 135670 ; \it(4) & 135766 ; \it(4) \\
		\hhline{~*{5}{-}} \hhline{~*{5}{-}} \hhline{~*{5}{-}}
		&\multirow{6}{*}{dTCN}
			& 107980 ; \it(1) & 108076 ; \it(1) & 120458 ; \it(1) & 120554 ; \it(1) \\
			&& 112204 ; \it(2) & 112300 ; \it(2) & 128906 ; \it(2) & 129002 ; \it(2) \\
			&& 116428 ; \it(3) & 116524 ; \it(3) & 137354 ; \it(3) & 137450 ; \it(3) \\
			&& 122540 ; \it(4) & 122636 ; \it(4) & 150026 ; \it(4) & 150122 ; \it(4) \\
			&& 139180 ; \it(5) & 139276 ; \it(5) & 183306 ; \it(5) & 183402 ; \it(5) \\
			&& 155820 ; \it(6) & 155916 ; \it(6) & 216586 ; \it(6) & 216682 ; \it(6) \\
		\hline 
	\end{tabular}
	\end{center}}	
}\end{table}

}

\newpage~\newpage
\setcounter{table}{0} \setcounter{figure}{0}
\section{Global performances} \label{sec:anx_clfperf}{

\begin{table}[htp!]
	\vspace{.5cm}
	\caption{
		Training convergence time (expressed in hours) of all models across three different datasets (B-band only [top], 
		and two variants of the combination of R- and B-bands [middle and bottom]). 
		Runs are performed on a CPU model \texttt{Intel Xeon E5-2643v3} using 4 cores per run at maximum capacity.
		The identifiers of the hyperparameters set configurations \textit{(1)} to \textit{(6)} are stated in the 
		Appendix Table \ref{table:networks_config1}.
		} 	
	\label{table:cputime_training}	
	\hspace{-1cm}
	\begin{minipage}[t]{1\linewidth}
		\includegraphics[width=1.1\textwidth]{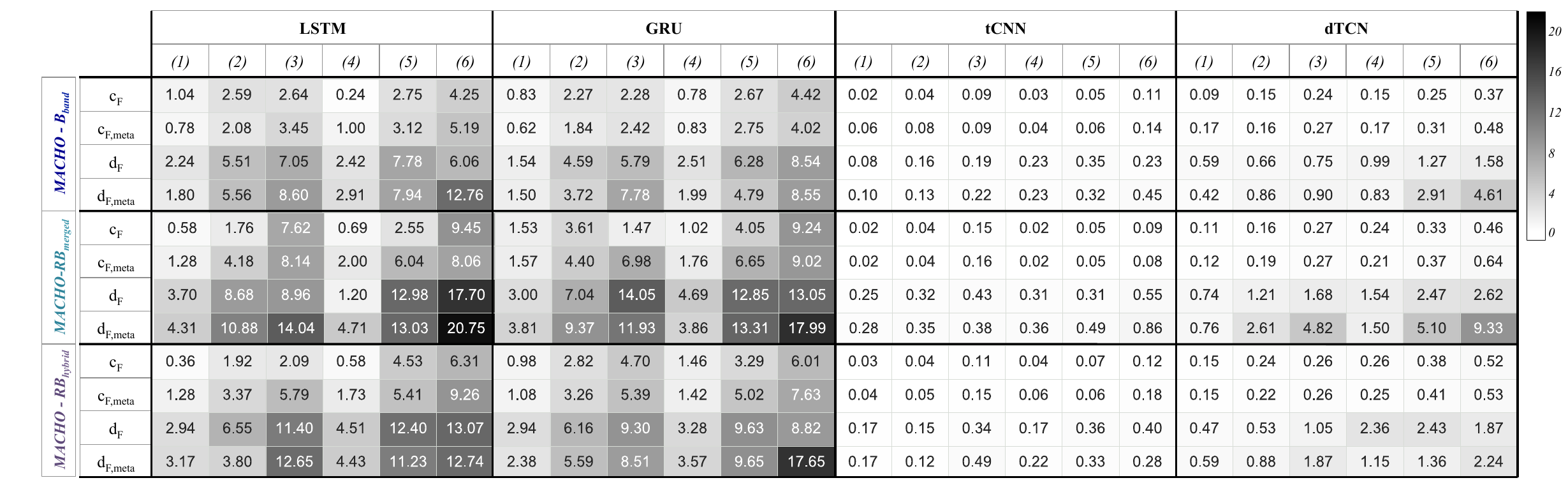}  
	\end{minipage}
\end{table}

\vspace{1cm}
\begin{table}[htp]{
	\scriptsize
	\caption{
			Hyperparameters set configurations identified for the best-performing networks based on minimum loss 
			obtained on the Test set across three different datasets (B-band only [top], and two variants of the combination 
			of R- and B-bands [middle and bottom]).
			The identifiers \textit{(1)} to \textit{(6)} are stated in the Appendix Table \ref{table:networks_config1}.
		}  
	\label{table:classif_perfoBestID}
	{\begin{center}{  
	\begin{tabular}{ c | c | cccc}
		\hline\hline 
		\multirow{5}{*}{\centering{\rotatebox[origin=c]{90}{\itbf{\color{blue!35!black} MACHO - B$_{band}$}}}}
			&{ \sc{Id net}}	
				&\multicolumn{1}{c}{LSTM} 
				&\multicolumn{1}{c}{GRU} 
				&\multicolumn{1}{c}{tCNN}
				&\multicolumn{1}{c}{dTCN} \\	
		\hhline{~-----}
		&$c_{F}$
			& \it (6)	
			& \it (6)	
			& \it (3)	
			& \it (1)	
			\\
		\hhline{~-----}
		&$c_{F,meta}$
			& \it (1)	
			& \it (5)	
			& \it (3)	
			& \it (1)	
			\\
		\hhline{~-----}
		&$d_{F}$
			& \it (5)	
			& \it (5)	
			& \it (4)	
			& \it (4)	
			\\
		\hhline{~-----}
		&$d_{F,meta}$
			& \it (6)	
			& \it (4)	
			& \it (5)	
			& \it (1)	
			\\
		\hline  
	\end{tabular}
	}\end{center}}
	{\begin{center}{ 	
	\begin{tabular}{c | c | cccc}
		\hline\hline 
		\multirow{5}{*}{\centering{\rotatebox[origin=c]{90}{\itbf{\color{cyan!50!black}MACHO - RB$_{merged}$ }}}} 
			&{ \sc{Id net}}	
				&\multicolumn{1}{c}{LSTM} 
				&\multicolumn{1}{c}{GRU} 
				&\multicolumn{1}{c}{tCNN}
				&\multicolumn{1}{c}{dTCN} \\											
		\hhline{~-----} 
		&$c_{F}$
			& \it (6)	
			& \it (6)	
			& \it (3)	
			& \it (3)	
			\\
		\hhline{~-----}
		&$c_{F,meta}$
			& \it (1)	
			& \it (5)	
			& \it (2)	
			& \it (1)	
			\\
		\hhline{~-----}
		&$d_{F}$
			& \it (5)	
			& \it (5)	
			& \it (5)	
			& \it (5)	
			\\
		\hhline{~-----}
		&$d_{F,meta}$
			& \it (6)	
			& \it (4)	
			& \it (6)	
			& \it (4)	
			\\
		\hline 	
	\end{tabular}
	}\end{center}}
	{\begin{center}{
	\begin{tabular}{c | c | cccc}
		 \hline\hline 	
		\multirow{5}{*}{\centering{\rotatebox[origin=c]{90}{\itbf{\color{violet!60!black}MACHO - RB$_{hybrid}$ }}}}				
			&{ \sc{Id net}}	
				&\multicolumn{1}{c}{LSTM} 
				&\multicolumn{1}{c}{GRU} 
				&\multicolumn{1}{c}{tCNN}
				&\multicolumn{1}{c}{dTCN} \\										
		\hhline{~-----}
		&$c_{F}$
			& \it (5)	
			& \it (3)	
			& \it (3)	
			& \it (1)	
			\\
		\hhline{~-----}
		&$c_{F,meta}$
			& \it (3)	
			& \it (2)	
			& \it (6)	
			& \it (1)	
			\\
		\hhline{~-----}
		&$d_{F}$
			& \it (5)	
			& \it (5)	
			& \it (5)	
			& \it (4)	
			\\
		\hhline{~-----}
		&$d_{F,meta}$
			& \it (5)	
			& \it (1)	
			& \it (4)	
			& \it (4)	
			\\
		\hline 	
	\end{tabular}
	}\end{center}}	
}\end{table} 

\begin{table}[htp]{
	\scriptsize
	\caption{
			Classification accuracy evaluated on the Test set for the best-performing networks (see text for a description) 
			across three different datasets (B-band only [top], and two variants of the combination of R- and B-bands 
			[middle and bottom]). 
		The classification accuracy is evaluated for the three main variability groups: short-period pulsators (group 1), 
		eclipsing binaries (group 2) and LPVs (group 3).
	} 
	\vspace{-.25cm}
	\label{table:perfoBest_acc_pergroup}
	 \setlength{\tabcolsep}{.1cm} 
	{\begin{center}{ 
 	\begin{tabular}{ c| c | c|ccc | c|ccc | c|ccc | c|ccc}
		\hline\hline 
			\multirow{6}{*}{\centering{\rotatebox[origin=c]{90}{\itbf{\color{blue!35!black} MACHO - B$_{band}$ }}}}
			& \multirow{2}{*}{\sc{Id net}}
				&\multicolumn{4}{c|}{LSTM} 
				&\multicolumn{4}{c|}{GRU} 
				&\multicolumn{4}{c|}{tCNN}
				&\multicolumn{4}{c}{dTCN} \\								
		\hhline{~~----------------} 
		& 	&Full & Group1& Group2& Group3
			&Full & Group1& Group2& Group3
			&Full & Group1& Group2& Group3
			&Full & Group1& Group2& Group3		
			\\
		\hhline{~-----------------} 
		&{$c_{F}$ }	 
			& 0.749	& 0.819	& 0.802	& 0.428	
			& 0.781	& 0.834	& 0.831	& 0.520	
			& 0.732	& 0.811	& 0.797	& 0.359	
			& 0.675	& 0.765	& 0.752	& 0.247	
			\\
		\hhline{~-----------------} 
		&{$c_{F,meta}$ }	 
			& 0.916	& 0.936	& 0.886	& 0.879	
			& 0.907	& 0.924	& 0.905	& 0.850	
			& 0.887	& 0.909	& 0.890	& 0.801	
			& 0.786	& 0.807	& 0.868	& 0.608	
			\\
		\hhline{~-----------------} 
		&{$d_{F}$ }	 
			& 0.730	& 0.798	& 0.787	& 0.407	
			& 0.739	& 0.811	& 0.772	& 0.431	
			& 0.701	& 0.781	& 0.760	& 0.337	
			& 0.689	& 0.782	& 0.777	& 0.239	
			\\
		\hhline{~-----------------} 
		&{$d_{F,meta}$ }	 
			& 0.905	& 0.930	& 0.887	& 0.833	
			& 0.886	& 0.943	& 0.907	& 0.652	
			& 0.900	& 0.929	& 0.838	& 0.870	
			& 0.802	& 0.848	& 0.819	& 0.611	
			\\
		\hline  
	\end{tabular}
	}\end{center}}
	{\begin{center}{ 
 	\begin{tabular}{ c| c | c|ccc | c|ccc | c|ccc | c|ccc}
		\hline\hline 
			\multirow{6}{*}{\centering{\rotatebox[origin=c]{90}{\itbf{\color{cyan!50!black}MACHO - RB$_{merged}$ }}}} 
			& \multirow{2}{*}{\sc{Id net}}
				&\multicolumn{4}{c|}{LSTM} 
				&\multicolumn{4}{c|}{GRU} 
				&\multicolumn{4}{c|}{tCNN}
				&\multicolumn{4}{c}{dTCN} \\								
		\hhline{~~----------------} 
		& 	&Full & Group1& Group2& Group3
			&Full & Group1& Group2& Group3
			&Full & Group1& Group2& Group3
			&Full & Group1& Group2& Group3		
			\\
		\hhline{~-----------------} 
		&{$c_{F}$ }	 
			& 0.737	& 0.808	& 0.755	& 0.455	
			& 0.780	& 0.834	& 0.819	& 0.532	
			& 0.722	& 0.794	& 0.785	& 0.379	
			& 0.667	& 0.729	& 0.768	& 0.315	
			\\
		\hhline{~-----------------} 
		&{$c_{F,meta}$ }	 
			& 0.890	& 0.913	& 0.856	& 0.845	
			& 0.910	& 0.933	& 0.888	& 0.852	
			& 0.815	& 0.848	& 0.805	& 0.707	
			& 0.747	& 0.783	& 0.772	& 0.582	
			\\
		\hhline{~-----------------} 
		&{$d_{F}$ }	 
			& 0.726	& 0.810	& 0.771	& 0.364	
			& 0.738	& 0.813	& 0.779	& 0.412	
			& 0.691	& 0.770	& 0.774	& 0.298	
			& 0.686	& 0.769	& 0.762	& 0.290	
			\\
		\hhline{~-----------------} 
		&{$d_{F,meta}$ }	 
			& 0.906	& 0.924	& 0.896	& 0.854	
			& 0.883	& 0.938	& 0.907	& 0.653	
			& 0.912	& 0.935	& 0.894	& 0.848	
			& 0.814	& 0.856	& 0.864	& 0.594	
			\\
		\hline  
	\end{tabular}
	}\end{center}}
	{\begin{center}{
 	\begin{tabular}{ c| c | c|ccc | c|ccc | c|ccc | c|ccc}
		\hline\hline 
			\multirow{6}{*}{\centering{\rotatebox[origin=c]{90}{\itbf{\color{violet!60!black}MACHO - RB$_{hybrid}$ }}}}
			& \multirow{2}{*}{\sc{Id net}}
				&\multicolumn{4}{c|}{LSTM} 
				&\multicolumn{4}{c|}{GRU} 
				&\multicolumn{4}{c|}{tCNN}
				&\multicolumn{4}{c}{dTCN} \\								
		\hhline{~~----------------} 
		& 	&Full & Group1& Group2& Group3
			&Full & Group1& Group2& Group3
			&Full & Group1& Group2& Group3
			&Full & Group1& Group2& Group3		
			\\
		\hhline{~-----------------} 
		&{$c_{F}$ }	 
			& 0.776	& 0.841	& 0.809	& 0.495	
			& 0.789	& 0.852	& 0.808	& 0.535	
			& 0.744	& 0.821	& 0.779	& 0.412	
			& 0.696	& 0.768	& 0.755	& 0.359	
			\\
		\hhline{~-----------------} 
		&{$c_{F,meta}$ }	 
			& 0.917	& 0.935	& 0.914	& 0.857	
			& 0.905	& 0.919	& 0.892	& 0.867	
			& 0.845	& 0.858	& 0.866	& 0.768	
			& 0.768	& 0.801	& 0.816	& 0.588	
			\\
		\hhline{~-----------------} 
		&{$d_{F}$ }	 
			& 0.748	& 0.819	& 0.789	& 0.434	
			& 0.749	& 0.821	& 0.772	& 0.456	
			& 0.706	& 0.788	& 0.779	& 0.311	
			& 0.726	& 0.803	& 0.777	& 0.379	
			\\
		\hhline{~-----------------} 
		&{$d_{F,meta}$ }	 
			& 0.905	& 0.921	& 0.894	& 0.859	
			& 0.880	& 0.939	& 0.890	& 0.648	
			& 0.904	& 0.933	& 0.881	& 0.825	
			& 0.818	& 0.873	& 0.872	& 0.545	
			\\
		\hline  
	\end{tabular}
	}\end{center}}		
}\end{table}

\begin{table}[htp]{
	\scriptsize
	\caption{
			Classification metrics evaluated on the Test set for the best-performing networks (see text for a description) 
			across three different datasets (B-band only [top], and two variants of the combination of R- and B-bands 
			[middle and bottom]).  
	} 
	\label{table:perfoBest_othermetrics}
	\setlength{\tabcolsep}{.095cm}
	\begin{center}{\hspace{-2cm}
 	\begin{tabular}{ c|  c | ccc| ccc| ccc| ccc}
		\hline\hline 
			\multirow{6}{*}{\centering{\rotatebox[origin=c]{90}{\itbf{\color{blue!35!black} MACHO - B$_{band}$ }}}}
			& \multirow{2}{*}{\sc{Id net}}
				&\multicolumn{3}{c|}{LSTM} 
				&\multicolumn{3}{c|}{GRU} 
				&\multicolumn{3}{c|}{tCNN}
				&\multicolumn{3}{c}{dTCN} \\								
		\hhline{~~------------} 
		& & Precision$_{M}$ 
				& Recall$_{M}$
				& F1-score$_{M}$
			& Precision$_{M}$ 
				& Recall$_{M}$
				& F1-score$_{M}$
			& Precision$_{M}$ 
				& Recall$_{M}$
				& F1-score$_{M}$
			& Precision$_{M}$ 
				& Recall$_{M}$
				& F1-score$_{M}$	
		\\
		\hhline{~-------------} 
		&{$c_{F}$ }	 
			& 0.503	& 0.515	& 0.504	
			& 0.559	& 0.564	& 0.552	
			& 0.510	& 0.487	& 0.488	
			& 0.447	& 0.424	& 0.425	
			\\
		\hhline{~-------------} 
		&{$c_{F,meta}$ }	 
			& 0.771	& 0.800	& 0.784	
			& 0.763	& 0.780	& 0.770	
			& 0.744	& 0.753	& 0.748	
			& 0.642	& 0.599	& 0.609	
			\\
		\hhline{~-------------} 
		&{$d_{F}$ }	 
			& 0.477	& 0.497	& 0.484	
			& 0.500	& 0.504	& 0.500	
			& 0.459	& 0.452	& 0.454	
			& 0.446	& 0.418	& 0.426	
			\\
		\hhline{~-------------} 
		&{$d_{F,meta}$ }	 
			& 0.765	& 0.774	& 0.768	
			& 0.679	& 0.705	& 0.689	
			& 0.748	& 0.796	& 0.768	
			& 0.597	& 0.627	& 0.603	
			\\
		\hline  
	\end{tabular}
	}\end{center}
	{\begin{center}{ \hspace{-2cm}
 	\begin{tabular}{ c|  c | ccc| ccc| ccc| ccc}
		\hline\hline 
			\multirow{6}{*}{\centering{\rotatebox[origin=c]{90}{\itbf{\color{cyan!50!black}MACHO - RB$_{merged}$ }}}} 
			& \multirow{2}{*}{\sc{Id net}}
				&\multicolumn{3}{c|}{LSTM} 
				&\multicolumn{3}{c|}{GRU} 
				&\multicolumn{3}{c|}{tCNN}
				&\multicolumn{3}{c}{dTCN} \\								
		\hhline{~~------------} 
		& & Precision$_{M}$ 
				& Recall$_{M}$
				& F1-score$_{M}$
			& Precision$_{M}$ 
				& Recall$_{M}$
				& F1-score$_{M}$
			& Precision$_{M}$ 
				& Recall$_{M}$
				& F1-score$_{M}$
			& Precision$_{M}$ 
				& Recall$_{M}$
				& F1-score$_{M}$	
		\\
		\hhline{~-------------} 
		&{$c_{F}$ }	 
			& 0.504	& 0.519	& 0.507	
			& 0.539	& 0.567	& 0.548	
			& 0.472	& 0.483	& 0.474	
			& 0.373	& 0.389	& 0.380	
			\\
		\hhline{~-------------} 
		&{$c_{F,meta}$ }	 
			& 0.744	& 0.757	& 0.749	
			& 0.759	& 0.794	& 0.775	
			& 0.667	& 0.673	& 0.668	
			& 0.565	& 0.566	& 0.562	
			\\
		\hhline{~-------------} 
		&{$d_{F}$ }	 
			& 0.481	& 0.475	& 0.477	
			& 0.499	& 0.498	& 0.496	
			& 0.450	& 0.423	& 0.433	
			& 0.444	& 0.431	& 0.435	
			\\
		\hhline{~-------------} 
		&{$d_{F,meta}$ }	 
			& 0.773	& 0.778	& 0.774	
			& 0.677	& 0.702	& 0.687	
			& 0.765	& 0.788	& 0.776	
			& 0.614	& 0.627	& 0.619	
			\\
		\hline  
	\end{tabular}
	}\end{center}}
	{\begin{center}{\hspace{-2cm}
 	\begin{tabular}{ c|  c | ccc| ccc| ccc| ccc}
		\hline\hline 
			\multirow{6}{*}{\centering{\rotatebox[origin=c]{90}{\itbf{\color{violet!60!black}MACHO - RB$_{hybrid}$ }}}}	
			& \multirow{2}{*}{\sc{Id net}}
				&\multicolumn{3}{c|}{LSTM} 
				&\multicolumn{3}{c|}{GRU} 
				&\multicolumn{3}{c|}{tCNN}
				&\multicolumn{3}{c}{dTCN} \\								
		\hhline{~~------------} 
		& & Precision$_{M}$ 
				& Recall$_{M}$
				& F1-score$_{M}$
			& Precision$_{M}$ 
				& Recall$_{M}$
				& F1-score$_{M}$
			& Precision$_{M}$ 
				& Recall$_{M}$
				& F1-score$_{M}$
			& Precision$_{M}$ 
				& Recall$_{M}$
				& F1-score$_{M}$	
		\\
		\hhline{~-------------} 
		&{$c_{F}$ }	 
			& 0.561	& 0.559	& 0.549	
			& 0.564	& 0.570	& 0.563	
			& 0.520	& 0.520	& 0.518	
			& 0.468	& 0.480	& 0.470	
			\\
		\hhline{~-------------} 
		&{$c_{F,meta}$ }	 
			& 0.784	& 0.790	& 0.785	
			& 0.770	& 0.785	& 0.776	
			& 0.704	& 0.691	& 0.694	
			& 0.615	& 0.600	& 0.606	
			\\
		\hhline{~-------------} 
		&{$d_{F}$ }	 
			& 0.505	& 0.518	& 0.508	
			& 0.516	& 0.522	& 0.516	
			& 0.457	& 0.443	& 0.447	
			& 0.478	& 0.481	& 0.478	
			\\
		\hhline{~-------------} 
		&{$d_{F,meta}$ }	 
			& 0.763	& 0.783	& 0.772	
			& 0.671	& 0.708	& 0.688	
			& 0.747	& 0.774	& 0.758	
			& 0.534	& 0.578	& 0.551	
			\\
		\hline  
	\end{tabular}
	}\end{center}}
}\end{table}

\renewcommand{\arraystretch}{1.05}
\begin{table}[htb]{
	\scriptsize
	\caption{Classification metrics computed on the Test set for the best-performing LSTM direct classifiers $c_{F}$ and $c_{F,meta}$ on the B-band.} 
	\label{table:exp_metrics}
	{\begin{center}{  \setlength{\tabcolsep}{.2cm} \hspace{-1.5cm}
	\begin{tabular}{ c | l | c|c | cccc  | ccc}
		\multicolumn{11}{r}{}{\itbf{Network} $c_{F}$; LSTM; \it best configuration (6); B$_{band}$}\\
		\hline\hline 
		\multirow{12}{*}{\centering{\rotatebox[origin=c]{90}{\itbf{\color{blue!35!black} MACHO - B$_{band}$}}}}
			& \multirow{2}{*}{\sc Classes $\{i\}_{i=1\rightarrow N_C}$}
			& \sc Nb of $\mathbf{Y}_{true}$
			& \sc Nb of  $\mathbf{Y}_{pred}$
			& \multirow{2}{*}{\sc TP$(i)$}	
			& \multirow{2}{*}{\sc FN$(i)$}	
			& \multirow{2}{*}{\sc FP$(i)$}	
			& \multirow{2}{*}{\sc TN$(i)$}
			& \multirow{2}{*}{\sc Precision$(i)$}	
			& \multirow{2}{*}{\sc Recall$(i)$}	
			&\multirow{2}{*}{\sc F1-score$(i)$}	\\
		&&\it (in counts) & \it (in counts)&&&&&&&\\
		\hhline{~----------}
			& RRL (type $ab$)	 	& 1430 & 1574 & 1370 & 60 & 204 & 1890 & 0.870 & 0.958 & 0.912 \\
			& RRL (type $c$)  		& 344 & 333 & 172 & 172 & 161 & 3019 & 0.517 & 0.500 & 0.508 \\
			& RRL (type $e$)  		& 60 & 0 & 0 & 60 & 0 & 3464 & -- & -- & --  \\
			& CEP (FU) 			& 229 & 202 & 174 & 55 & 28 & 3267 & 0.861 & 0.760 & 0.807 \\
			& CEP (FO)			& 133 & 153 & 82 & 51 
										& 71 
										& 3320
										&  0.536 
										& 0.617 
										& 0.573 \\ 
			& LPV (Wood seq. A) 	& 62 & 0 & 0 & 62 & 0 & 3462 & -- & -- & --  \\
			& LPV (Wood seq. B) 	& 160 & 289 & 87 & 73 & 202 & 3162 & 0.301 & 0.544 & 0.388 \\
			& LPV (Wood seq. C) 	& 220 & 164 & 64 & 156 & 100 & 3204 & 0.390 & 0.291 & 0.333 \\
			& LPV (Wood seq. D) 	& 152 & 158 & 103 & 49 & 55 & 3317 & 0.652 & 0.678 & 0.665 \\
			& Eclipsing binaries           & 734 & 651 & 589 & 145 & 62 & 2728 & 0.905 & 0.802 & 0.851 \\
		\hline 
		\multicolumn{2}{r}{}{$N_C=10$}
				& \multicolumn{1}{c}{}{$N_s=3524$}
				& \multicolumn{5}{c}{}{\sc Accuracy} & {\sc Precision$_M$} & {\sc Recall$_M$} & {\sc F1-score$_M$} \\
			\multicolumn{8}{c}{}  {0.749} 
				&  0.503 
				& 0.515 
				&  0.504 
			\\[1em]
	\end{tabular}
	}\end{center}}	
	{\begin{center}{  \setlength{\tabcolsep}{.2cm} \hspace{-1.5cm}
	\begin{tabular}{ c | l | c|c | cccc  | ccc}
		\multicolumn{11}{r}{}{\itbf{Network} $c_{F,meta}$; LSTM; \it best configuration (1); B$_{band}$}\\
		\hline\hline 
		\multirow{12}{*}{\centering{\rotatebox[origin=c]{90}{\itbf{\color{blue!35!black} MACHO - B$_{band}$}}}}
			& \multirow{2}{*}{\sc Classes $\{i\}_{i=1\rightarrow N_C}$}
			& \sc Nb of $\mathbf{Y}_{true}$
			& \sc Nb of  $\mathbf{Y}_{pred}$
			& \multirow{2}{*}{\sc TP$(i)$}	
			& \multirow{2}{*}{\sc FN$(i)$}	
			& \multirow{2}{*}{\sc FP$(i)$}	
			& \multirow{2}{*}{\sc TN$(i)$}
			& \multirow{2}{*}{\sc Precision$(i)$}	
			& \multirow{2}{*}{\sc Recall$(i)$}	
			&\multirow{2}{*}{\sc F1-score$(i)$}	\\
		&&\it (in counts) & \it (in counts)&&&&&&&\\
		\hhline{~----------}
			& RRL (type $ab$) 		& 1430 & 1460 & 1422 & 8 & 38 & 2056 & 0.974 & 0.994 & 0.984 \\
			& RRL (type $c$)  		& 344 & 388 & 316 & 28 & 72 & 3108 & 0.814 & 0.919 & 0.863 \\
			& RRL (type $e$)  		& 60 & 0 & 0 & 60 & 0 & 3464 & -- & -- & --  \\
			& CEP (FU) 			& 229 & 232 & 204 & 25 & 28 & 3267 & 0.879 & 0.891 & 0.885 \\
			& CEP (FO) 			& 133 & 141 & 113 & 20 & 28 & 3363 & 0.801 & 0.850 & 0.825 \\
			& LPV (Wood seq.A) 	& 62 & 65 & 49 & 13 & 16 & 3446 & 0.754 & 0.790 & 0.772 \\
			& LPV (Wood seq. B) 	& 160 & 157 & 131 & 29 & 26 & 3338 & 0.834 & 0.819 & 0.826 \\
			& LPV (Wood seq. C) 	& 220 & 230 & 195 & 25 & 35 & 3269 & 0.848 & 0.886 & 0.867 \\
			& LPV (Wood seq. D) 	& 152 & 175 & 147 & 5 & 28 & 3344 & 0.840 & 0.967 & 0.899 \\
			& Eclipsing binaries           & 734 & 676 & 650 & 84 & 26 & 2764 & 0.962 & 0.886 & 0.922 \\			
		\hline 
			\multicolumn{2}{r}{}{$N_C=10$}
				& \multicolumn{1}{c}{}{$N_s=3524$}
				& \multicolumn{5}{c}{}{\sc Accuracy} & {\sc Precision$_M$} & {\sc Recall$_M$} & {\sc F1-score$_M$} \\
			\multicolumn{8}{c}{}{0.916} & 0.771 & 0.800 & 0.784 \\  
	\end{tabular}
	}\end{center}}
}\end{table}

\begin{table}[htb]{
	\scriptsize
	\caption{
		{ 
			Classification metrics computed on the Test set using the metadata via the classifier module.} 
	}
	\label{table:exp_metrics_metaonly}
	{\begin{center}{  \setlength{\tabcolsep}{.2cm} \hspace{-1.5cm}
	\begin{tabular}{ c | l | c|c | cccc  | ccc}
		\multicolumn{11}{r}{}{\itbf{Network} classifier module (MLP); metadata}\\
		\hline\hline 
		\multirow{12}{*}{\centering{\rotatebox[origin=c]{90}{\itbf{ MACHO - metadata}}}}
			& \multirow{2}{*}{\sc Classes $\{i\}_{i=1\rightarrow N_C}$}
			& \sc Nb of $\mathbf{Y}_{true}$
			& \sc Nb of  $\mathbf{Y}_{pred}$
			& \multirow{2}{*}{\sc TP$(i)$}	
			& \multirow{2}{*}{\sc FN$(i)$}	
			& \multirow{2}{*}{\sc FP$(i)$}	
			& \multirow{2}{*}{\sc TN$(i)$}
			& \multirow{2}{*}{\sc Precision$(i)$}	
			& \multirow{2}{*}{\sc Recall$(i)$}	
			&\multirow{2}{*}{\sc F1-score$(i)$}	\\
		&&\it (in counts) & \it (in counts)&&&&&&&\\
		\hhline{~----------}
			& RRL (type $ab$) 		& 1430 & 1457 & 1414 & 16 & 43 & 2051 & 0.970 & 0.989 & 0.980    \\
			& RRL (type $c$)  		& 344 & 397 & 323 & 21 & 74 & 3106 & 0.814 & 0.939 & 0.872    \\
			& RRL (type $e$)  		& 60 & 0 & 0 & 60 & 0 & 3464 & -- & -- & -- \\
			& CEP (FU) 			& 229 & 254 & 213 & 16 & 41 & 3254 & 0.839 & 0.930 & 0.882    \\
			& CEP (FO) 			& 133 & 114 & 95 & 38 & 19 & 3372 & 0.833 & 0.714 & 0.769    \\
			& LPV (Wood seq. A) 	& 62 & 50 & 42 & 20 & 8 & 3454 & 0.840 & 0.677 & 0.750    \\
			& LPV (Wood seq. B) 	& 160 & 265 & 158 & 2 & 107 & 3257 & 0.596 & 0.988 & 0.744    \\
			& LPV (Wood seq. C) 	& 220 & 0 & 0 & 220 & 0 & 3304 & -- & -- & -- \\
			& LPV (Wood seq. D) 	& 152 & 0 & 0 & 152 & 0 & 3372 & -- & -- & -- \\
			& Eclipsing binaries          & 734 & 987 & 686 & 48 & 301 & 2489 & 0.695 & 0.935 & 0.797    \\
		\hline 
			\multicolumn{2}{r}{}{$N_C=10$}
				& \multicolumn{1}{c}{}{$N_s=3524$}
				& \multicolumn{5}{c}{}{\sc Accuracy} & {\sc Precision$_M$} & {\sc Recall$_M$} & {\sc F1-score$_M$} \\
			\multicolumn{8}{c}{}{0.832}  & 0.559 & 0.617 & 0.579  \\  
	\end{tabular}
	}\end{center}}
}\end{table}


\newpage
\begin{table}[htp!]
	\caption{
			Total loss (weighted MAE for reconstruction and categorical cross-entropy for classification) computed for all 
			trained models across three different 
			datasets (B-band only [top], and two variants of the combination of R- and B-bands [middle and bottom]).
			Solid underlines highlight to the best-performing models associated to minimum loss obtained on the Test set.
			The identifiers \textit{(1)} to \textit{(6)} refer to the hyperparameters set configurations.
			} 	
	\label{table:totloss_perfoAll}
	\hspace{-1cm}
	\begin{minipage}[t]{1\linewidth}
		\includegraphics[width=1.1\textwidth, page=1]{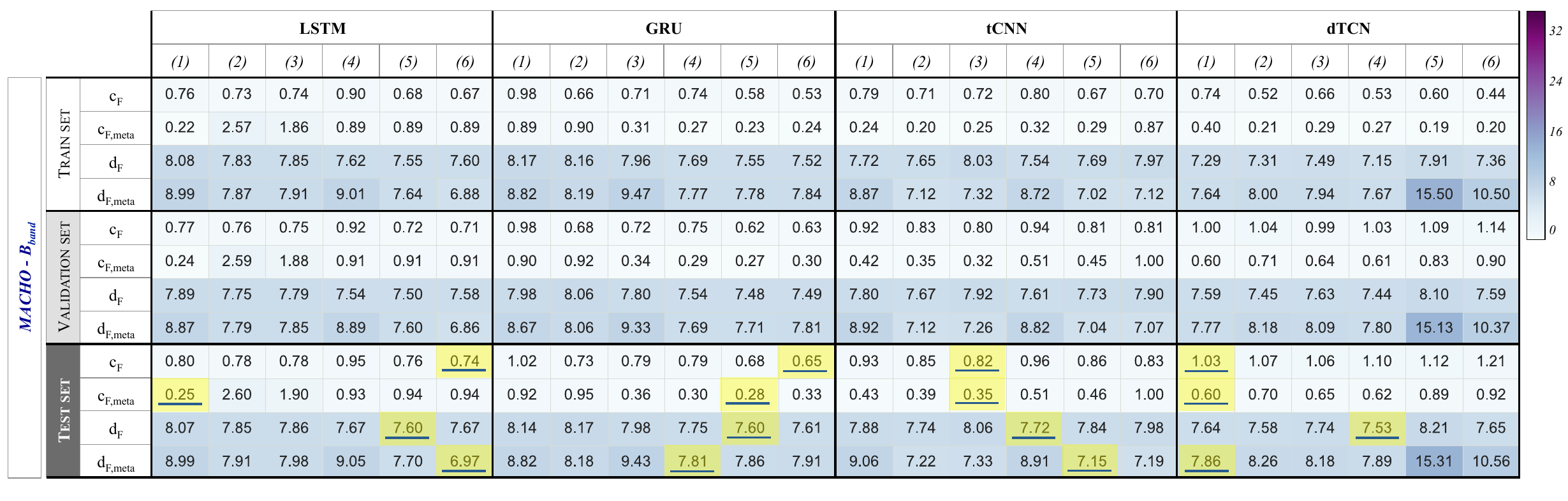}  
		\includegraphics[width=1.1\textwidth, page=2]{result_table_totloss_wMAE.pdf}  
		\includegraphics[width=1.1\textwidth, page=3]{result_table_totloss_wMAE.pdf}  
	\end{minipage}
\end{table}

\begin{table}[htp!]
	\caption{
			Classification accuracy, defined as the fraction of true positives within the sample, computed for all trained 
			models across three different datasets 
			(B-band only [top], and two variants of the combination of R- and B-bands [middle and bottom]). 
			Solid underlines highlight the best-performing models associated to minimum loss obtained on the Test set.
			The identifiers \textit{(1)} to  \textit{(6)} refer to the hyperparameters set configurations.
			} 
	\label{table:classif_perfoAll}	 
	\hspace{-1cm}
	\begin{minipage}[t]{1\linewidth}
		\includegraphics[width=1.1\textwidth, page=1]{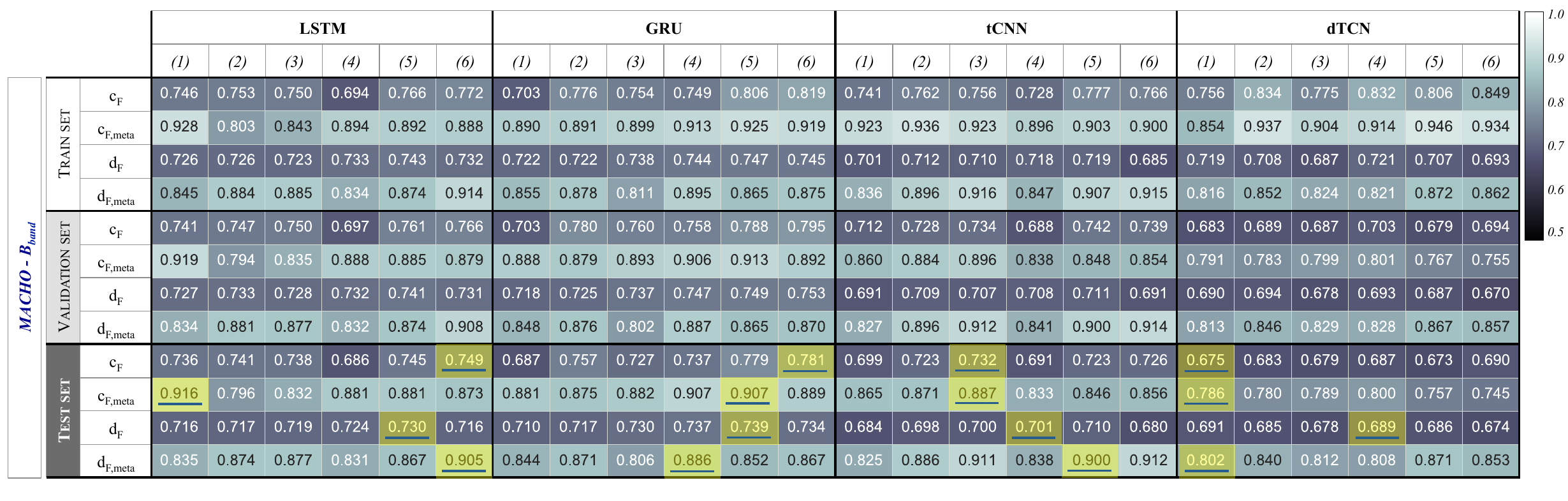}  
		\includegraphics[width=1.1\textwidth, page=2]{result_table_accuracy_wMAE.pdf}  
		\includegraphics[width=1.1\textwidth, page=3]{result_table_accuracy_wMAE.pdf}  
	\end{minipage}
\end{table}

\begin{figure}[htp!]
	\centering
	\includegraphics[width=0.8\textwidth]{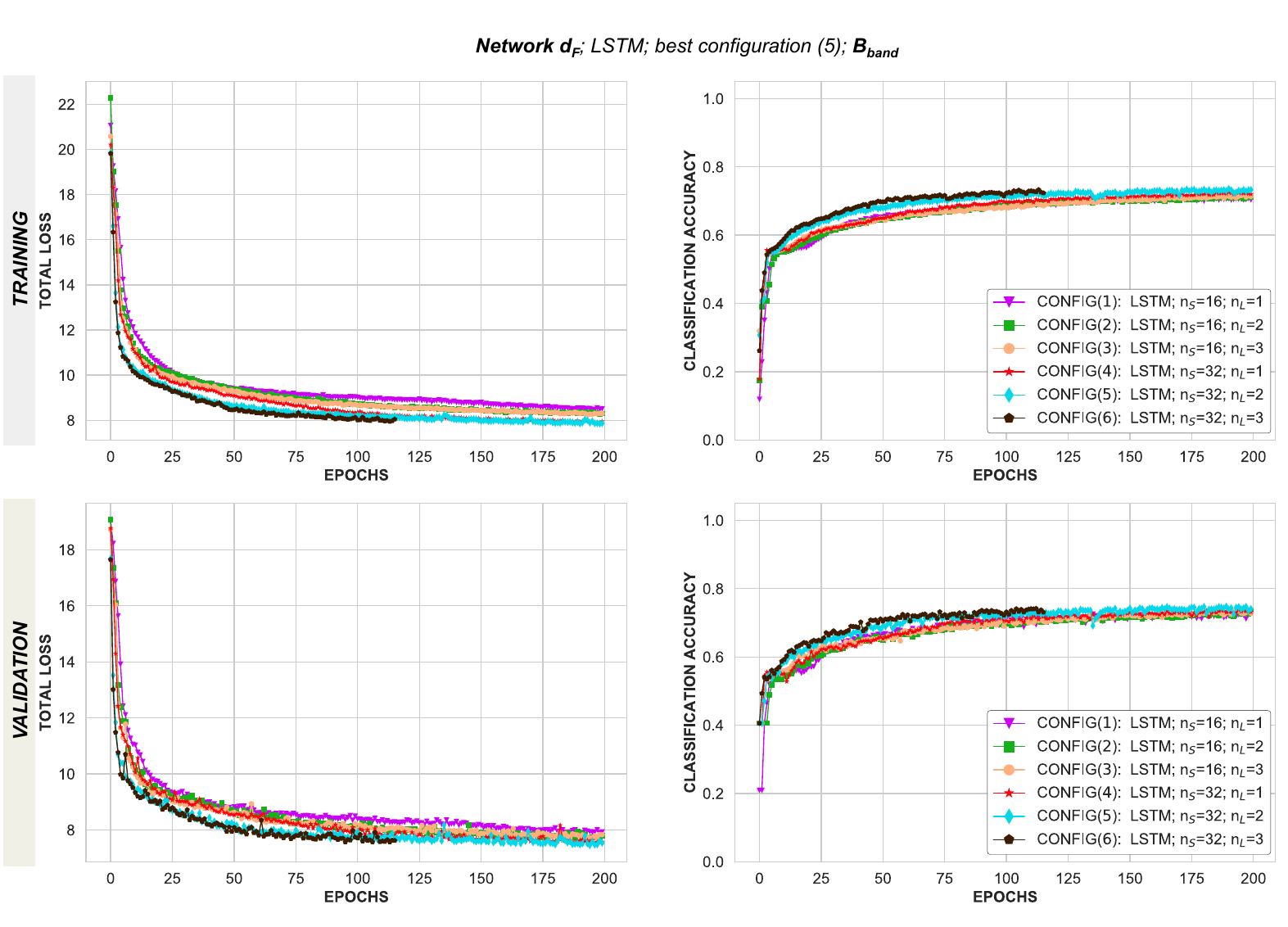}  
	\caption{
			Total loss and classification accuracy for the LSTM composite $d_{F}$ on the B-band.}
	\label{fig:loss_acc_display_lstm_df}
	 \vspace{.25cm}\hrule\vspace{.25cm}
	\includegraphics[width=0.8\textwidth]{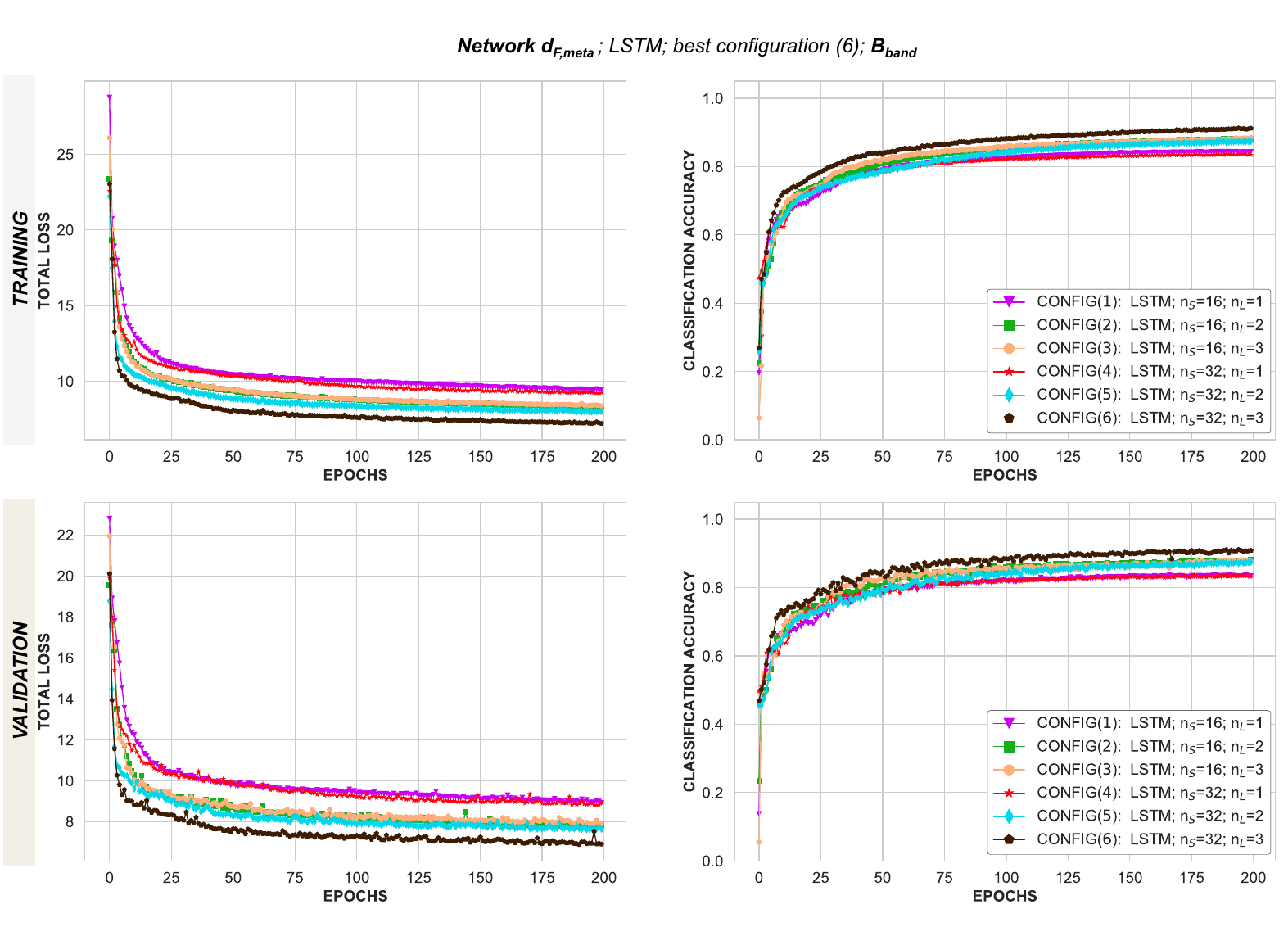} 
	\caption{
			Total loss and classification accuracy for the LSTM composite $d_{F,meta}$ using metadata as a secondary input on the B-band.}
	\label{fig:loss_acc_display_lstm_dfmeta}
\end{figure}

\newpage 
\begin{sidewaysfigure}[htp]
	\centering 
	\begin{minipage}[t]{1\linewidth}
		\hspace{-1cm}
		\includegraphics[width=1.07\textwidth]{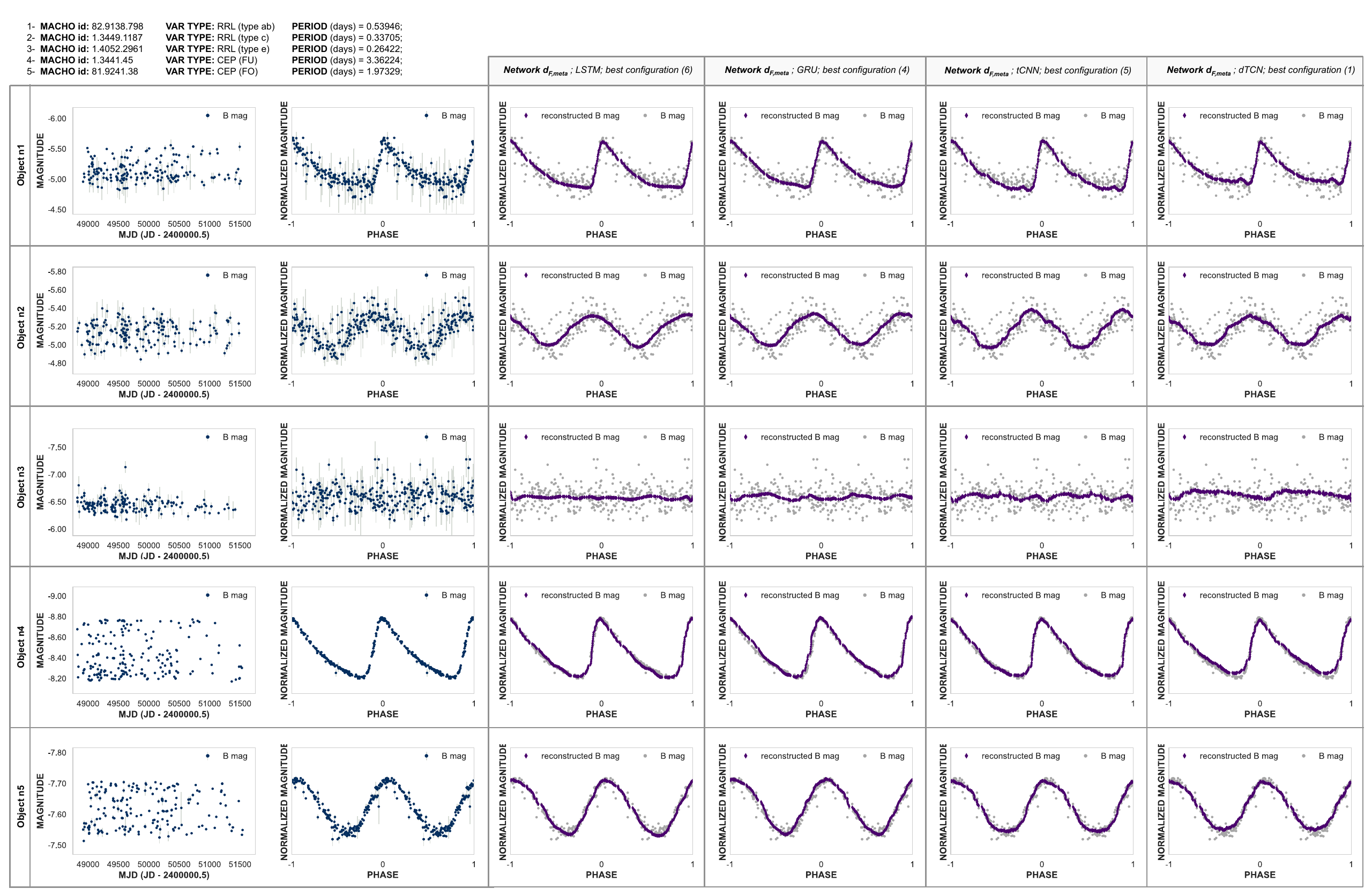}  
	\end{minipage}
	\caption{
			Displays (1) of reconstructed light-curves from the Test set for the best-performing composite $d_{F,meta}$ on 
			the B-band (\textit{left to right}: the best-performing LSTM, GRU, tCNN and dTCN).
			For visualization purposes, the 1$-\sigma$ error measurements of the input data are not displayed in the 
			reconstruction results.}
	\label{fig:reconstruction_displays_set1}
\end{sidewaysfigure}
\newpage
\begin{sidewaysfigure}[htp] 
	\centering 
	\begin{minipage}[t]{1\linewidth}
		\hspace{-1cm}
		\includegraphics[width=1.07\textwidth]{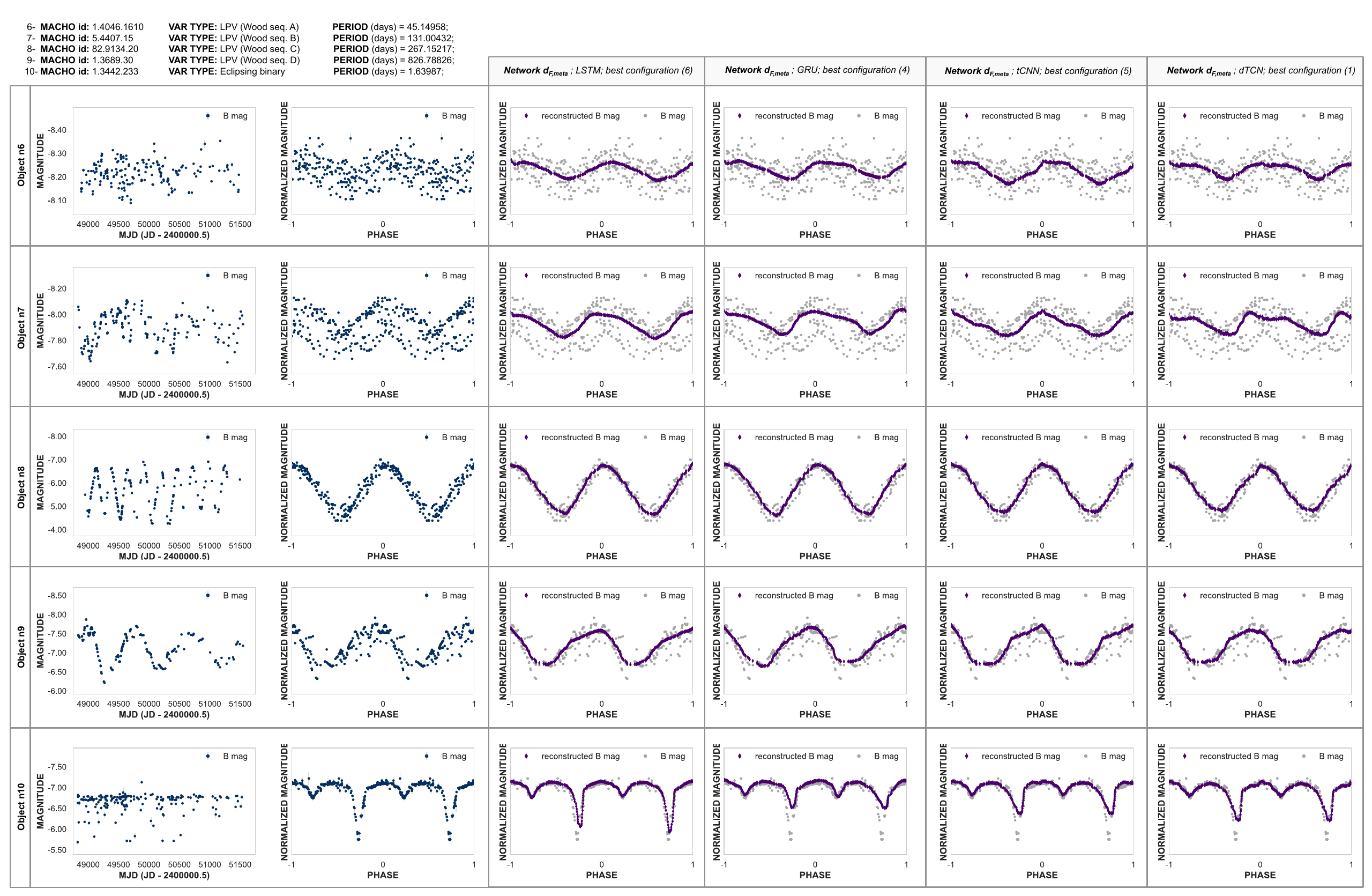}  
	\end{minipage}
	\caption{
			Displays (2) of reconstructed light-curves from the Test set for the best-performing composite $d_{F,meta}$ 
			on the B-band (\textit{left to right}: the best-performing LSTM, GRU, tCNN and dTCN).
			For visualization purposes, the 1$-\sigma$ error measurements of the input data are not displayed in the 
			reconstruction results.}
	\label{fig:reconstruction_displays_set2}
\end{sidewaysfigure}

\newpage
\begin{sidewaysfigure}[htp]
	\centering 
	\includegraphics[width=0.40\textwidth]{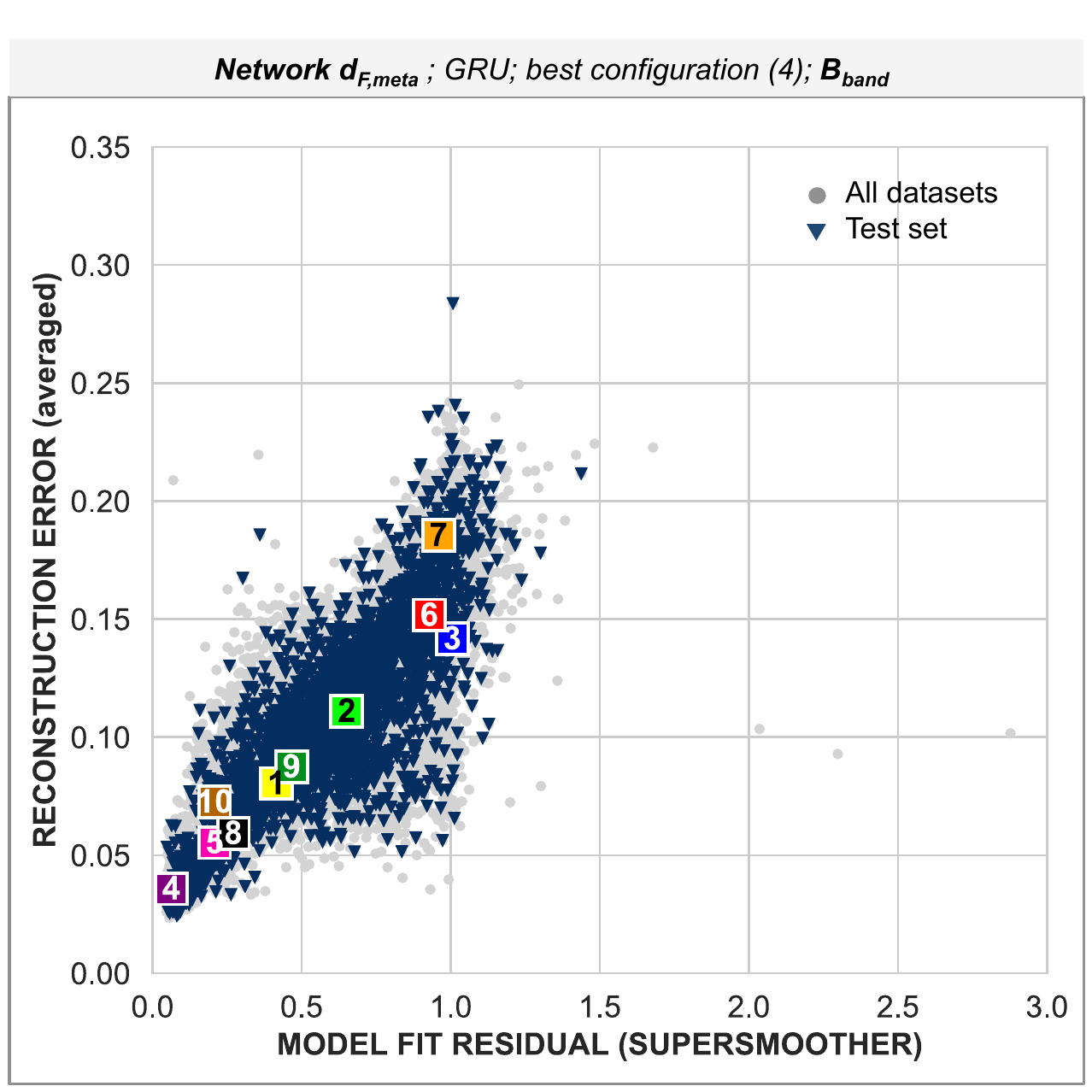}  
	\hspace{.5cm}
	\includegraphics[width=0.40\textwidth]{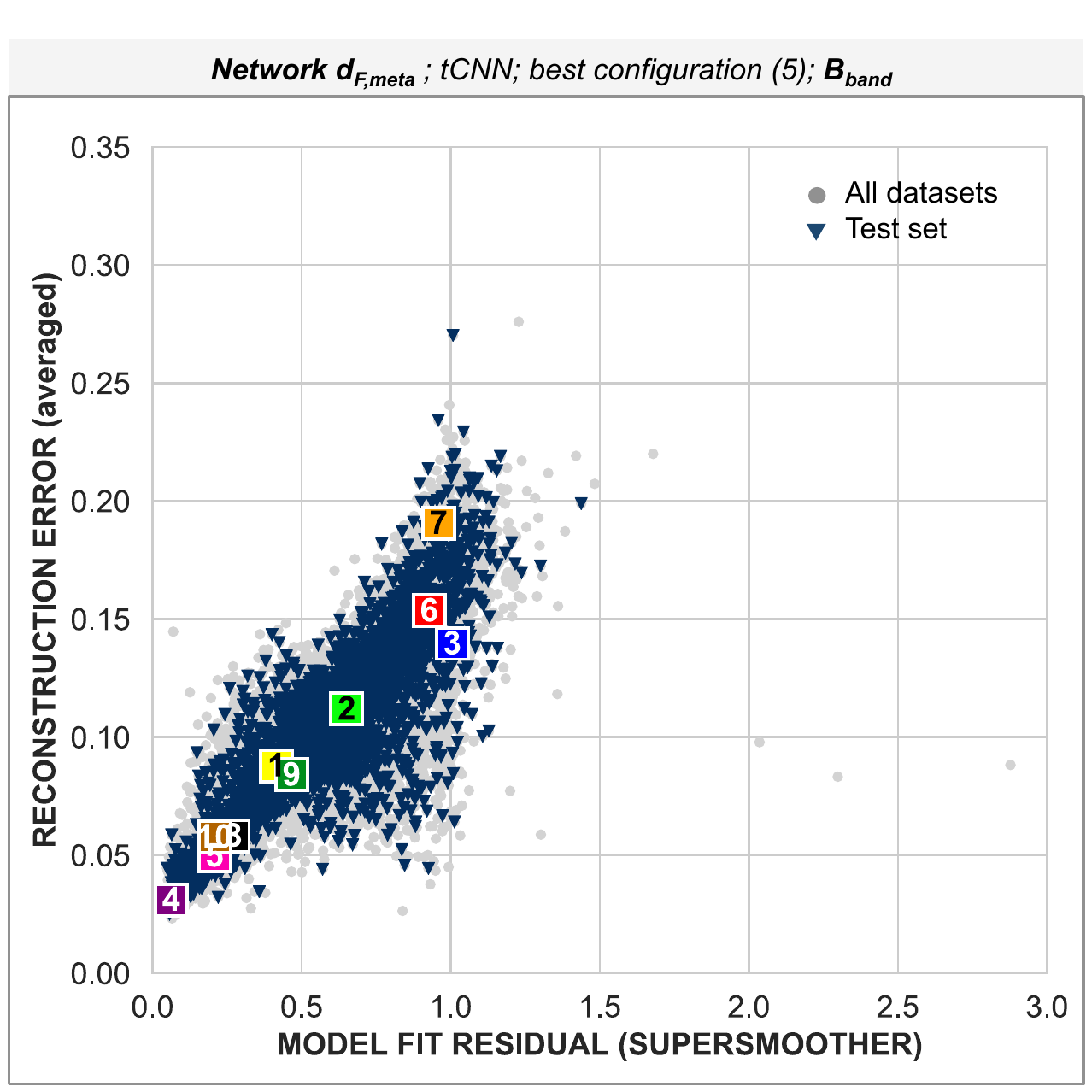}  
	\includegraphics[width=0.40\textwidth]{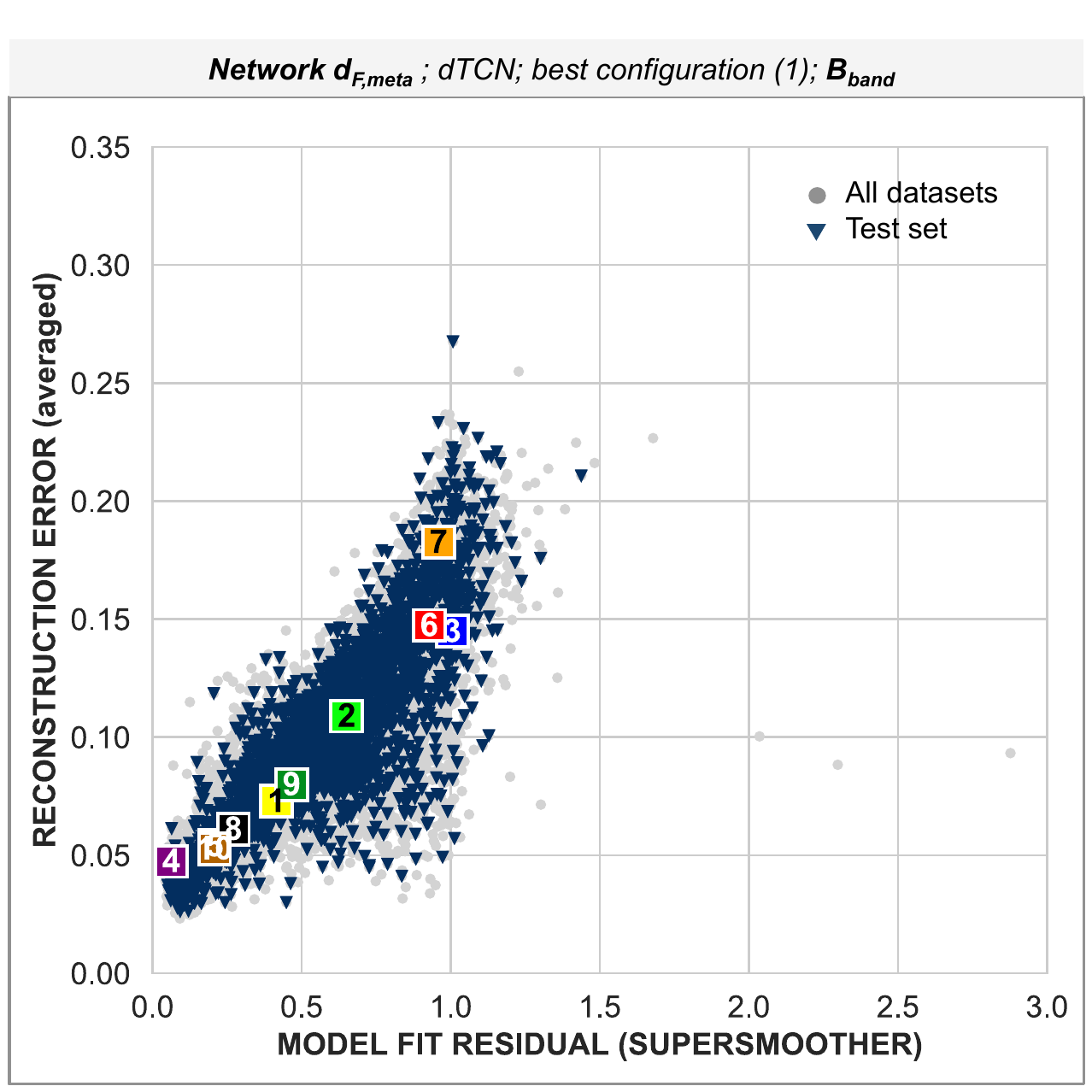}  
	\caption{
			Reconstruction error (MAE) as a function of the model fit residuals from the \texttt{SuperSmoother} algorithm 
			\citep{friedman_variable_1984} for the best-performing GRU, tCNN [top, left to right] and dTCN [bottom] 
			$d_{F,meta}$ on the B-band.
			The highlighted numbers (1 to 10) refer to the subset of selected objects from the Test set used to showcase the 
			reconstruction quality of the autoencoder branch.}
	\label{fig:rec_error_distribution_} 
\end{sidewaysfigure}

\newpage
\begin{figure}[htp]
	\centering
	\includegraphics[width=0.95\textwidth, page=1]{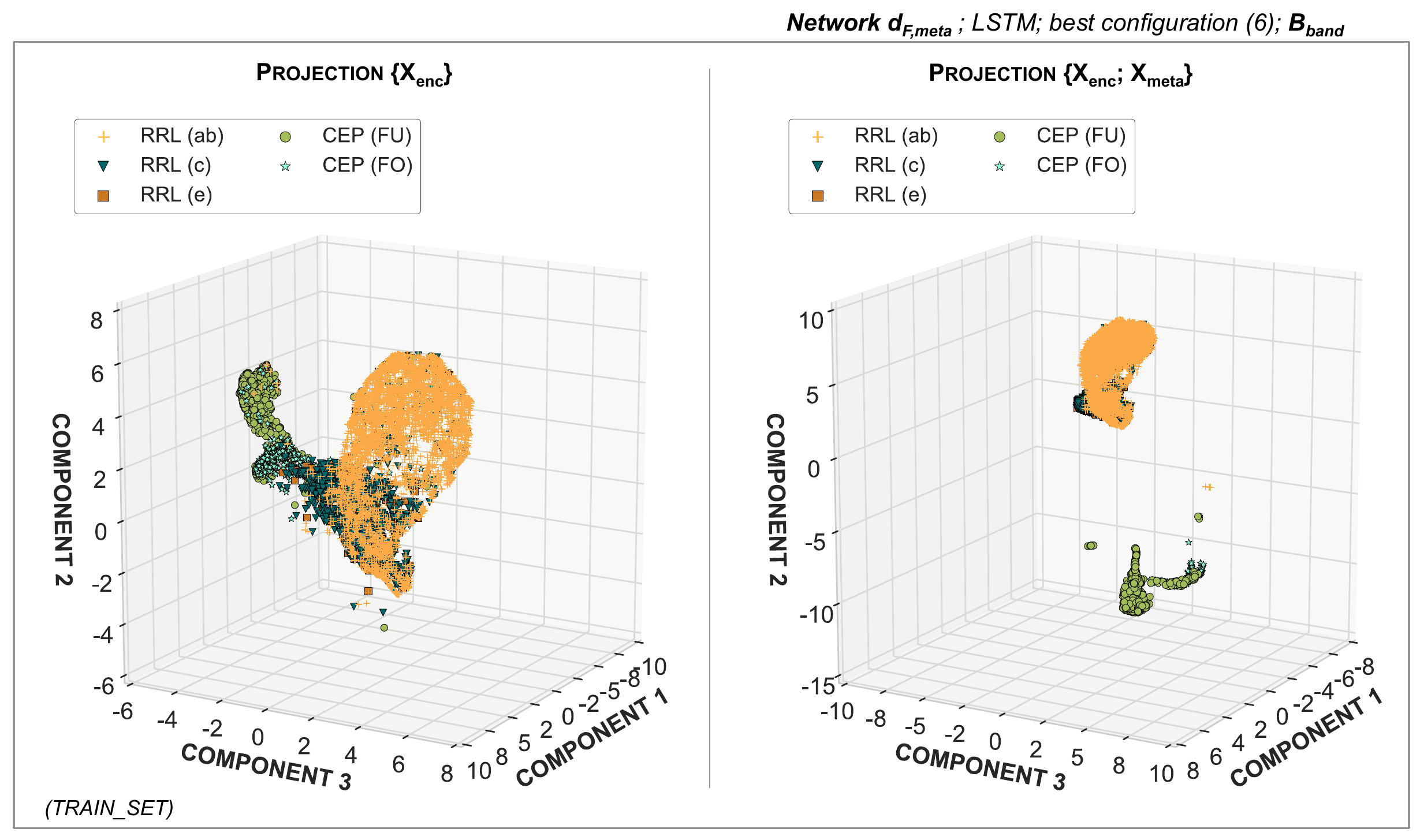}
	\caption{
		3-d representation of encoded features for the short-period pulsators in MACHO. 
		The generated features from the best-performing LSTM {composite network} 
		$d_{F,meta}$ on the B-band are projected into a reduced 3-d representation using the UMAP algorithm.
	}
	\label{fig:umap_projection_short}
	\vspace{1cm}
	\includegraphics[width=0.95\textwidth, page=2]{umap_projection_LSTM_dFmeta_perclass_wMAE.pdf}
	\caption{
		3-d representation of encoded features for the LPVs in MACHO. 
		The generated features from the best-performing LSTM {composite network}
		$d_{F,meta}$ on the B-band are projected into a reduced 3-d representation using the UMAP algorithm.
	}
	\label{fig:umap_projection_long}
\end{figure}

\begin{figure}[htp]
	\centering
	\includegraphics[width=0.95\textwidth, page=3]{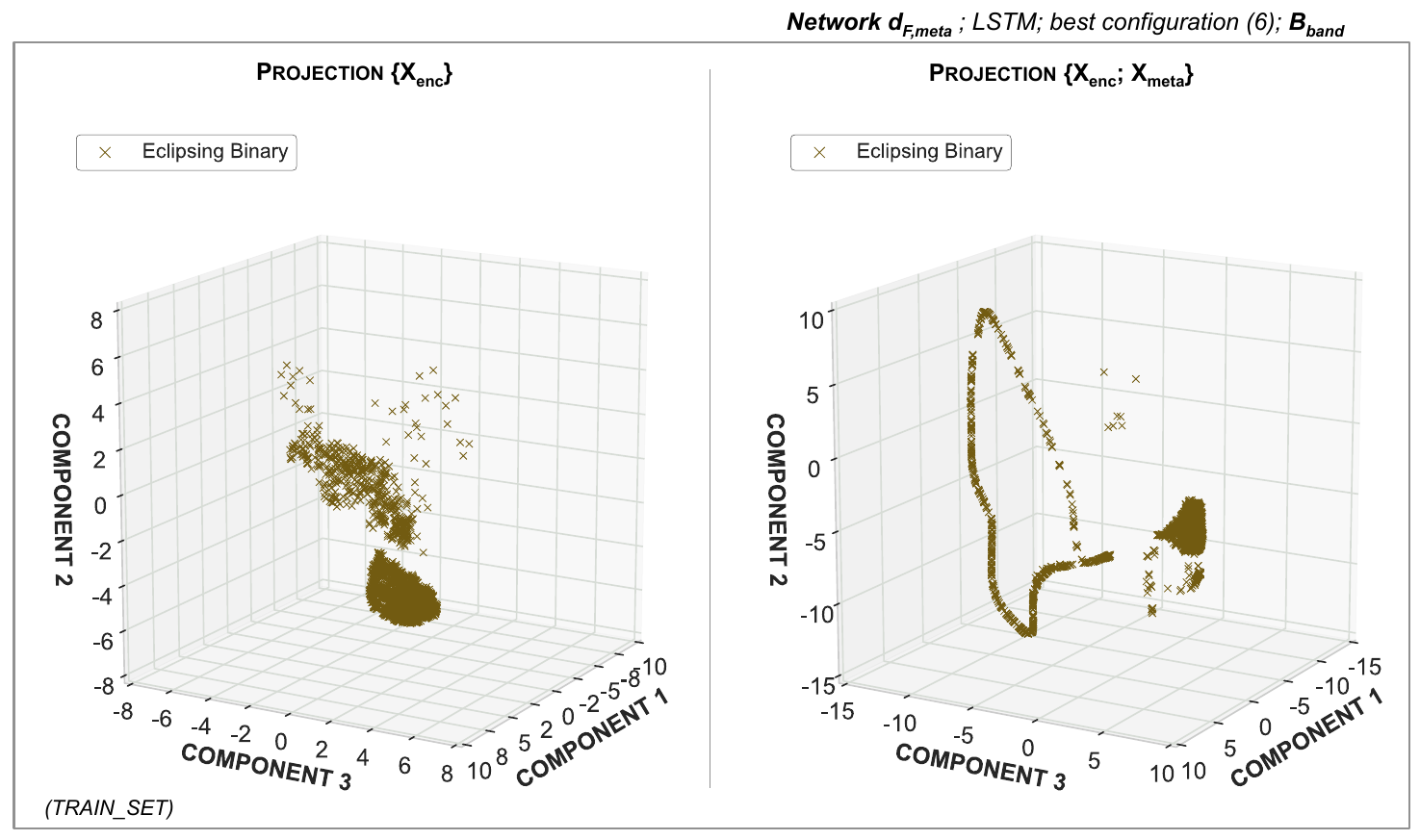}
	\caption{
		3-d representation of encoded features for the eclipsing binaries in MACHO. 
		The generated features from the best-performing LSTM {composite network} 
		$d_{F,meta}$ on the B-band are projected into a reduced 3-d representation using the UMAP algorithm.
	}
	\label{fig:umap_projection_eclipsing}
\end{figure}

\newpage
\begin{figure}[htp]
	\centering
	\includegraphics[width=1.0\textwidth]{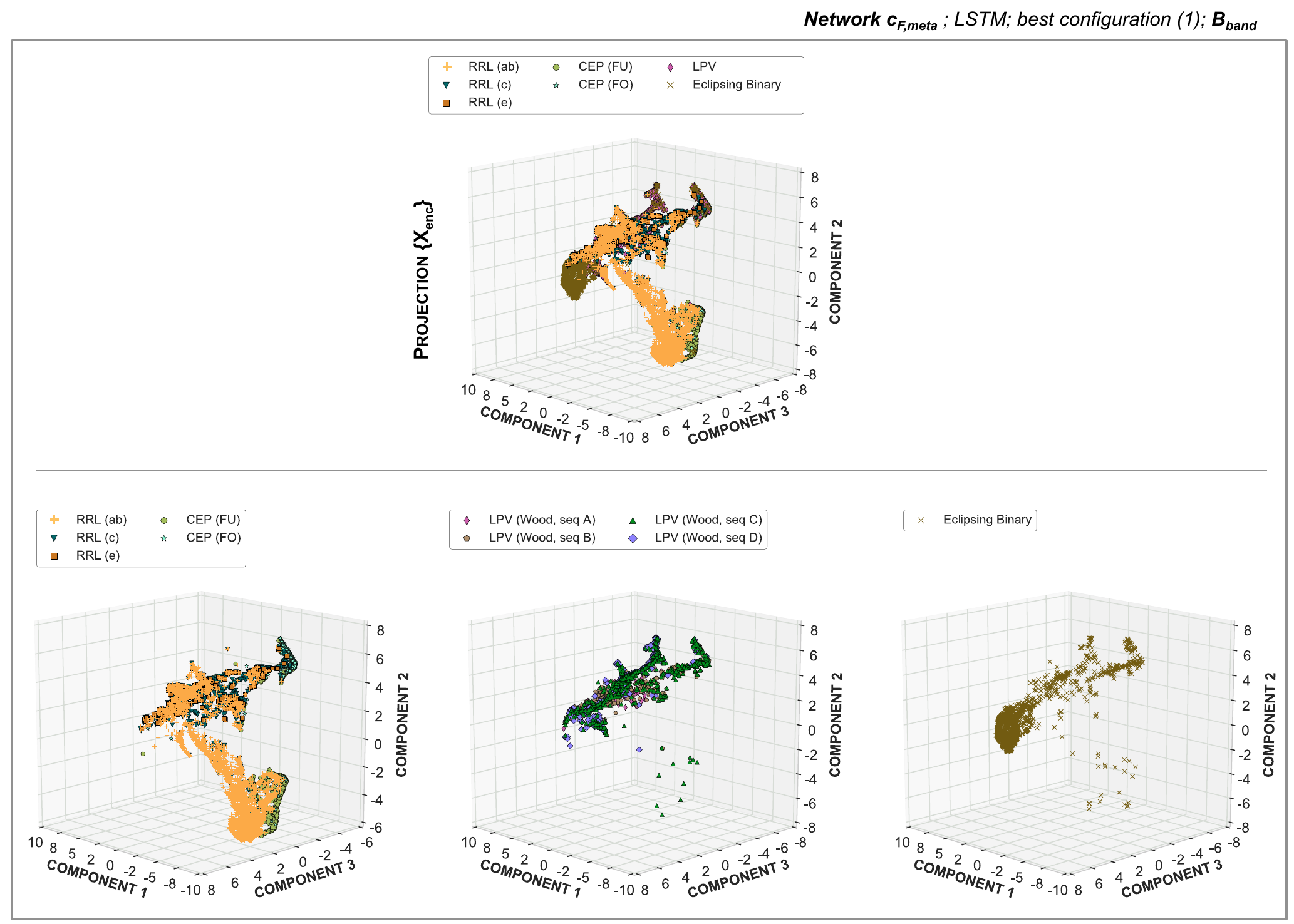}  
	\caption{
		{
			3-d representation of encoded features for the best-performing LSTM direct classifier $c_{F, meta}$ 
			on the B-band. Generated encodings are projected into a reduced 3-d representation using the UMAP 
			algorithm.}
	}
	\label{fig:umap_projection_classifier}
\end{figure}

}

\end{document}